\documentclass[preprint]{aastex63}
\usepackage{xspace}
\usepackage{booktabs}
\usepackage{longtable}
\usepackage{morefloats}
\usepackage{blindtext}
\usepackage{placeins}
\usepackage{makecell}
\usepackage[shortlabels]{enumitem}
\usepackage[caption=false]{subfig}
\usepackage{float}
\usepackage{gensymb}
\usepackage{longtable}

\usepackage{graphicx}
\usepackage{changepage}

\usepackage{color}

\begin{document}


 
\title{On the investigation of the closure relations for Gamma-Ray Bursts observed by Swift in the post-plateau phase and the GRB fundamental plane}
\author{G.P. Srinivasaragavan}
\affiliation{Cahill Center for Astrophysics, California Institute of Technology, 1200 E. California Blvd. Pasadena, CA 91125, USA}
\author{M.G. Dainotti$^{*}$}
\affiliation{Interdisciplinary Theoretical \& Mathematical Science Program,RIKEN (iTHEMS), 2-1 Hirosawa, Wako, Saitama, Japan 351-0198}
\affiliation{Physics Department, Stanford University, 382 Via Pueblo Mall, Stanford, USA}
\affiliation{Space Science Institute, Boulder, Co}
\affiliation{Obserwatorium Astronomiczne, Uniwersytet Jagiello\'nski, ul. Orla 171, 31-501 Krak{\'o}w, Poland.}
\email{*corresponding author=mdainott@stanford.edu, First and second author share the same contribution}
\author{N. Fraija}
\affiliation{Instituto de Astronomia, Universidad Nacional Autonoma de Mexico, Apartado Postal 70264, C.P. 04510, Mexico D.F., Mexico}
\author{X. Hernandez}
\affiliation{Instituto de Astronomia, Universidad Nacional Autonoma de Mexico, Apartado Postal 70264, C.P. 04510, Mexico D.F., Mexico}
\author{S. Nagataki}
\affiliation{RIKEN Cluster for Pioneering Research, Astrophysical Big Bang Laboratory (ABBL), 2-1 Hirosawa, Wako, Saitama, 351-0198, Japan}
\affiliation{Interdisciplinary Theoretical \& Mathematical Science Program,RIKEN (iTHEMS), 2-1 Hirosawa, Wako, Saitama, 351-0198, Japan }
\author{A. Lenart}
\affiliation{Faculty of Physics, Astronomy and Applied Computer Science, Jagiellonian University, Krak{\'o}w, Poland}
\author{L. Bowden}
\affiliation{Astronomy Department, Cornell University, 616-A Space Sciences Building, Ithaca, USA.}
\author{R. Wagner}
\affiliation{Physics Department, The College of New Jersey, 2000 Pennington Rd, Ewing, NJ, USA}

\begin{abstract}
Gamma-ray Bursts (GRBs) are the most explosive phenomena in the universe after the big bang. A large fraction of GRB lightcurves (LCs) shows X-ray plateaus. We perform the most comprehensive analysis of all GRBs (with known and unknown redshifts) with plateau emission observed by The Neil Gehrels Swift Observatory from its launch until 2019 August. We fit 455 LCs showing a plateau and explore whether these LCs follow closure relations, relations between the temporal and spectral indices of the afterglow, corresponding to two distinct astrophysical environments and cooling regimes within the external forward shock (ES) model, and find that the ES model works for the majority of cases. The most favored environments are a constant-density interstellar or wind medium with slow cooling.  We also confirm the existence of the fundamental plane relation between the rest-frame time and luminosity at the end of the plateau emission and the peak prompt luminosity for this enlarged sample, and test this relation on groups corresponding to the astrophysical environments of our known redshift sample. The plane becomes a crucial discriminant corresponding to these environments in terms of the best-fitting parameters and dispersions. Most GRBs for which the closure relations are fulfilled with respect to astrophysical environments have an intrinsic scatter $\sigma$ compatible within 1$\sigma$ of that of the ``Gold" GRBs, a subset of long GRBs with relatively flat plateaus. We also find that GRBs satisfying closure relations indicating a fast cooling regime have a lower $\sigma$ than ever previously found in literature.\end{abstract}


\date{\today}
\section{Introduction}
\label{Intro}
Gamma-ray bursts (GRBs) are short-lived bursts of $\gamma$-ray photons originating from high-energy astrophysical phenomena, and are spectacular events due to their energy emission mechanism. The Neil Gehrels Swift Observatory \citep{Gehrels2004}, launched in 2004 November, has observed GRBs within a wide range of redshifts. More specifically, Swift, with its onboard instruments - the Burst Alert Telescope; (BAT; 15 - 150 keV; \cite{Barthelmy:05}), the X-ray Telescope (XRT; 0.3 - 10 keV;  \cite{2005SSRv..120..165B}) and Ultra-Violet and Optical Telescope (UVOT; 170 - 650 nm; \cite{2005SSRv..120...95R}) -- provides rapid localization of many GRBs and enables fast follow-up of the afterglows in several wavelengths. The afterglow of GRBs is likely due to the external forward shock (ES) where the relativistic ejecta impacts the external medium \citep{Paczynski1993,Katz+97,meszaros97}. It has already been shown that Swift GRB lightcurves (LCs) have more complex features than a simple power law (PL, \cite{Tagliaferri2005,Nousek2006,OBrien2006,Zhang2006,sakamoto07,Zhang2019}. 
\citet{OBrien2006} and \citet{sakamoto07} discovered the existence of a flat part in the X-ray LCs of GRBs, called the plateau, which is present soon after the decaying phase of the prompt emission. The Swift plateaus generally last from hundreds to a few thousands of seconds \citep{Willingale2007}. Physically, this plateau emission has been associated either due to continuous energy injection from the central engine \citep{dai98,rees98,sari2000,zhang2001,Zhang2006,liang2007}, with millisecond newborn spinning neutron stars \citep[e.g.,][]{zhang2001,troja07,dallosso2011,rowlinson2013,rowlinson14,rea15,BeniaminiandMochkovitch2017,Toma2007,Metzger2018,Stratta2018,Fraija2020} or with mass fall-back accretion onto a black hole \citep{Kumar2008,Cannizzo2009,cannizzo2011,Beniamini2017,Metzger2018}. 

In previous literature, some theoretical models ascribe the X-ray plateau to the continuous, long-lasting, energy injection into the ES \citep{zhang2001,zhang06,macfadyen01}. 
The study of the ES emission within the standard fireball model has already been tested in X--rays by  \citet{Willingale2007} through a set of relationships called closure relations \citep{Zhang2004, gao13}. These relations are theoretical relationships between the temporal PL index of the afterglow ($\alpha$) and the spectral index derived from the electron spectral index of synchrotron emission ($\beta$). Each relation indicates different possible astrophysical environments and cooling regimes for GRBs. 
The first study related to the closure relationships dates back to 2004: \citet{Zhang2004} collected the available closure relations (and derived some more) in a comprehensive review. Their review was later used by several authors including \citet{Racusin+09}, whose work on these relations is highly relevant to this current paper, and \citet{gao13}, who has collected the most complete set of closure relations in literature. Here we test an extended set of closure relations taken from \citet{Racusin+09} with a large sample of GRBs observed by Swift from 2005 January until 2019 August, an analysis of 15 yr of observations. 
 
 
GRBs are fascinating events not only for the challenge of understanding their emission mechanism, but also because they are observed at very high redshifts ($z$) up to $z=10$ \citep{cucchiara11}. Thus, they have the potential to be used as standard candles, with known redshifts much larger than the most accredited standard candles, Supernova Ia, with the furthest one observed at $z = 2.26$ \citep{Rodney2015}. Because {\it Swift} has observed GRBs in such a wide range of redshifts, it has contributed enormously to this process of possibly making GRBs standard candles. However, with their prompt luminosities ranging over eight orders of magnitudes, the process of standardizing GRBs has proven to be a challenging task. Classifying GRBs into their type-specific classes has been one of the first steps in starting this process for the afterglow emission, as detailed in \citet{dainotti2010}. The mixing of GRBs of different classes leads to comparing observables and relations between phenomena with fundamentally different intrinsic physics, which is why it is not advisable to standardize GRBs as a whole. Thus, it is better to instead hunt specific classes in order to work with observationally homogeneous samples that could lead to tighter correlations \citep{cardone09,cardone10, dainotti16c,  Dainotti2017c,Dainotti2017a, Dainotti2017b}.

This new information has helped with the search for finding meaningful correlations between intrinsic parameters of GRBs. \cite{dainotti16c} made use of the discovery of the plateau in order to isolate a sub-class of GRBs detailed as the “Gold” class that defines a very tight 3D fundamental plane relation, called the gold fundamental plane (the so-called Dainotti 3D relation). This relation is an extension of the 2D relation given by $L_X-T_X$ (the so-called Dainotti 2D relation) and previously discovered by \cite{Dainotti2008, dainotti2010, dainotti11a, Dainotti2013a,Dainotti2015b, Dainotti15a,Dainotti2017a}. The ``Gold" class presents the smallest intrinsic scatter among all of the subclasses, showing it has the highest potential to be used as a standard candle. In this paper, we define a slightly different ``Gold" class, called the ``Gold 2", to see if the tight relations found in \citet{Dainotti2017a} continue to hold despite using a different fitting of prompt spectral parameters, a cutoff power law (CPL) model when possible rather than a PL model, and with the addition of two extra years of GRB measurements. Furthermore, to the best of our knowledge, there has not been an analysis done on this fundamental plane relation with respect to the astrophysical environments of GRBs derived from their fulfillment of the aforementioned theoretical closure relations, which is another main aim of our paper. 

\begin{figure}[h!]
\begin{center}
\includegraphics[width=0.8\hsize,height=0.65\textwidth,angle=0,clip]{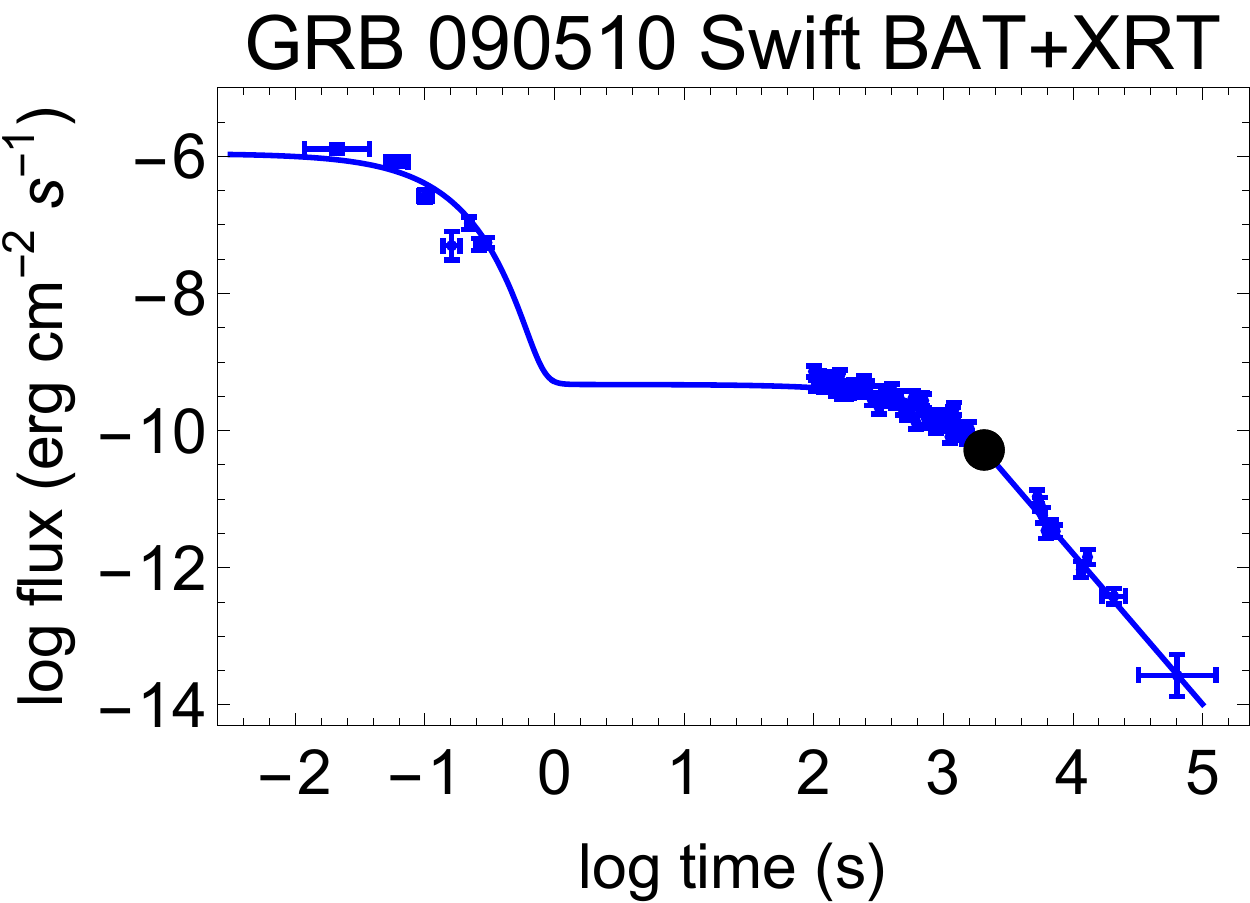}
\caption{Flux vs. time in the observer frame of GRB 090510 fitted by the W07 model defined in Equation \ref{W07eq} with the blue solid line by employing the BAT+XRT data. The black dot shows the best-fit parameters of the end time of the plateau and its respective flux within the same model. \label{fig1}}
\end{center}
\end{figure}

In the current analysis, we investigate the following questions:

\begin{enumerate}

\item How many GRBs observed by Swift showing a plateau (similar to the one presented in Figure \ref{fig1}) obey the closure relations in the post-plateau phase, and what can we infer about their astrophysical environments?\newline

\item Does the 3D fundamental plane relation hold with the addition of two extra years of GRB measurements?\newline

\item Do the closure relations reveal new fundamental planes which show tighter correlations than the ones already found?

\end{enumerate}

The paper has the following structure: in \S \ref{Swift Sample} we present the sample selection for the Swift GRBs, in \S \ref{Methodology} we explain the details of the fitting with the \citet{Willingale2007} model (W07 model) and the spectral data analysis, along with the computation of the luminosity in X-rays, in \S \ref{the closure relations} we present the X-ray closure relations, in \S \ref{Interpretation} their interpretations, in \S \ref{Results from SWIFT} we show the 3D correlation, compare and contrast it with a new definition of the ``Gold" sample, and test the 3D correlation for groups with respect to their astrophysical environments and cooling regimes, and in \S \ref{conclusion} we summarize the analysis and the results.
\begin{center}
\section{Sample Selection}
\label{Swift Sample}
\end{center}

We analyze 222 GRBs detected by {\it Swift} from 2005 January  up to 2019 August with known redshifts, and 233 GRBs from the same time range with unknown redshifts, giving us a total sample of 455 GRBs. All these GRBs have a well-defined plateau in the afterglow phase. The LCs have been downloaded from the {\it Swift} webpage repository\footnote{http://www.swift.ac.uk/burst\_analyser}, and have a signal to noise ratio of 4:1 with the {\it Swift} XRT bandpass $(E_{\mathrm{min}},E_{\mathrm{max}})$ = (0.3,10) keV. We then fit these GRBs according to the phenomenological W07 model. These GRBs all satisfy the fitting procedure and satisfy the Avni $\chi^2$ prescriptions (see the XSPEC manual)\footnote{http://heasarc.nasa.gov/xanadu/xspec/manual/XspecSpectralFitting.html} at the 1$\sigma$ confidence interval. The 222 GRBs with known redshifts have their spectroscopic or photometric redshifts available through \citet{xiao09}, on the Greiner webpage \footnote{http://www.mpe.mpg.de/~jcg/grbgen.html}, and have a known redshift range of $0.0331 \leq z \leq 9.4$. Due to the fitting, each GRB has an associated time, $T_a$, and flux, $F_a$, at the end of the plateau phase, as well as a PL index for the temporal evolution of the afterglow after the plateau, $\alpha$.

We divide the 222 GRBs with redshift in accordance to \citet{Dainotti2017a} into their respective subclasses: Short (sGRBs), Short with Extended Emission (SEE), Supernovae (SNe), X-ray Flashes (XRFs), Ultra Long (UL), and Long (lGRBs). sGRBs have $T_{90} \leq 2$s \citep{kouveliotou93}. Some of these sGRBs also have an extended emission hereafter called the SEE category \citep{Norris:06, Levan2007, Norris2010}. The GRB-SNe is a class of GRBs that has an associated supernova which has been clearly detected. XRFs are peculiar GRBs with an X-ray fluence (2 - 30 keV) $>$ $\gamma$- ray fluence (30 - 400 keV). GRBs with $T_{90} \geq 1000$ s are categorized as UL \citep{Nakauchi2013, Stratta2013, levan14}.
Finally, all GRBs with $T_{90} > 2s$ that do not fall into any of the other classes are categorized as purely lGRBs, with the aims of creating a single observationally homogeneous class.
\newline
\begin{center}
\section{Methodology}
\label{Methodology}
\end{center}
We fit the {\it Swift} LCs with the W07 model, which is able to reveal the presence of a plateau and is versatile enough to fit both the prompt and afterglow LCs.
Following the phenomenological W07 model, each function $f_{i}(t)$ can be written as:
\begin{equation}
f_i(t) = \left \{
\begin{array}{ll}
\displaystyle{F_i \exp{\left ( \alpha_i \left( 1 - \frac{t}{T_i} \right) \right )} \exp{\left (
- \frac{\tau_i}{t} \right )}} & {\rm for} \ \ t < T_i \\
~ & ~ \\
\displaystyle{F_i \left ( \frac{t}{T_i} \right )^{-\alpha_i}
\exp{\left ( - \frac{\tau_i}{t} \right )}} & {\rm for} \ \ t \ge T_i \,, \\
\end{array}
\right .
\label{W07eq}
\end{equation}

\noindent which contains four parameters for the prompt emission ($T_p$,$F_p$,$\alpha_p$,$\tau_p$), and four parameters for the afterglow emission ($T_a$,$F_a$,$\alpha_a$,$\tau_a$), $\alpha_i$ is the late time PL decay index, $\tau_i$ is the initial rise timescale and $T_i$ gives the plateau duration, which is well defined when $T_i \gg \tau_i$. The subscript $i=(p,a)$ denotes either the prompt emission, $p$, or the afterglow emission, $a$. When there is a paucity of data we set $\tau_i=0$ or $T_i=T_p$, where $T_p$ is the end of the prompt emission. 
 
Regarding the plateau in X-rays, we compute the rest-frame luminosity, $L_a$, in units of erg s$^{-1}$ at the end of the afterglow plateau as follows:
\begin{equation}
L_a = 4 \pi D_L^2(z) F_a(E_{\mathrm{min}}, E_{\mathrm{max}}, T_a) * K \, ,
\label{equplateaulum}
\end{equation}
where $D_L$ is the distance luminosity calculated assuming a flat $\Lambda$CDM cosmological model with $\Omega_m= 0.3$ and $H_0 = 0.7$ km s$^{-1}$ Mpc$^{-1}$; $F_a$ is the flux given at the end of the plateau determined by the fitting according to the W07 model; ($E_{\mathrm{min}}$,$E_{\mathrm{max}}$) is the given energy band; and {\it K} is the correction factor that accounts for cosmic expansion, which is given by following \citet{Bloom2001}:
\begin{equation}
K=\frac{\int_{0.3/(1+z)}^{10/(1+z)} \Phi(E)dE}{\int_{0.3}^{10} \Phi(E) dE} \, ,
\label{kcorrection}
\end{equation}
where $\Phi(E)$ is the functional form for the spectrum, which, in our case is either a simple PL or a CPL. 
Given the definition of the {\it K}-correction, the energy band depends on the redshift of the GRB. Since the energy band of XRT is narrow, we also have added the calculation of the bolometric luminosity for which the {\it K}-correction is usually computed over the total energy range between 1 and $10^4$ keV, see \citet{schaefer2007}:
\begin{equation}
K=\frac{\int_{1/(1+z)}^{10^4/(1+z)} \Phi(E)dE}{\int_{1}^{10^4} \Phi(E) dE} \, .
\label{kcorrection}
\end{equation}
A comparison of the results with the bolometric and the K-corrected luminosities in the XRT energy range is given in \S \ref{Results from SWIFT}.

We also calculate a time-sliced spectrum of each GRB from the time at the end of the plateau $T_a$ until the end of the LC, T$_{end}$. This analysis has been performed making use of the {\it Swift} BAT+XRT online repository\footnote{https://www.swift.ac.uk/xrt\_spectra/} \citep{evans09}. In our calculations, the photon index is calculated by first taking values from the windowed timing mode and photon counting mode of the XRT. We then average the two values and propagate their errors. However, if one of the indices does not exist, has larger error bars than the index itself, or has an extremely high value (a photon index greater than $6$), we do not consider that index for the analysis. This way of computing the photon indices leads the majority of them to be consistent within 1$\sigma$. This differs from the calculation of spectral parameters from \citet{dainotti16c} and \citet{Dainotti2017a}, where only the photon counting mode was taken into account. The spectral index $\beta$, which is used in calculating the closure relations detailed in \S \ref{the closure relations}, is the photon index subtracted by one.

The peak prompt luminosity L$_{peak}$, also in units of  erg \, s$^{-1}$, is computed over a 1 s time interval, with the exception of GRBs 150821A and 170405A, whose luminosities are computed over a full time-averaged interval due to a lack of a 1 s peak flux. L$_{peak}$ is computed as:
\begin{equation}
L_{\mathrm{peak}} = 4 \pi D_L^2(z) F_{\mathrm{peak}}(E_{\mathrm{min}}, E_{\mathrm{max}}, T_X) * K \,,
\label{equpeakluminosity}
\end{equation}
where F$_{peak}$ is the measured energy flux over the 1 s interval (erg cm$^{-2}$ s$^{-1}$). \citet{Dainotti2017a} considered GRBs whose spectrum computed at 1 s have a smaller $\chi^2$ value for a simple PL fit rather than a CPL. The difference between the two calculations lies in their K corrections, due to the difference in the functional forms of the spectrum. Here, we instead always use a CPL fit when necessary parameters are available, since the criterion presented in \citet{Sakamoto+11} states that if the $\Delta \chi^2$ difference between the PL and CPL fittings is less than 6, the fitting results are equivalent. In total, there are 65 Swift GRBs with known redshifts fitted with a CPL, and 157 with a PL. To check if this new procedure leads to an increase in the errors of luminosities, we also check the scatter of the fundamental plane relation, and find that it slightly increases (more details are presented in \S \ref{Results from SWIFT}).
\newline
\subsection{The definition of the ``Gold" and ``Gold 2" samples}
The ``Gold 2" class is defined here for the first time in literature, and differs from the ``Gold" in \citet{dainotti16c,Dainotti2017a}. The ``Gold" must have at least five points in the beginning of the plateau and a plateau angle of $< 41\degree$, while the ``Gold 2" considers LCs that have at least one data point in the beginning of the plateau and a plateau angle of $< 41\degree$, allowing for a larger sample size. We here stress that the identification of both the ``Gold" and ``Gold 2" samples rely on a robust phenomenological analysis on the number of data points shown in the beginning of the plateau. The ``Gold" sample classes are created from lGRBs and on the analysis of the LCs, in order to check if the correlations found considering these different samples are more robust using a phenomenological approach based on the LCs rather than an analysis based on type-specific classes. We show explicitly in \S \ref{Results from SWIFT} that the ``Gold" classes are not due to selection effects because of the sample size.
\FloatBarrier
\section{The closure relations}\label{the closure relations}
The closure relations are relationships between the temporal and spectral indices, respectively ($\alpha$ and $\beta$), that test the effectiveness of the ES fireball model \citep{Cavallo,g86,paczynski86}, assuming that synchrotron radiation is the dominant mechanism in the afterglow. The parameter $\alpha$ is taken from the W07 fitting for the afterglow detailed in \S \ref{Methodology}, while the parameter $\beta$ is the spectral index (details on its calculation are presented in \S \ref{Methodology}). Here, similarly to \citet{Racusin+09}, we consider LCs divided in several segments that follow a similar form as described by \citet{Zhang2006} and \citet{Nousek2006}. 

According to \citet{Racusin+09} these four segments are: I: the initial steep decay phase, which is attributed to the high-latitude emission or the curvature effect for the majority of the time \citep{Kumar+00,Qin2004,liang06,zhang07b}; II: the plateau phase whose origin and features have already been discussed in the introduction; III: the normal decay phase occurring due to the deceleration of an adiabatic fireball \citep{m02,Zhang2006}; IV: the post-jet break phase \citep{Rhoads99,sari99,m02,piran04}. Flares are seen in around one-third of all Swift GRB X-ray afterglows during any phase ($I-IV$), and may stem from the sporadic emission from the central engine \citep{Burrows05,Zhang2006,Falcone2007}.
In our analysis we only consider phase III for the LCs that present plateau emission. We also include LCs with flaring activities, since we manually remove flares in our fits, thereby not influencing the determination of the $\alpha$ or $\beta$ parameters. 

The closure relations assume that breaks in the LCs are sharp, when in reality they are often smooth. However, this simplified assumption is necessary because smooth spectral breaks are very difficult to measure. They are derived through assuming that $F_{\nu} \propto t^{-\alpha}\nu^{-\beta}$, where $F_{\nu}$ is the flux seen at a particular frequency \citep{Sari+98, Granot+02}, and both $\alpha$ and $\beta$ are related to the spectral index $p$ of the electron distribution, varying depending on the different astrophysical environments GRBs originate from. The electron spectral index is related to the spectral index $\beta$ used in the closure relations through $\beta = (p-1)/2$ for $\nu_m < \nu < \nu_c$, and $\beta = p/2$ for $(\nu_m, \nu_c) < \nu$ where $\nu_m$ and $\nu_c$ are the characteristic and the cooling spectral breaks of the synchrotron emission frequencies \citep{Sari+98,Zhang2004,zhang06, Racusin+09}.
In the ES model, the electron distribution is usually described through a PL function $\frac{dn_e}{d\gamma_e} \propto \gamma_e^{-p}$, where $p$ is the spectral index and $\gamma_e$ is the electron Lorentz factor, which must be greater than $\gamma_m.$ The term is represented as  
 \begin{equation}
     \gamma_m \simeq 610\epsilon_e\gamma \, ,
 \end{equation}
 and indicates the minimum Lorentz factor needed to create a PL distribution of electrons \citep{Sari+98}. In this paper, we specifically investigate  a subset of cases given in \citet{Racusin+09}. It is worth noting that differently from \citet{Racusin+09}, we do not consider jet models in the ES, since these relations are only relevant when the post-jet-break phase occurs (Phase IV). Indeed, in order to apply such models, the segment should follow a ``normal decay phase" as defined by \citet{zhang06}. The transition from the normal decay phase (Phase III) to the post-jet-break phase (Phase IV) is where the jet break occurs. In rare cases, a plateau phase (Phase II) may be followed immediately by Phase IV. In this scenario, one must invoke that the energy injection ending time coincides with the jet break time. We also do not consider energy injection into the ES for the relations that we test, and focus on determining the GRB during Phase III of its LC, and whether it is either in a constant-density ISM or wind environment with slow or fast cooling during this phase.
 
We here stress that though we test the closure relations after the end of the plateau emission (Phase III), we are aware of the importance of testing the closure relations for the plateau phase (Phase II) as done by \citet{lu2014} for lGRBs,  by \citet{lu2015} for sGRBs, and by \citet{wang15} for both groups. However, this analysis is out of the scope of the current paper, and we plan to perform an analogous analysis of the closure relations using Phase II of our LCs in a forthcoming paper.
 
Below we describe the environments and regimes we test:

\begin{enumerate}
\item Early GRB models that assumed the relativistic ejecta from the blast wave expanding into a constant-density ($\propto R^{0}$) ISM \citep{Sari+98} were highly compatible with LC observations up to the early 2000s. However, it was determined afterwards that some GRBs have massive star progenitors, and may be the result of core collapse supernovae, which ascertains that the relativistic ejecta expands into the stellar wind environment ($\propto R^{-2}$) of the progenitor source \citep{Chevalier+00}. 

\item In both these environments, the cooling regime can be either fast or slow. The calculations of the closure relations depend on the particle energy distribution, in particular, whether electrons have gone through significant cooling \citep{Sari+98}. Electrons are significantly cooled via synchrotron radiation if they have a Lorentz factor $\gamma_e > \gamma_c$, where $\gamma_c$ is the critical Lorentz factor where synchrotron cooling becomes significant. Therefore, two different regimes exist in relation to $\gamma_m$. When $\gamma_m > \gamma_c$, all electrons in the shocked ejecta will be able to cool down to $\gamma_c$. This regime is called the fast cooling regime. In contrast, if $\gamma_m < \gamma_c$, then only some electrons whose $\gamma_e > \gamma_c$ will be able to cool. This leaves the electrons in the region where $\gamma_m < \gamma_e < \gamma_c$ unaffected.  This region of the spectrum is where the majority of electrons in the shock lies, and they therefore will not be able to cool in a given $t$, leading to the slow cooling regime \citep{Sari+98}.
\end{enumerate}

In order to answer question 1) from \S\ref{Intro} and check the validity of the ES model and infer features of the GRBs' astrophysical environments, we determine how many GRBs in our sample composed of 15 yr of Swift data with both known and unknown redshifts follow a set of closure relations. We first group GRBs into their frequency ($\nu$) range (see Table \ref{CR}), through recreating  the top panels of Figure 3 in \citet{Uhm13}, which are theoretical models of the afterglow spectra assuming the ES model (without taking into account curvature effects or an energy injection mechanism), at different observational times and cooling regimes.  We convert the flux from units of mJy in Figure 3 in \citet{Uhm13} to $ergs\, cm^{-2} \, s^{-1}$. We check the regime during which $F_a$ (derived from our fitting procedures) is found, and then categorize every GRB into either $\nu_m < \nu < \nu_c$ or $\nu > \nu_c$ for slow cooling, or $\nu_c < \nu < \nu_m$ or $\nu > \nu_m$ for fast cooling. Then, through using the relations between $p$ and $\beta$ depending on the regime we are in, we determine whether $p>2$ or $1 < p < 2$ for each GRB.

\begin{figure}[h!]
\centering
{
  \includegraphics[width=0.45\columnwidth]{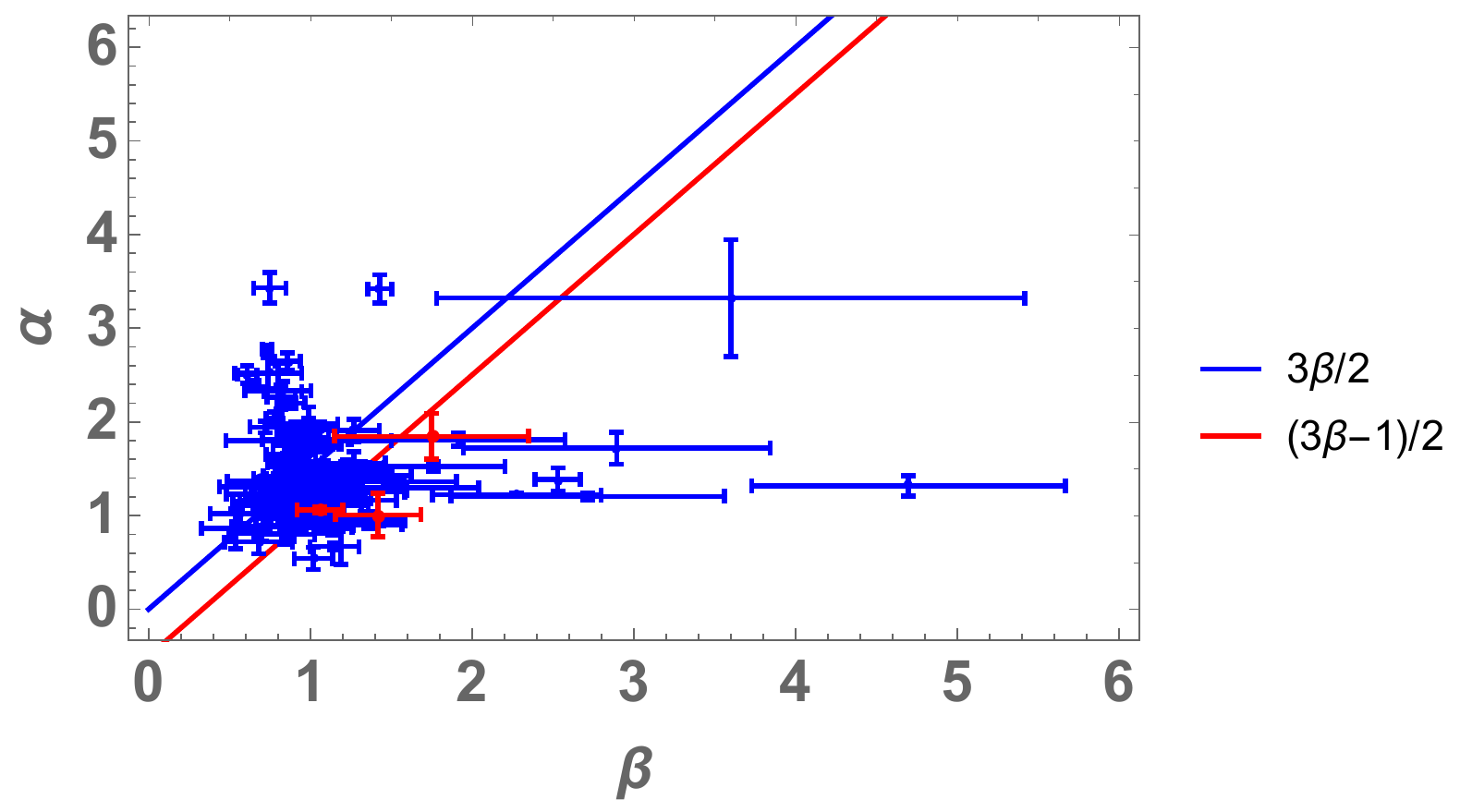}
  \label{fig:sub1}
}\qquad
{
  \includegraphics[width=0.45\columnwidth]{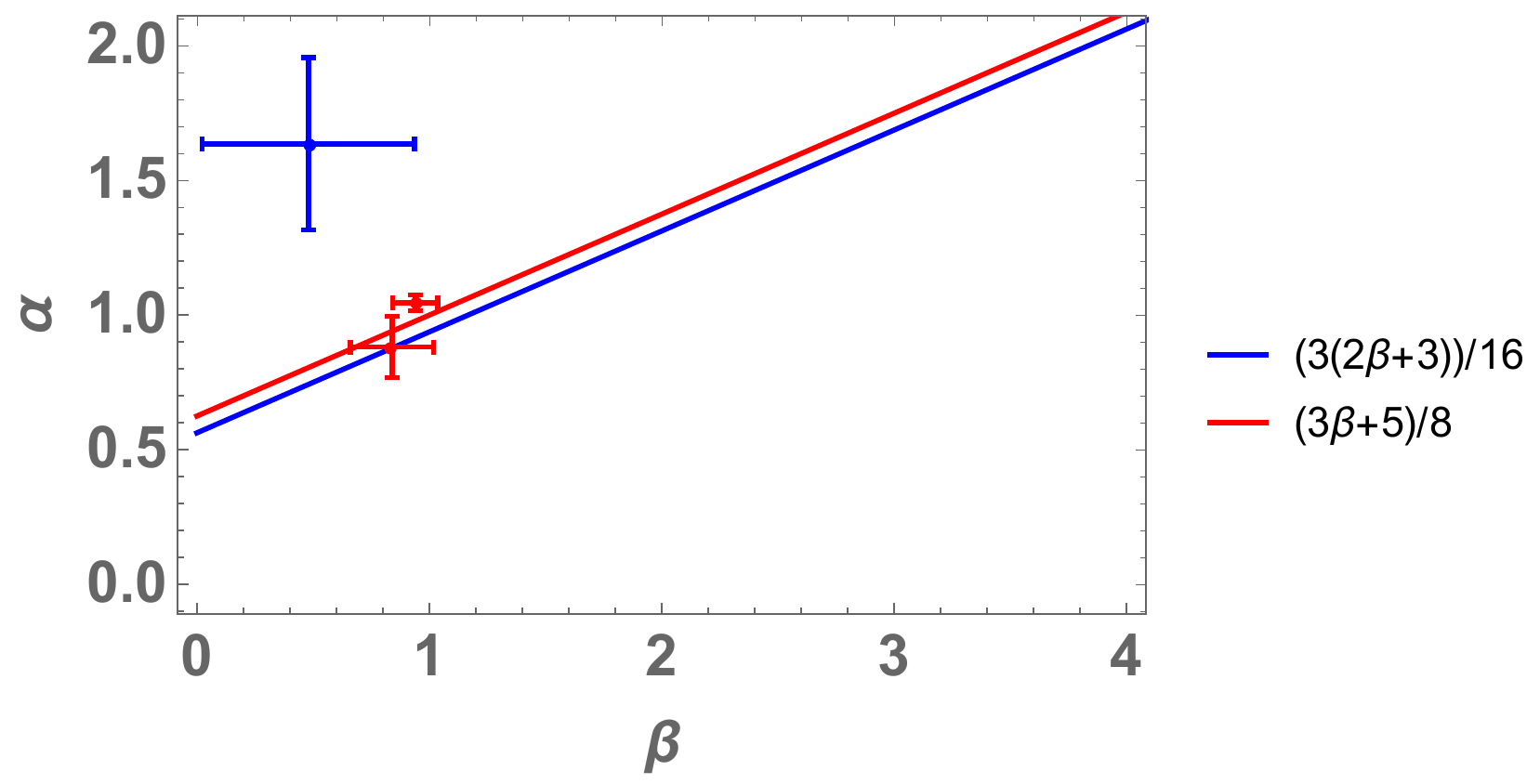}
  \label{fig:sub2}
}\\
{
  \includegraphics[width=0.45\columnwidth]{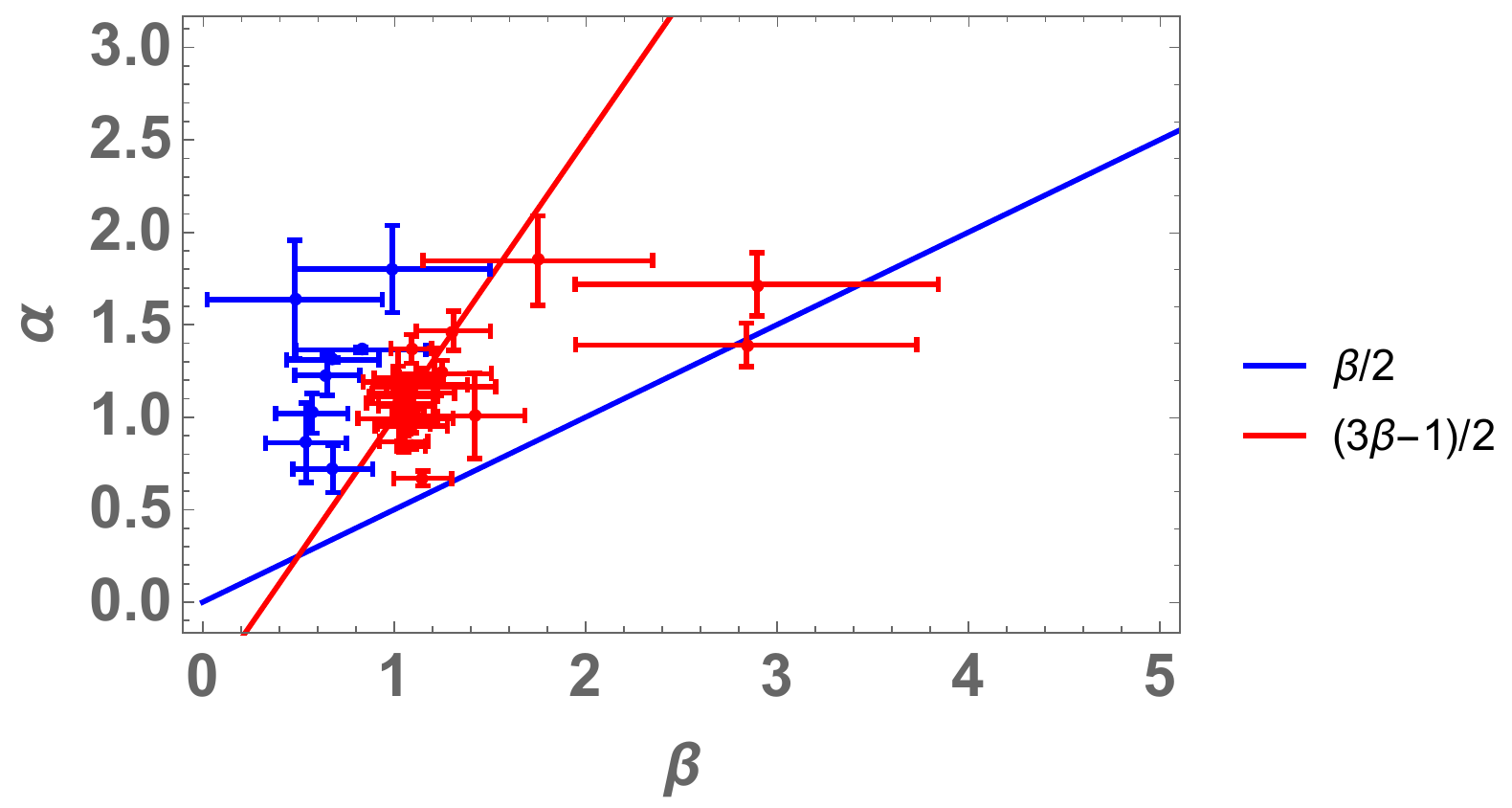}
  \label{fig:sub3}
}\qquad
{
  \includegraphics[width=0.45\columnwidth]{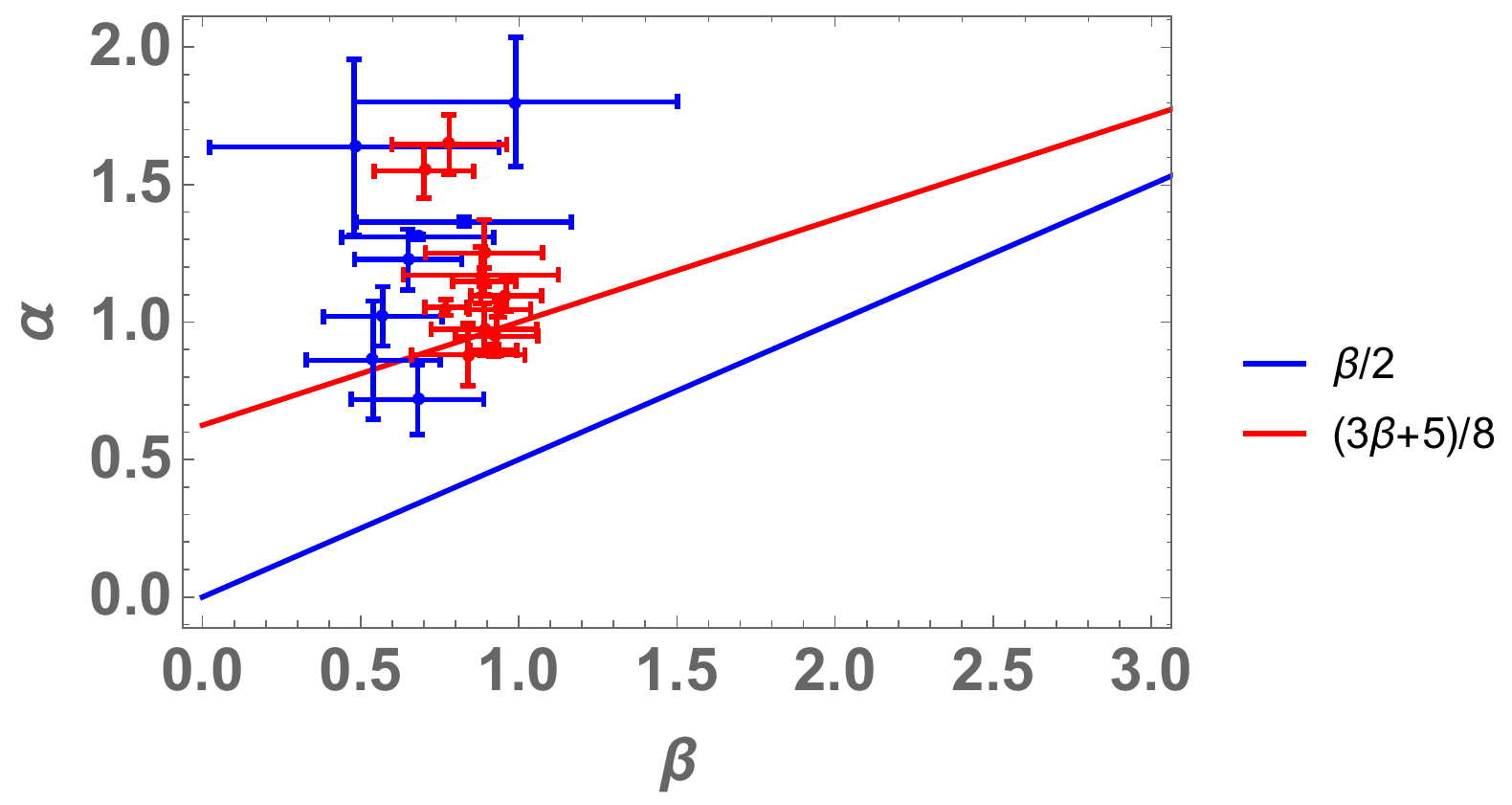}
  \label{fig:sub4}
}\\
{
  \includegraphics[width=0.45\columnwidth]{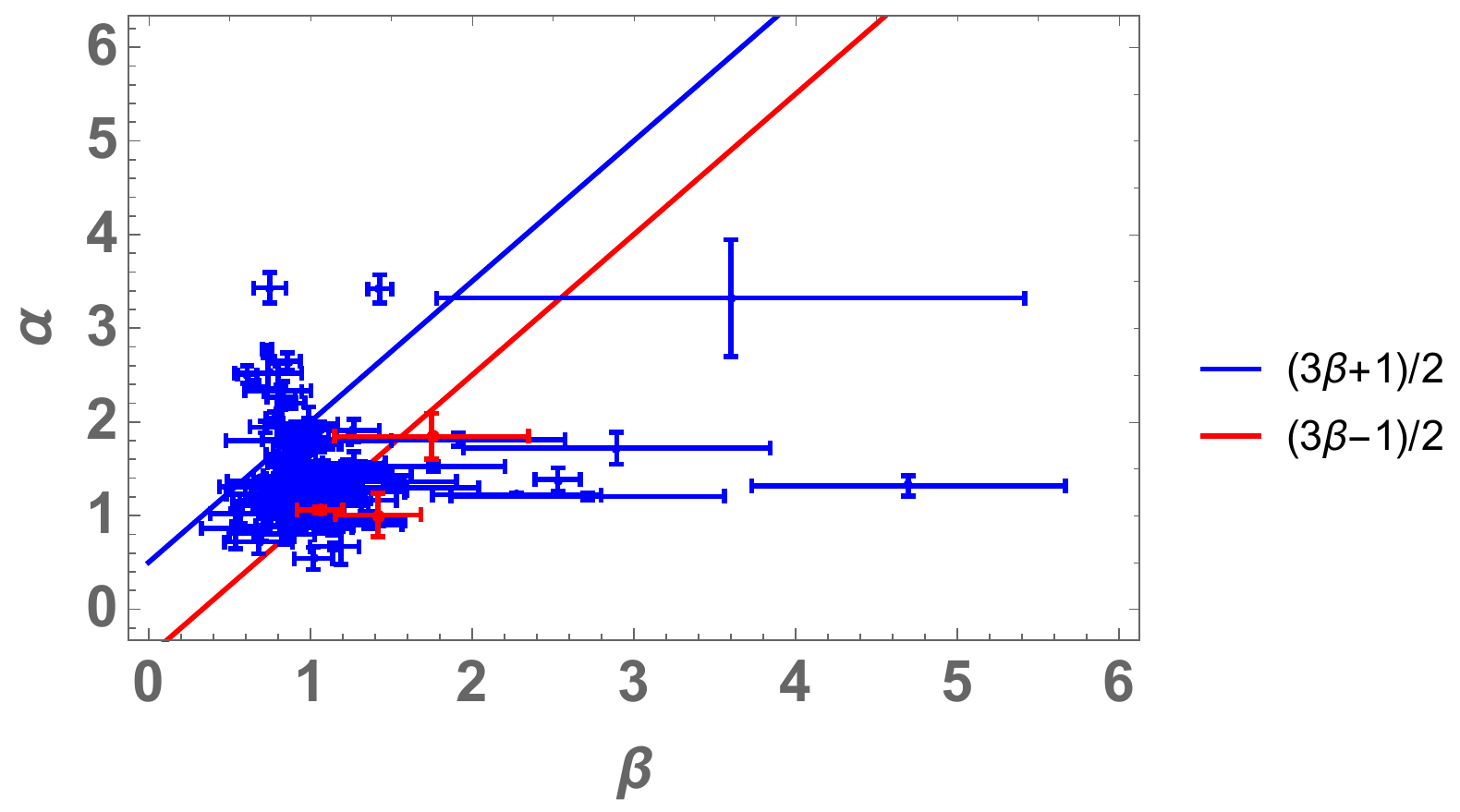}
  \label{fig:sub5}
}\qquad
{
  \includegraphics[width=0.45\columnwidth]{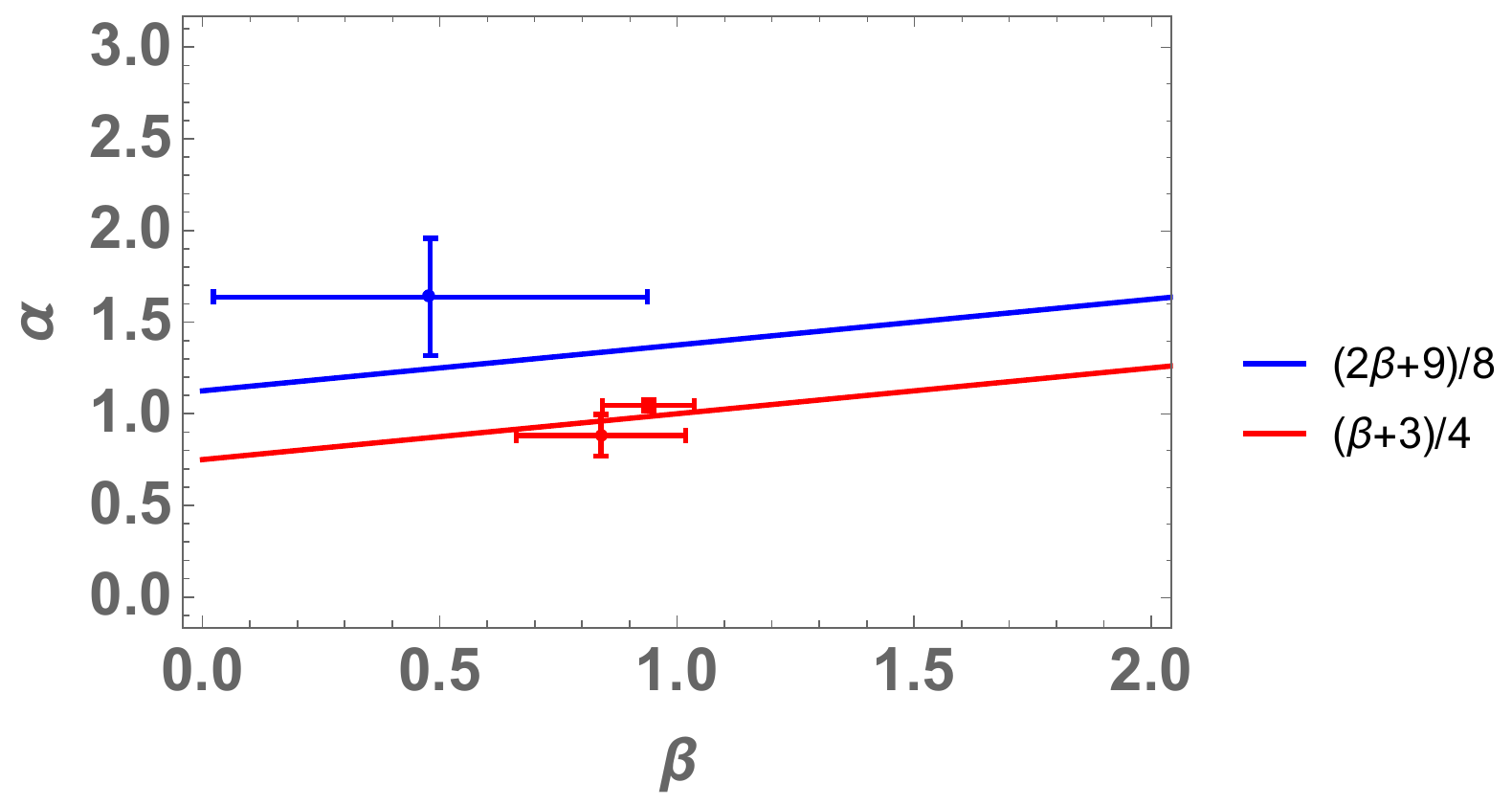}
  \label{fig:sub6}
}\\
{
  \includegraphics[width=0.45\columnwidth]{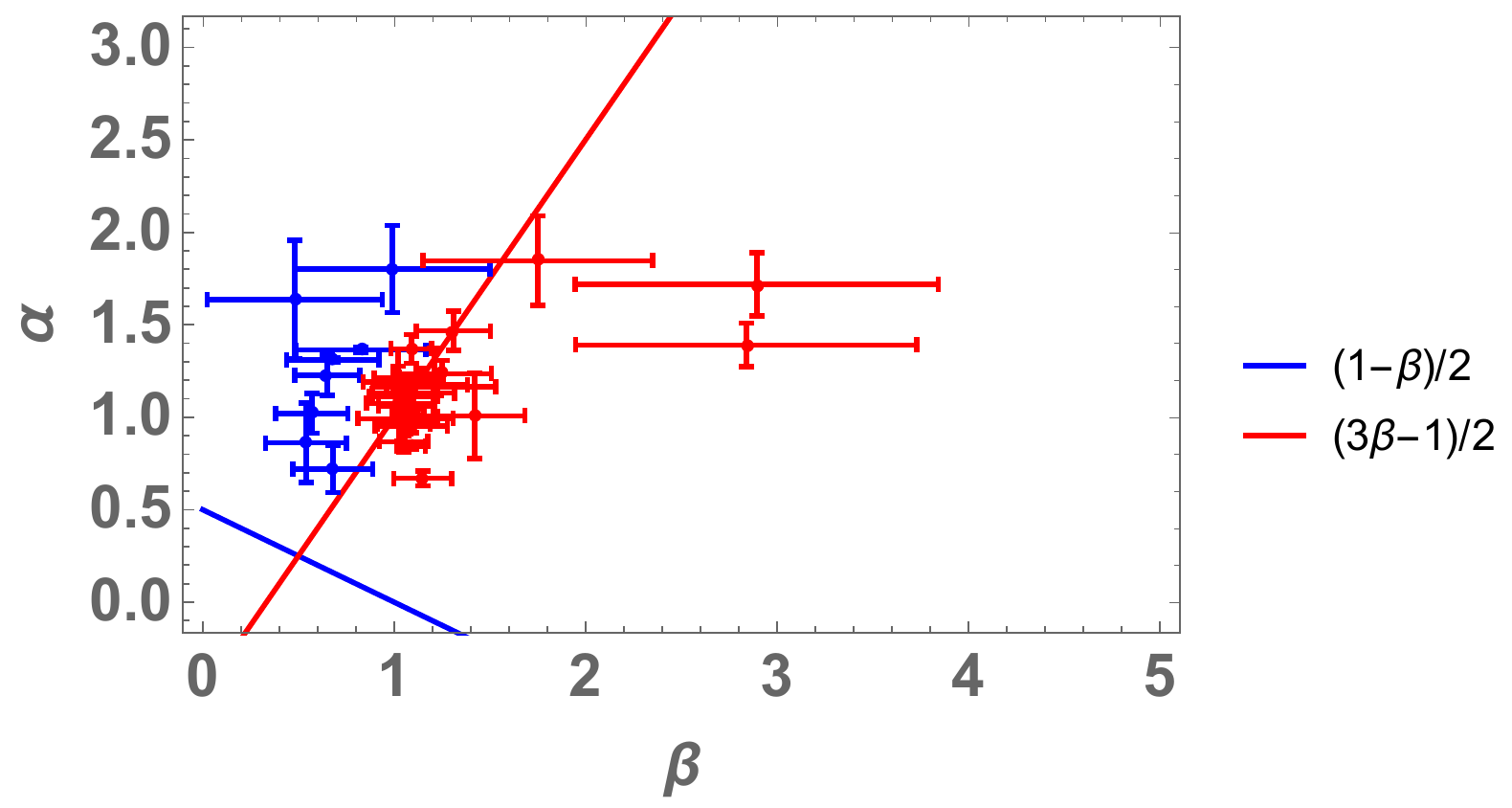}
  \label{fig:sub7}
}\qquad
{
  \includegraphics[width=0.45\columnwidth]{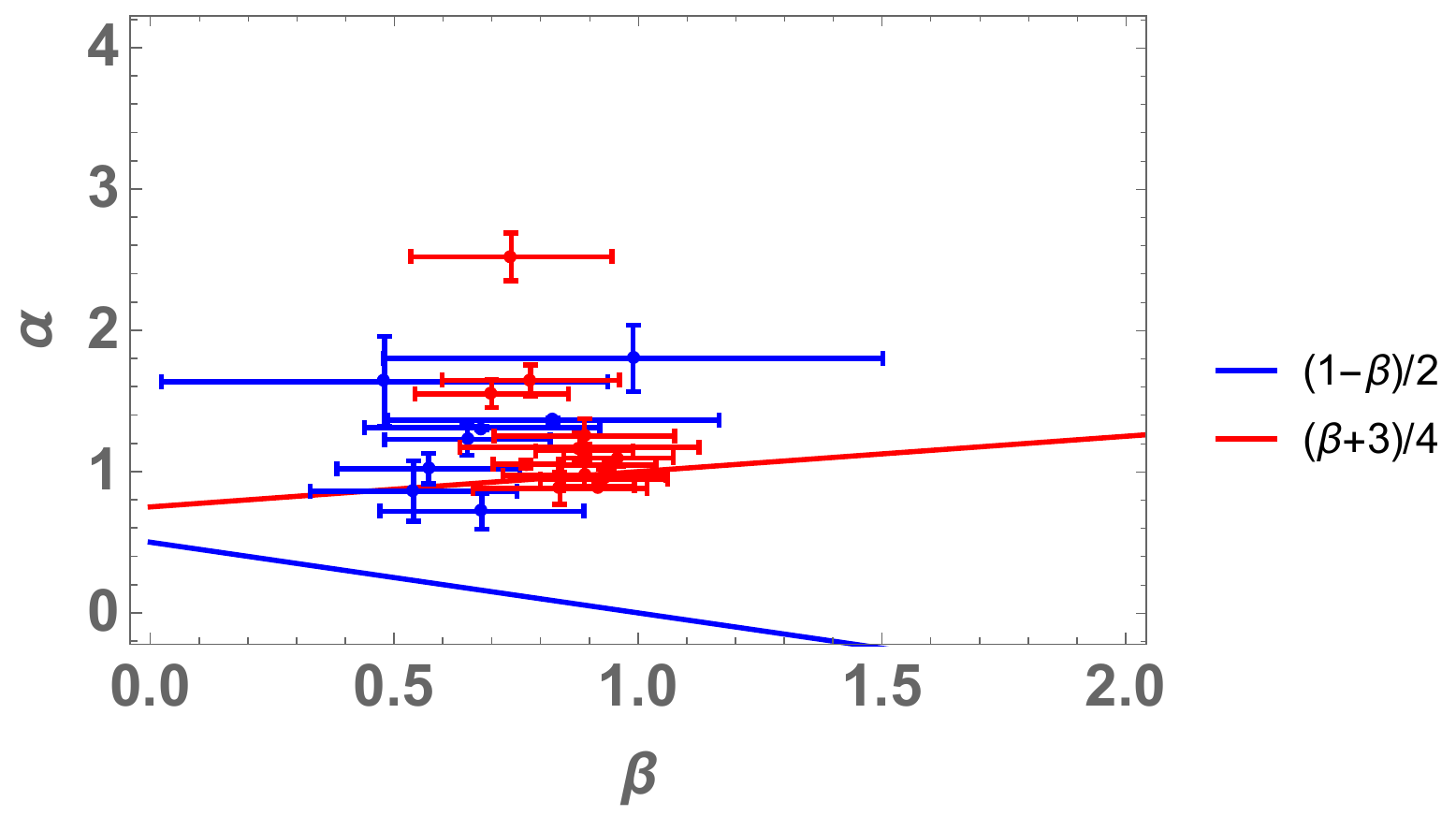}
  \label{fig:sub8}
}\\
\caption{Closure relations extracted from the ES for the GRBs with known redshifts from $T_a$ to $T_{\mathrm{end}}$. GRB sets are color-coded (blue or red) according to their frequency range given in Table \ref{CR} and the closure relations. The equality line corresponds to the two different relations and are color-coded in the same color. Relations corresponding to the same environment and electron spectral index ($p$) range are grouped together.}
\label{RedshiftClosure}
\end{figure}

\begin{figure}[h!]
\centering
{
  \includegraphics[width=0.45\columnwidth]{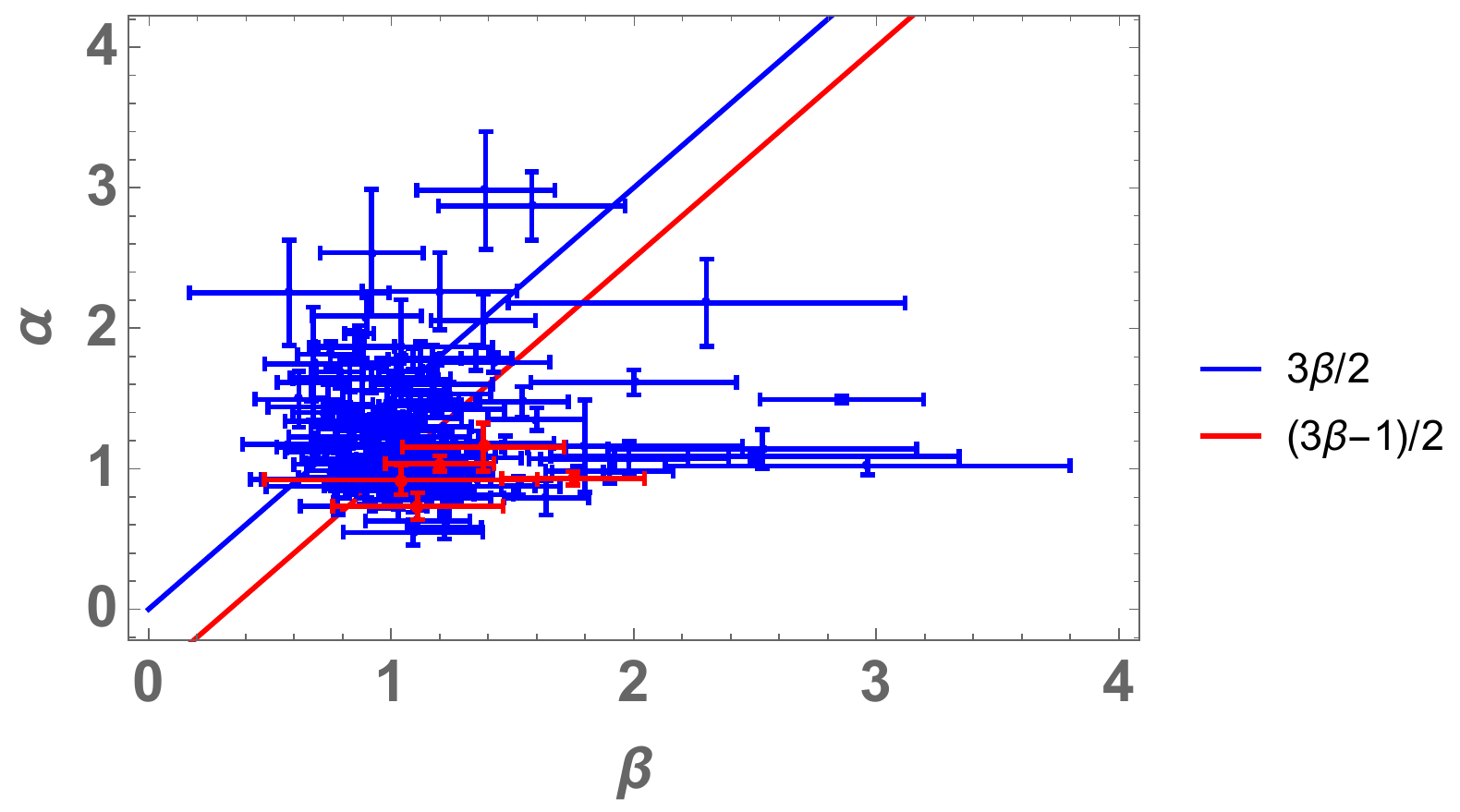}
  \label{fig1:sub1}
}\qquad
{
  \includegraphics[width=0.45\columnwidth]{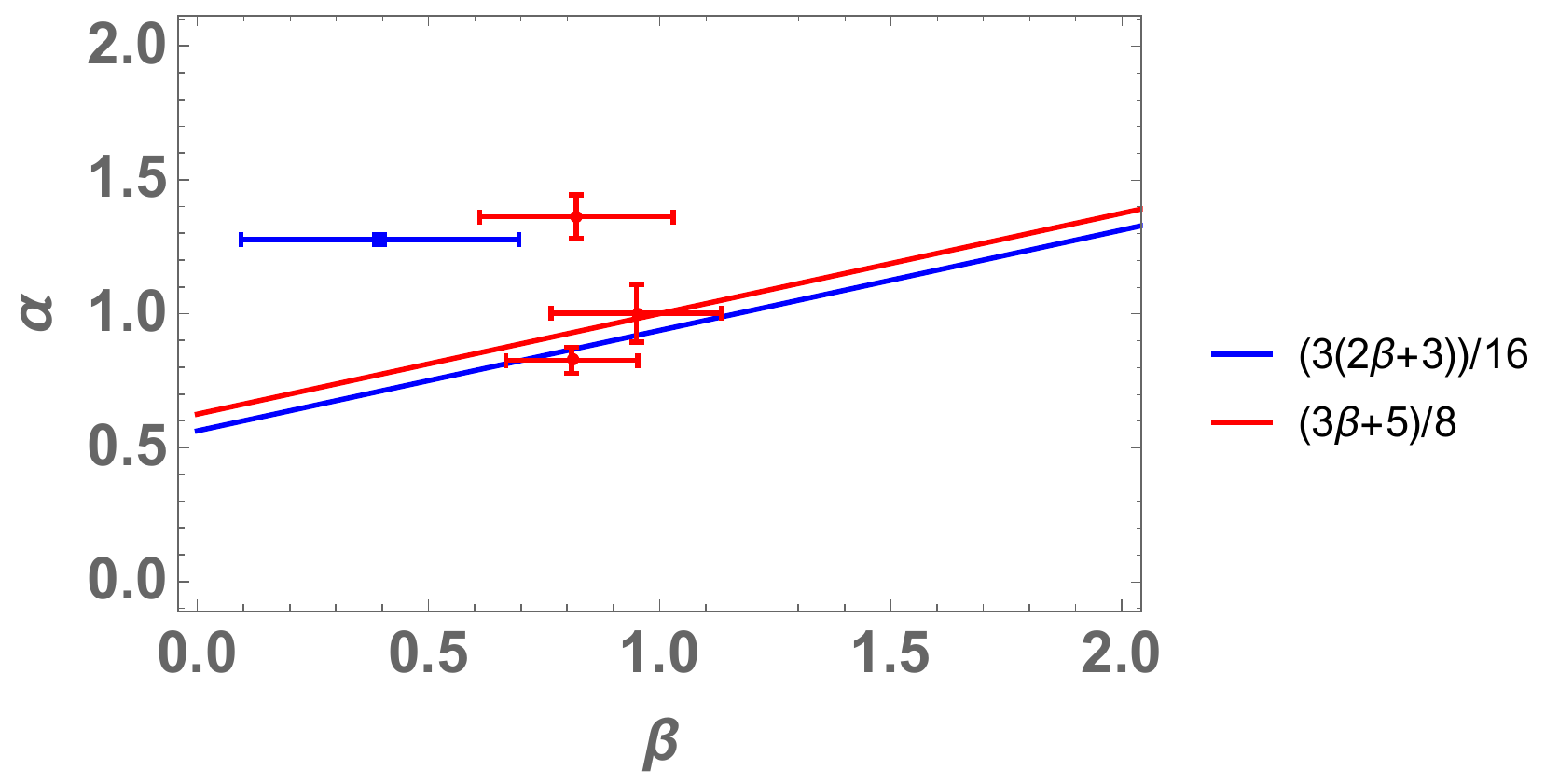}
  \label{fig1:sub2}
}\\

{
  \includegraphics[width=0.45\columnwidth]{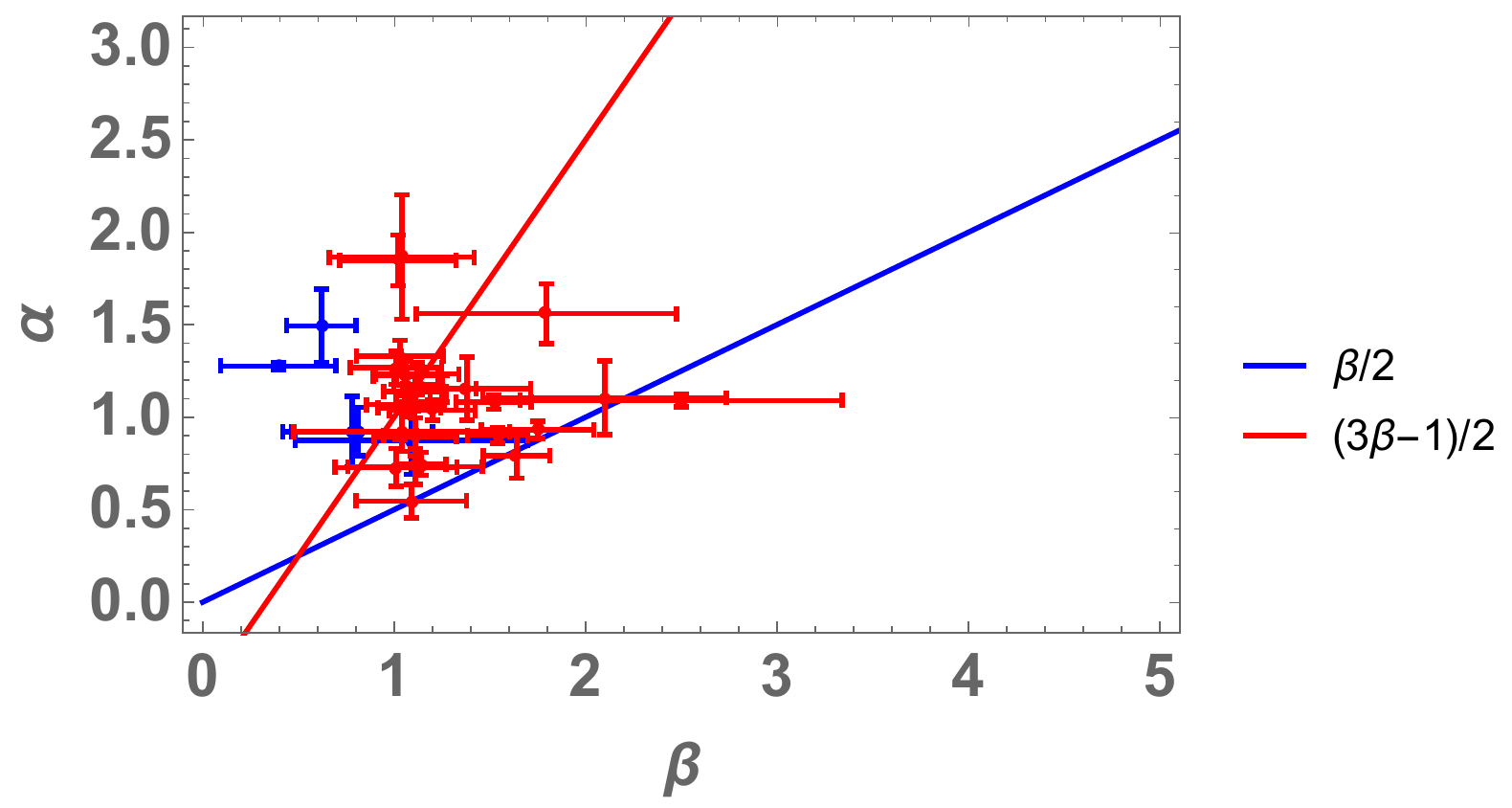}
  \label{fig1:sub3}
}\qquad
{
  \includegraphics[width=0.45\columnwidth]{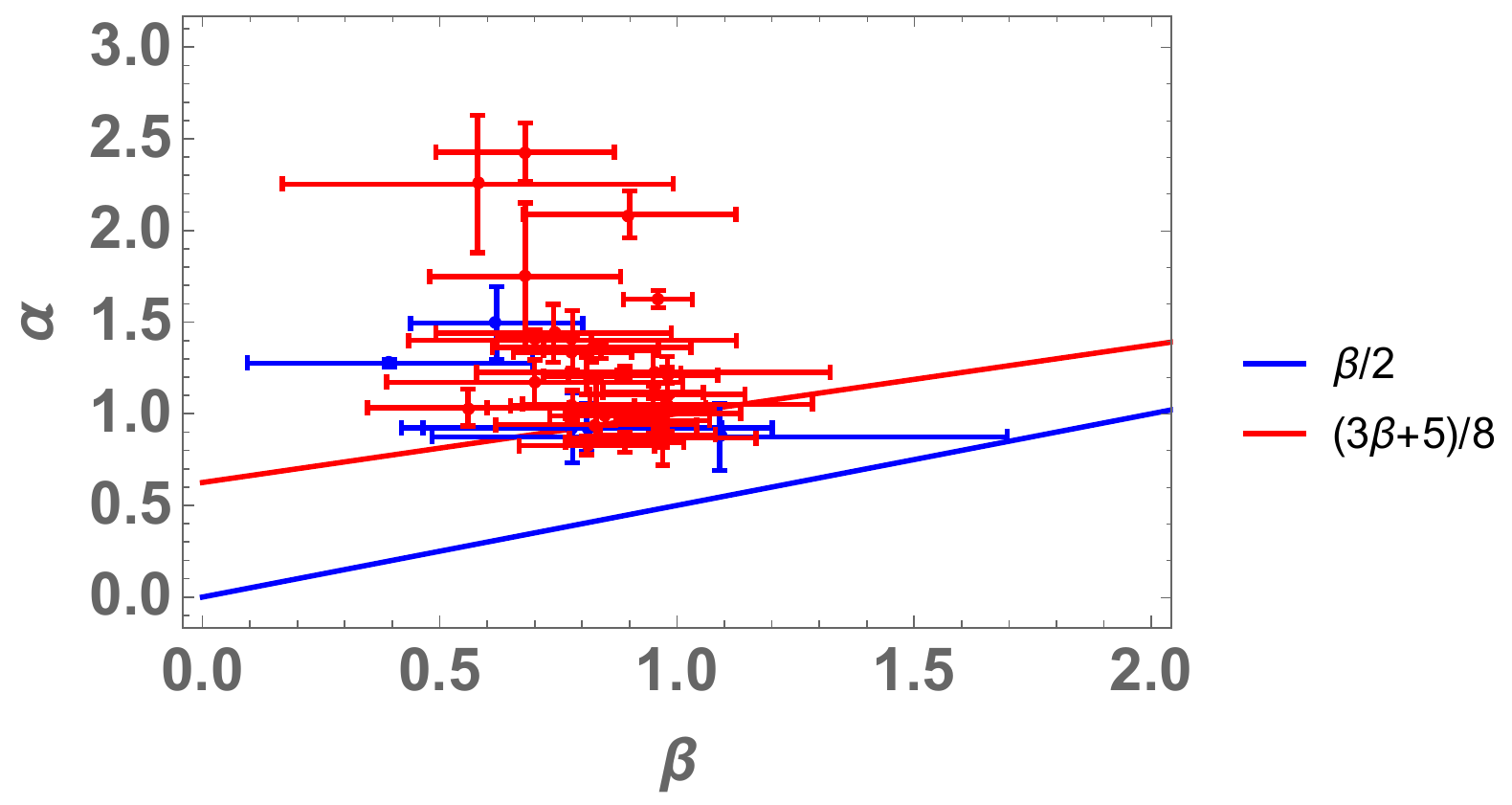}
  \label{fig1:sub4}
}\\
{
  \includegraphics[width=0.45\columnwidth]{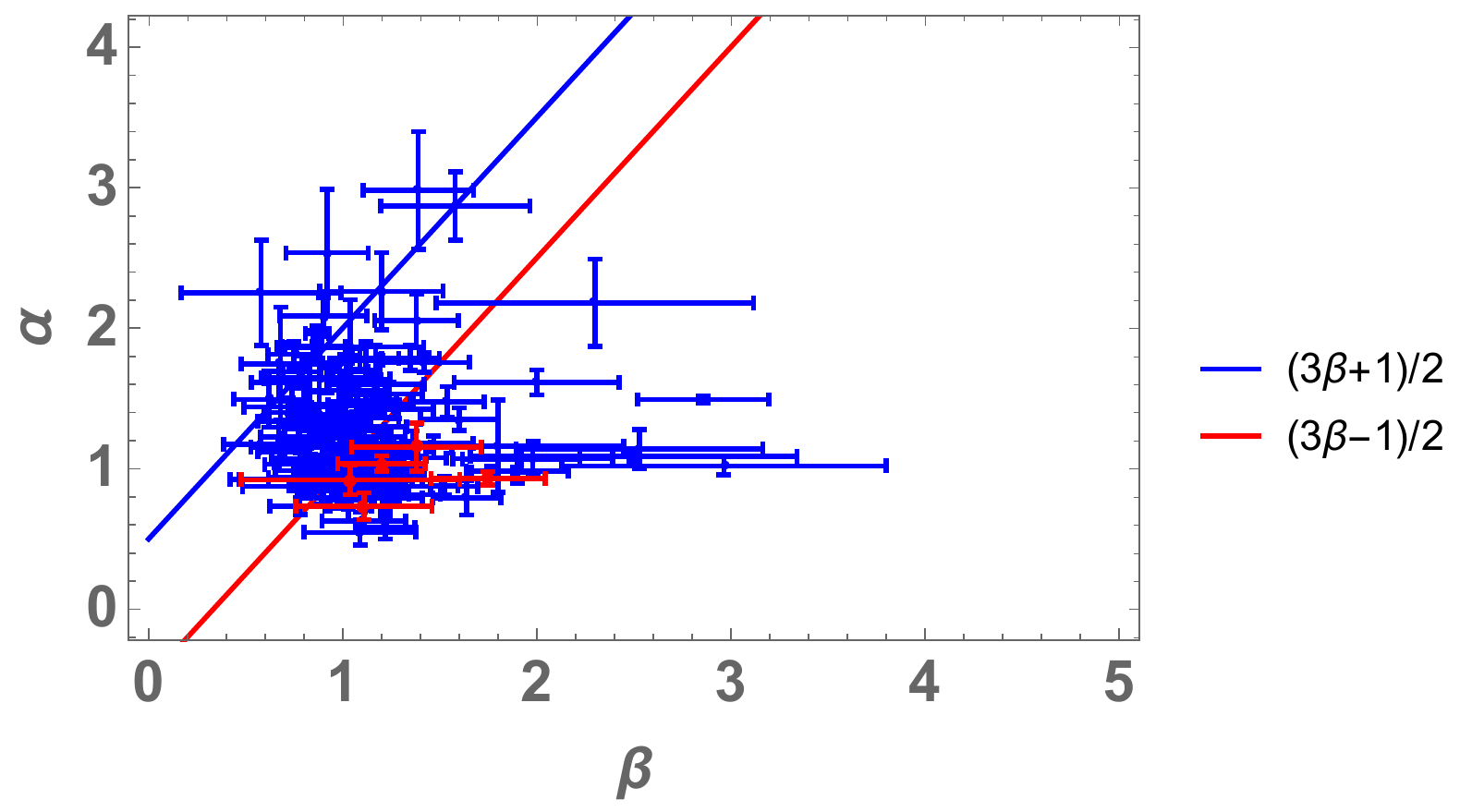}
  \label{fig1:sub5}
}\qquad
{
  \includegraphics[width=0.45\columnwidth]{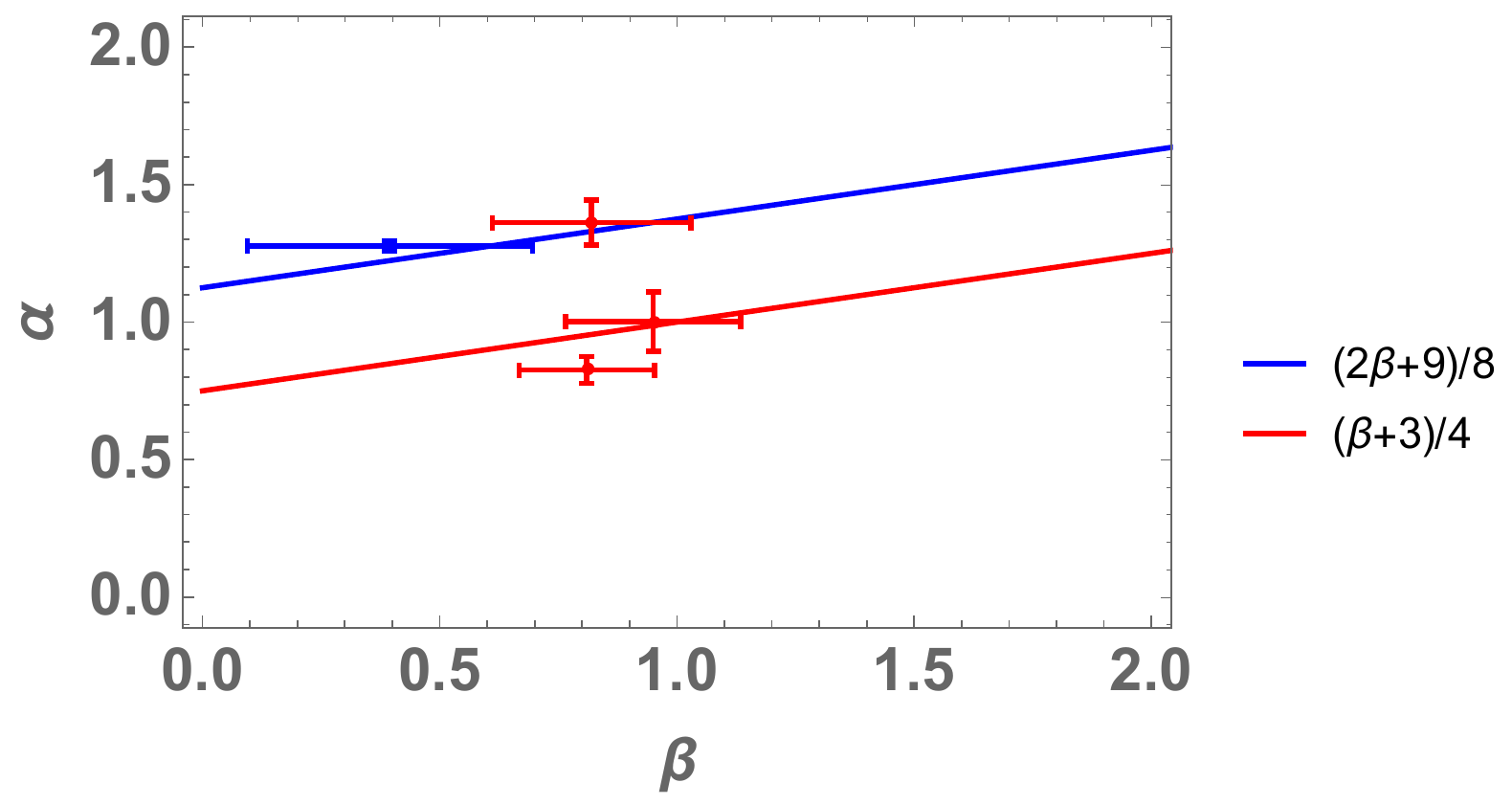}
  \label{fig1:sub6}
}\\
{
  \includegraphics[width=0.45\columnwidth]{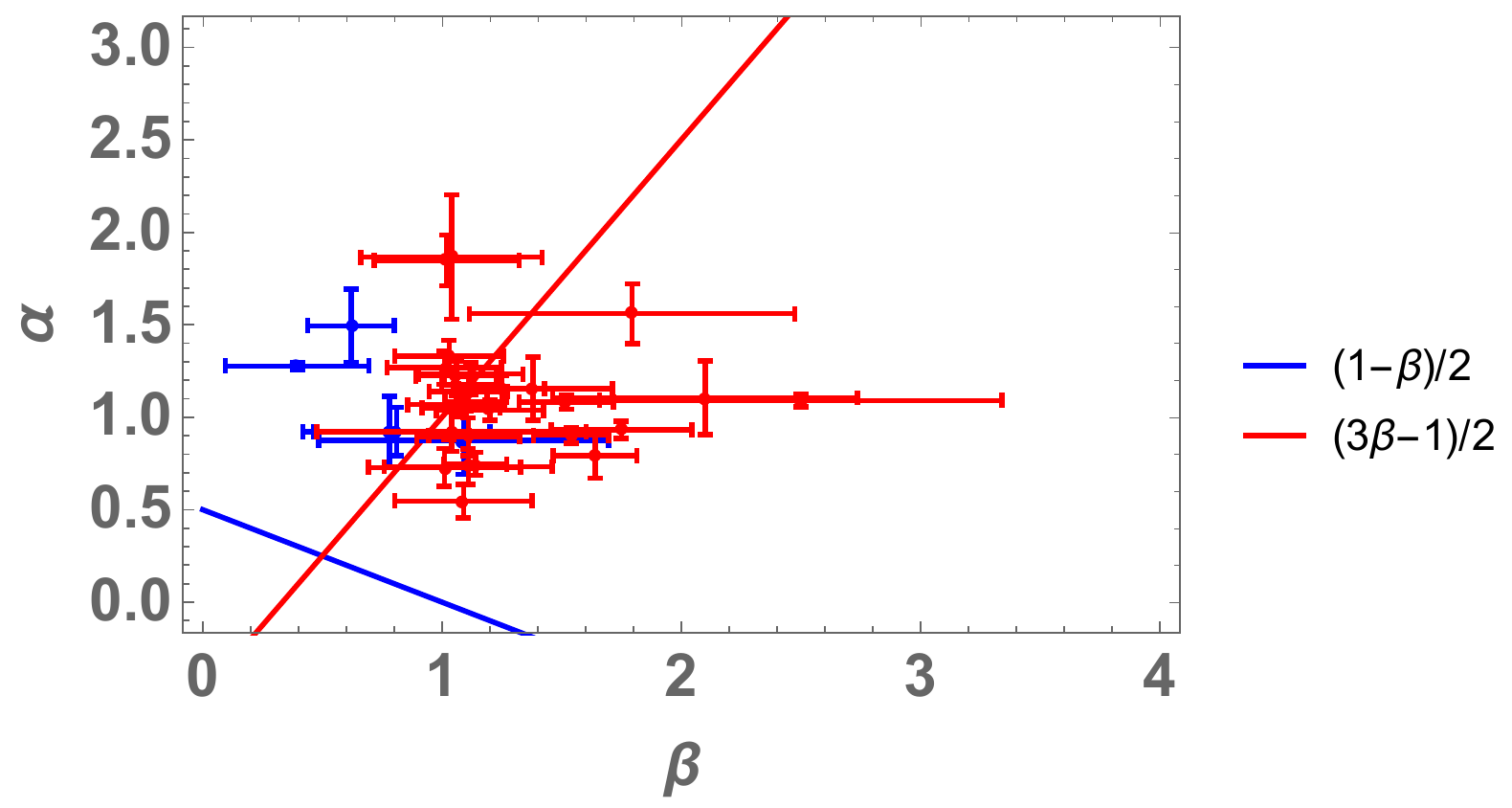}
  \label{fig1:sub7}
}\qquad
{
  \includegraphics[width=0.45\columnwidth]{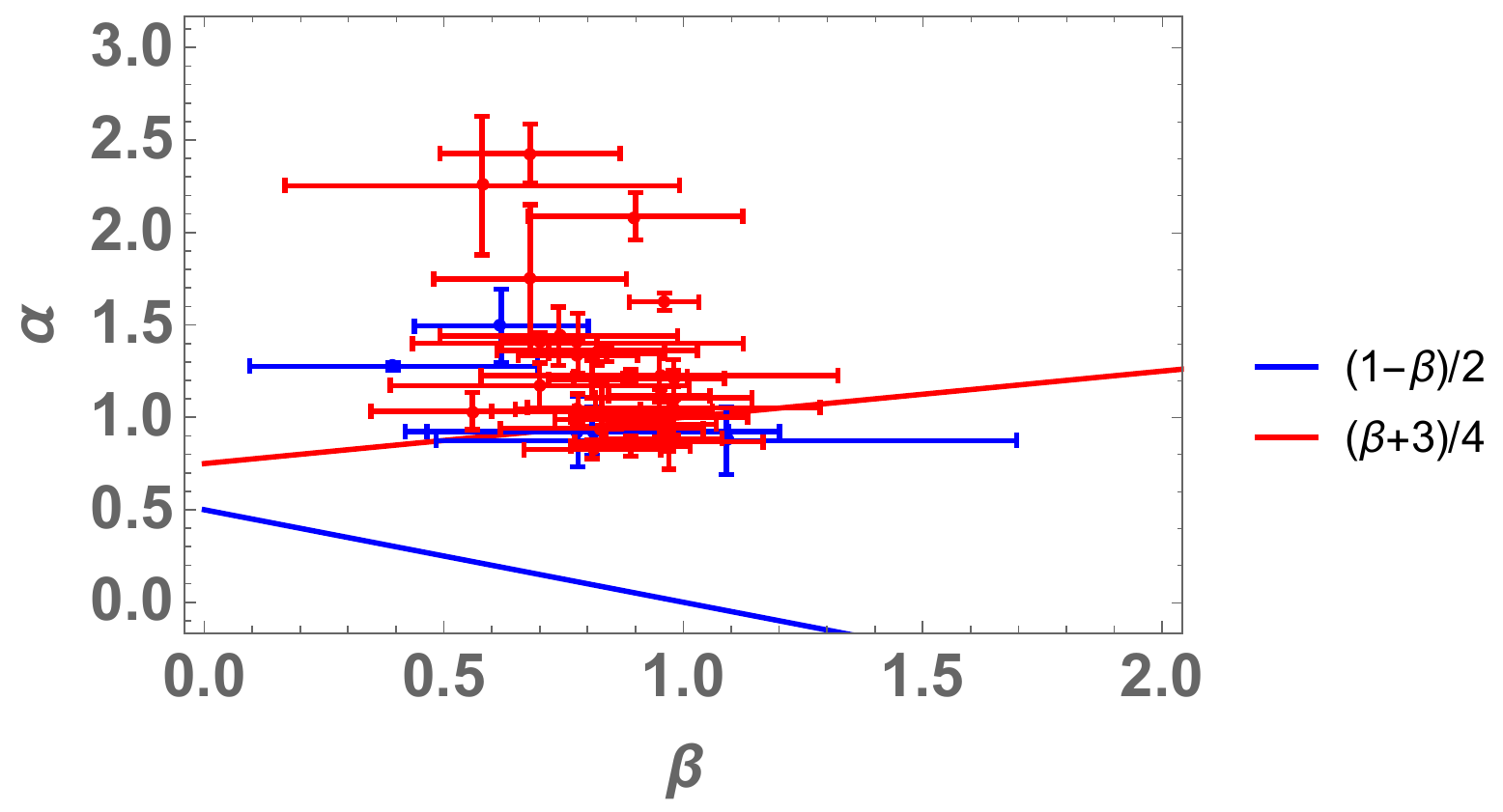}
  \label{fig1:sub8}
}\\
\caption{Closure relations extracted from the ES for the GRBs with unknown redshifts with the same scheme as Figure \ref{RedshiftClosure}.}
\label{NoRedshiftClosure}
\end{figure}

\begin{figure}[h!]
\begin{adjustwidth}{-50pt}{-50pt}
\centering
\gridline{\fig{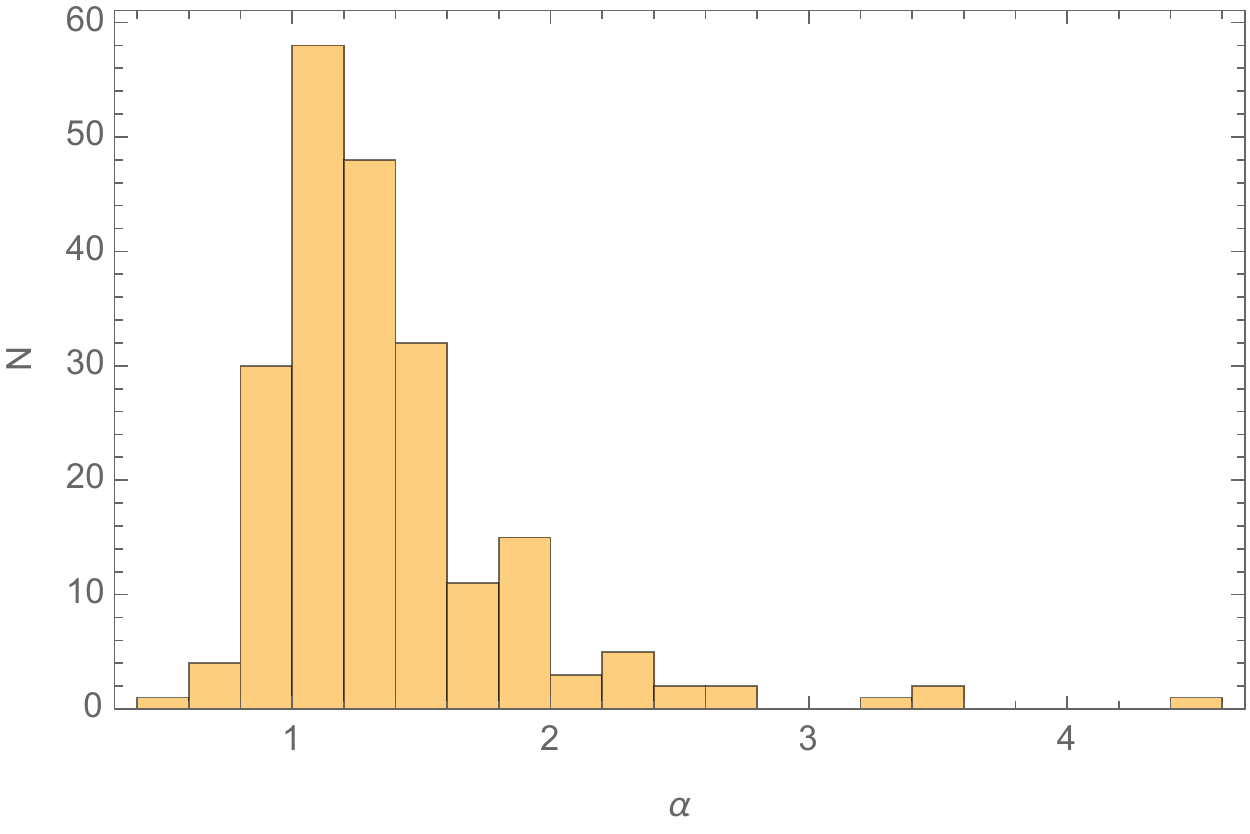}{0.5\textwidth}{(a) $\alpha$ for GRBs with known redshifts}
          \fig{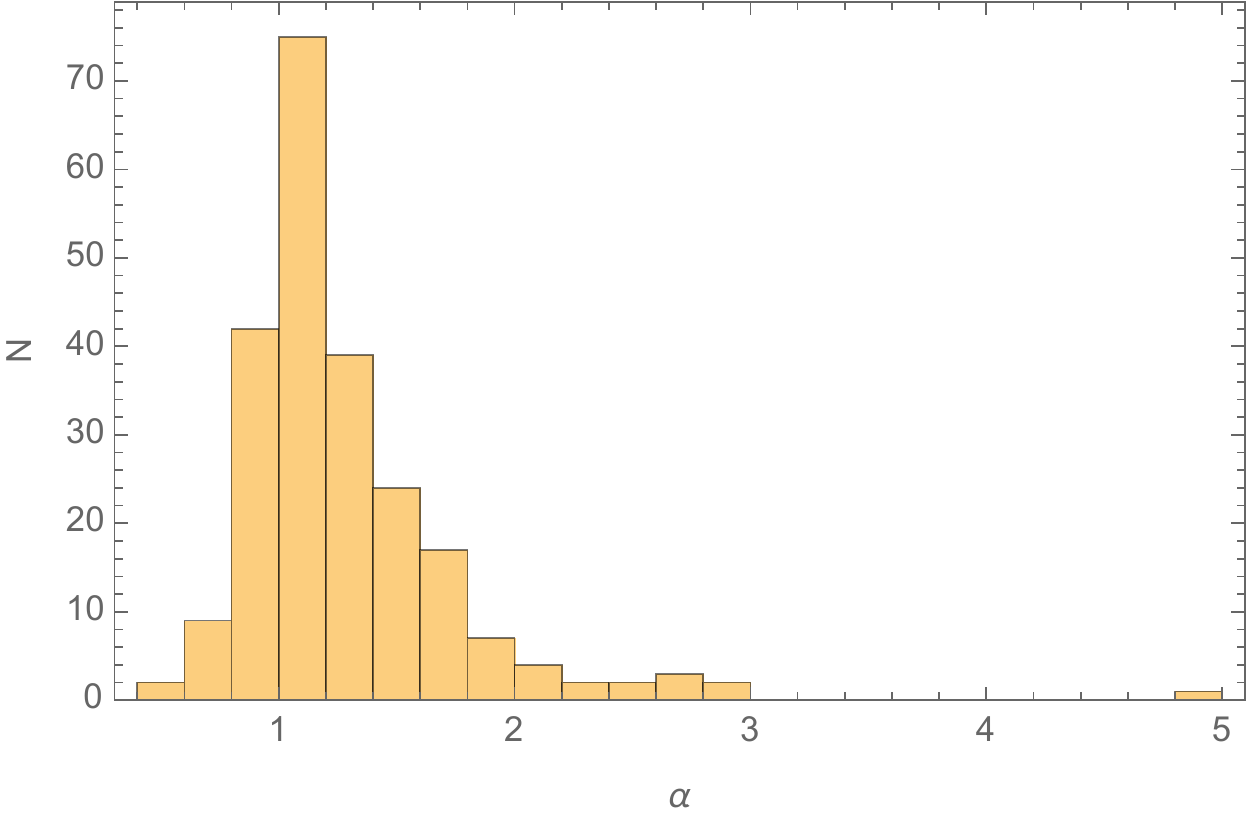}{0.5\textwidth}{(b) $\alpha$ for GRBs with unknown redshifts}}
\gridline{\fig{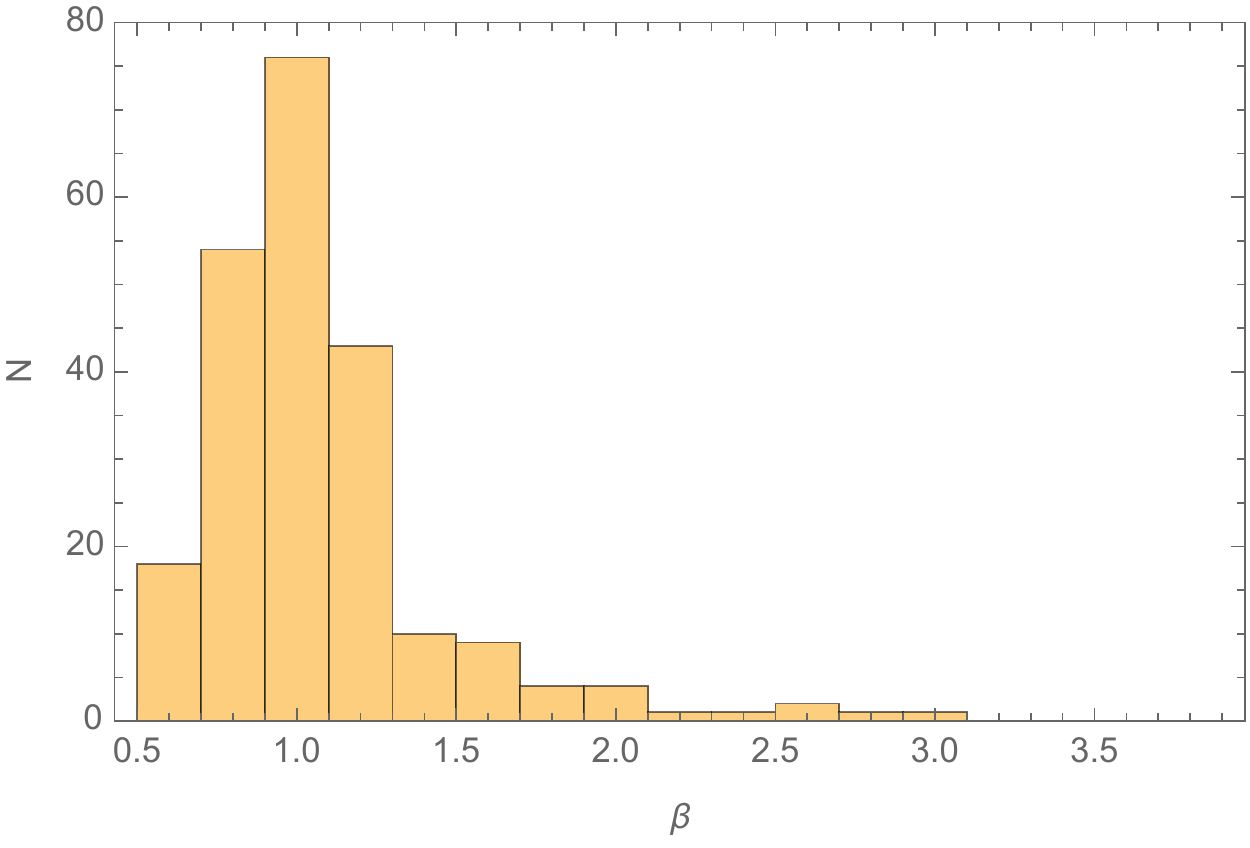}{0.5\textwidth}{(c) $\beta$ for GRBs with known redshifts from $T_a$ to $T_{\mathrm{end}}$}
          \fig{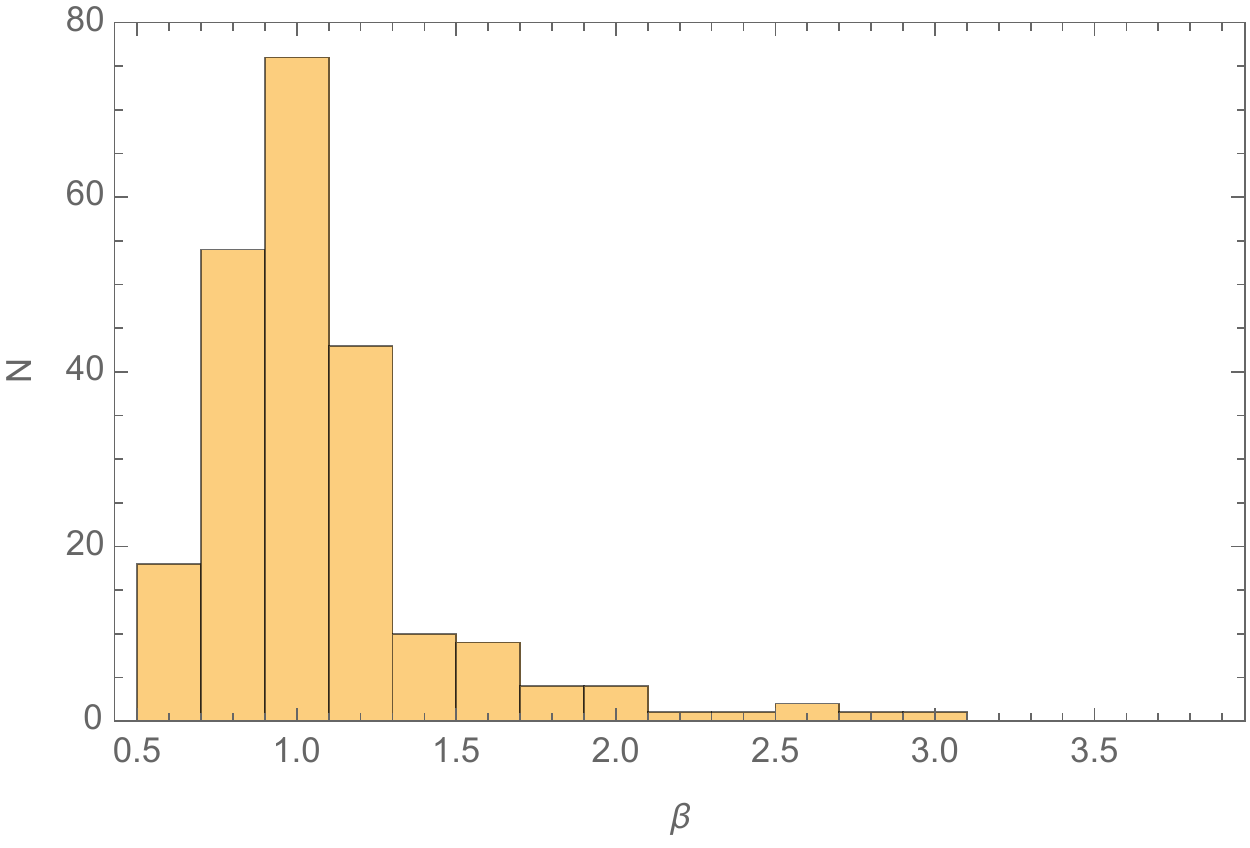}{0.5\textwidth}{(d) $\beta$ for GRBs with unknown redshifts from $T_a$ to $T_{\mathrm{end}}$}}
\caption{Histograms of $\alpha$ and $\beta$ parameters.}\label{alphabetahistoreal}
\end{adjustwidth}
\end{figure}

We then test each closure relation corresponding to the correct $\nu$ and $p$ range for every GRB in our sample, through plotting the $\alpha$ and $\beta$ parameters along with their error bars at a 1$\sigma$ level, as well as the equations of the closure relations. We group together relations characterized by the same $p$ range, astrophysical environment, and cooling regime on the same plot to create the so-called ``gray-region", a zone between two relations within the same environment. GRBs that lie between the two lines should be regarded as consistent cases, see Figure \ref{RedshiftClosure}, and \ref{NoRedshiftClosure}, along with Tables  \ref{Redshifttable} and \ref{NoRedshifttable}. In these Figures, we detail the specific closure relationships calculated for the time range $T_a$ to $T_{\mathrm{end}}$ in the afterglow, along with the error bars and the lines corresponding to the closure relation equations. We only consider GRBs with $\delta_x/x < 1$ where $x$ indicates either $\alpha$ or $\beta$ and $\delta_x$ indicates the error measurement. In addition, we also discard GRBs for which the errors on the closure relations are greater than the values of the closure relations themselves. Furthermore, histograms detailing the distribution of the $\alpha$ and $\beta$ parameters are presented in Figure \ref{alphabetahistoreal}.

\begin{deluxetable}{cccccc}[h!]
\tablecolumns{4} 
\tablewidth{0pc} 
\captionsetup{justification=centering}
\caption{Closure relations (Part of the table is taken from \citet{Racusin+09})}
\label{CR}
\tabletypesize{\footnotesize}
\tablehead{ 
\multicolumn{4}{c}{No Energy Injection} \\ 
\cline{1-4}
\colhead{$\nu$ range} & $\beta(p)$ & \colhead{$\alpha(\beta)$} & \colhead{$\alpha(\beta)$} & \\
& & \colhead{$(p > 2)$} & \colhead{$(1<p<2)$}}
\startdata 
\multicolumn{4}{c}{ISM, Slow Cooling} \\[0cm]
\cline{1-4} $\nu_m < \nu < \nu_c$ & $\frac{p-1}{2}$ &
$\alpha=\frac{3\beta}{2}$  & $\alpha=\frac{3(2\beta+3)}{16}$\\ 
$\nu > \nu_c$& $\frac{p}{2}$ & $\alpha=\frac{3\beta-1}{2}$ &  $\alpha=\frac{3\beta+5}{8}$
\\
\cline{1-4}
\multicolumn{4}{c}{ISM, Fast Cooling} \\[0cm]
\cline{1-4}
$\nu_c < \nu < \nu_m$ & $\frac{1}{2}$ &
 $\alpha=\frac{\beta}{2}$ &
 $\alpha=\frac{\beta}{2}$ &   \\
$\nu > \nu_m$ & $\frac{p}{2}$&
 $\alpha=\frac{3\beta-1}{2}$ &
 $\alpha=\frac{3\beta+5}{8}$ &  \\
\cline{1-4}
\multicolumn{4}{c}{Wind, Slow Cooling} \\[0cm]
\cline{1-4}
$\nu_m < \nu < \nu_c$ & $\frac{p-1}{2}$ &
 $\alpha=\frac{3\beta+1}{2}$ &
 $\alpha=\frac{2\beta+9}{8}$ \\
$\nu > \nu_c$ & $\frac{p}{2}$ &
 $\alpha=\frac{3\beta-1}{2}$ &  $\alpha=\frac{\beta+3}{4}$
\\
\cline{1-4}
\multicolumn{4}{c}{Wind, Fast Cooling} \\[0cm]
\cline{1-4}
$\nu_c < \nu < \nu_m$ & $\frac{1}{2}$ &
$\alpha=\frac{1-\beta}{2}$ &
 $\alpha=\frac{1-\beta}{2}$ \\
$\nu > \nu_m$ & $\frac{p}{2}$ &
 $\alpha=\frac{3\beta-1}{2}$ &  $\alpha=\frac{\beta+3}{4}$\\
\cline{1-4}
\enddata
\end{deluxetable}

\FloatBarrier
\FloatBarrier

\begin{table}
\begin{center}
 \begin{tabular}{||c c c c c c||} 
 \hline
 $\nu$ Range & $p$ Range & Closure Relation &  GRBs Total & GRBs Satisfying Relation & Percentage \\ [0.5ex]
 \hline\hline
 $\nu_m < \nu < \nu_c$ & $p>2$ & $3\beta/2$ & 150 & 109 & 72.67\% \\ 
 $\nu > \nu_c$ & $p>2$ & $(3\beta-1)/2$ &  &  & \\
 \hline
 $\nu_m < \nu < \nu_c$ & $1<p<2$ &$(3(2\beta+3))/16$ & 3 & 1 & 33.3\% \\
 $\nu > \nu_c$ & $1<p<2$ &$(3\beta+5)/8$ &  &  &  \\
 \hline
 $\nu_c < \nu < \nu_m$ & $p>2$ &$\beta/2$ & 32 & 26 & 81.25\% \\
 $\nu > \nu_m$ & $p>2$ &$(3\beta-1)/2$ &  &  &  \\
 \hline
 $\nu_c < \nu < \nu_m$ & $1<p<2$& $\beta/2$ & 21 & 6 & 28.6\% \\
 $\nu > \nu_m$ & $1<p<2$& $(3\beta+5)/8$ &  &  &  \\
 \hline
 $\nu_m < \nu < \nu_c$ & $p>2$ & $(3\beta+1)/2$ & 150 & 124 & 82.7\% \\
 $\nu > \nu_c$ & $p>2$ & $(3\beta-1)/2$ &  &  &  \\
 \hline
 $\nu_m < \nu < \nu_c$ & $1<p<2$ & $(2\beta+9)/8$ & 3 & 3 & 100\% \\
 $\nu > \nu_c$ & $1<p<2$ &$(\beta+3)/4$ &  &  &  \\
 \hline
 $\nu_c < \nu < \nu_m$ & $p>2$ & $(1-\beta)/2$ & 32 & 26 & 81.25\% \\
 $\nu > \nu_m$ & $p>2$ &$(3\beta-1)/2$ &  &  &  \\
 \hline
 $\nu_c < \nu < \nu_m$ & $1<p<2$ & $(1-\beta)/2$ & 21 & 7 & 33.3\% \\
 $\nu > \nu_m$ & $1<p<2$ & $(\beta+3)/4$ &  &  &  \\
 \hline
\end{tabular}
\captionsetup{justification=centering}
\end{center}
\caption{Closure relations for the 222 redshift GRBs. }\label{Redshifttable}.
\end{table}
 
\begin{table}
\begin{center}
  \begin{tabular}{||c c c c c c||} 
 \hline
 $\nu$ Range & $p$ Range & Closure Relation & GRBs Total & GRBs Satisfying Relation & Percentage \\ [0.5ex]
 \hline\hline
 $\nu_m < \nu < \nu_c$ & $p>2$ & $3\beta/2$ & 144 & 100 & 69.4\% \\ 
 $\nu > \nu_c$ & $p>2$ & $(3\beta-1)/2$ &  &  &  \\
 \hline
 $\nu_m < \nu < \nu_c$ & $1<p<2$ & $(3(2\beta+3))/16$ & 4 & 2 & 50.0\% \\
 $\nu > \nu_c$ & $1<p<2$ & $(3\beta+5)/8$ &  &  &  \\
 \hline
 $\nu_c < \nu < \nu_m$ & $p>2$ & $\beta/2$ & 33 & 30 & 90.9\% \\
 $\nu > \nu_m$ & $p>2$ &$(3\beta-1)/2$ &  &  & \\
 \hline
 $\nu_c < \nu < \nu_m$ & $1<p<2$& $\beta/2$ & 36 & 16 & 44.4\% \\
 $\nu > \nu_m$ & $1<p<2$& $(3\beta+5)/8$ &  &  & \\
 \hline
 $\nu_m < \nu < \nu_c$ & $p>2$ & $(3\beta+1)/2$ & 144 & 114 & 79.2\% \\
 $\nu > \nu_c$ & $p>2$ & $(3\beta-1)/2$ &  &  &  \\
 \hline
 $\nu_m < \nu < \nu_c$ & $1<p<2$ & $(2\beta+9)/8$ & 4 & 3 & 75.0\% \\
 $\nu > \nu_c$ & $1<p<2$ & $(\beta+3)/4$ &  &  &  \\
 \hline
 $\nu_c < \nu < \nu_m$ & $p>2$ & $(1-\beta)/2$ & 33 & 30 & 90.9\% \\
 $\nu > \nu_m$ & $p>2$ & $(3\beta-1)/2$ &  &  &  \\
 \hline
 $\nu_c < \nu < \nu_m$ & $1<p<2$ & $(1-\beta)/2$ & 36 & 16 & 44.4\% \\
 $\nu > \nu_m$ & $1<p<2$ & $(\beta+3)/4$ &  &  & \\
 \hline
\end{tabular}
\captionsetup{justification=centering}
\end{center}
\caption{Closure relations for the 233 GRBs with unknown redshifts.}\label{NoRedshifttable}
\end{table}

\FloatBarrier
\section{Interpretation of the closure relations}
\label{Interpretation}
Through our analysis, we see that overall, the ES model is quite successful in modeling the afterglows of GRBs (see Table \ref{Redshifttable} and Table \ref{NoRedshifttable}).
\citet{wang15} tested the ES model with a set of closure relations by using a sample of 85 GRBs with both X-ray and optical afterglow data, and concluded that the ES model is able to account for at least half of GRB afterglows. Through our analysis, we see that even with a greater number of GRB X-ray LCs used (455), this result still holds. Our results are discrepant with that of \citet{Willingale2007}, who through an analysis of 107 Swift LCs, reached to the conclusion that the ES model works for less than 50\% of Swift GRBs. This difference is due to the fact that we took into account the ``gray-region", differently from \citet{Willingale2007}. Furthermore, through their analysis of 318 Swift LCs,  \citet{evans09} came to the conclusion that closure relations corresponding to the ES model without energy injection are fulfilled by a reasonable amount of GRBs, though there are also a number of GRBs for which energy injection mechanisms are needed. If there is a source of continuous energy injection, then the forward shock will continue to be ``refreshed" with continuous bursts of energy such that the fireball decelerates at a slower timescale than it would in the normal ES scenario \citep{Zhang2006}. A previous analysis in relation to this energy injection mechanism and its relation to the 2D Dainotti relation has been investigated in \citet{delvecchio16}. 

We can interpret the results of the closure relations within these main scenarios:
\begin{enumerate}[label=(\roman*)]
\item Out of the 16 closure relation groups for the standard fireball model tested for both the known and unknown redshift data sets, 11 have at least 50\% of their GRBs fulfilled.  The closure relations are a quick check to assess the reliability of the ES scenario, thus it is possible that for cases where the closure relations are not fulfilled, more complex physical processes must be taken into account.
\item The W07 model is an empirical model for the plateau and afterglow emission, but in its current formulation, we have removed flares manually. Thus, we cannot control how much this manual removal of flares influences results for cases where they are present in the plateau emission. 

\item According to Table \ref{Redshifttable} and Table \ref{NoRedshifttable}, the most favored closure relation set for the known redshift data set is $(2\beta+9)/8$ and $(\beta+3)/4$, which corresponds to a wind environment with slow cooling and $1<p<2$, with all possible GRBs in the correct $\nu$ and $p$ range fulfilling these relations. However, there are only three GRBs within these ranges. Another set of note that has a high fulfillment rate (82.7\%) in the known redshift data set is $(3\beta+1)/2$ and $(3\beta-1)/2$, also corresponding to a wind environment with slow cooling but with $1<p<2$. The most favored sets for the unknown redshift data set are $\beta/2$ and $(3\beta-1)/2$, as well as $(1-\beta)/2$ and $(3\beta-1)/2$, which correspond to a constant-density ISM or wind environment with fast cooling respectively, also both with $p>2$. The fulfillment rate of both these sets is 90.9\%. However, it is important to note that these percentages are relative to the number of GRBs that satisfy the $\nu$ ranges for each cooling regime. As seen in Table \ref{Redshifttable} and Table \ref{NoRedshifttable} there are less GRBs that satisfy the fast cooling regime than do the slow cooling, which has an effect on these percentages.

With regards to the actual astrophysical environments and cooling regimes of the GRBs in our sample with known redshifts, the most fulfilled environments are a wind and constant-density ISM environment with slow cooling. There are a total of 127 GRBs that satisfy a closure relation indicating a wind environment with slow cooling, and 110 that indicate a constant-density ISM environment with slow cooling. The environments that are fulfilled the least are a wind and constant-density ISM environment with fast cooling, with 32 and 31 GRBs fulfilling those environments, respectively. Furthermore, regarding the sample of GRBs with unknown redshifts, the most fulfilled environments again are either a wind or constant-density ISM environment with slow cooling. There are 117 GRBs that indicate a wind environment with slow cooling, and 110 pointing toward a constant-density ISM environment with slow cooling. Again, the environments that are fulfilled the least are a wind and constant-density ISM environment with fast cooling, with 43 GRBs fulfilling both groups respectively.  Therefore, we can conclude that a constant-density ISM or wind environment with slow cooling is the most likely scenario for our sample of GRBs during Phase III of their LCs, and that the fast cooling regime is disfavored.
\end{enumerate}

This interpretation (i-iii) constitutes the answer to question 1) in \S \ref{Intro}. The ES model seems to be a good explanation of the high-energy LCs presenting a plateau emission. However, we are open to exploring new possibilities which allow us to take into account cases that do not follow the ES model. These cases include more complex physical processes such as nonlinear particle acceleration, or an energy injection mechanism such as the one obtained with a magnetar or mass accretion onto a black hole.

Given that it is necessary to explain the minority of GRBs for which standard closure relations are not a viable explanation, more complex evolution of afterglows, alternative models, and corresponding closure relations have been so far investigated \citep{Meszaros1998, Sari+98, Chevalier+00, Dai&Cheng2001, Zhang2004, zhang06}. Furthermore, in our analysis, we consider linear particle acceleration, although the nonlinear particle acceleration scenario \citep{2017ApJ...835..248W} cannot be ruled out. \citet{2017ApJ...835..248W} studied the time evolution of LCs of afterglows by taking into account the effects of non-linear particle acceleration for the first time. They found that the temporal and spectral evolution is much different from the simplistic formulation of the afterglow model mentioned above.
They also showed that very high energy $\gamma$-rays can be produced by synchrotron self-Compton emission, especially at the early phase of the afterglow \citep{zhang2001}. Analyzing more data can help us shed more light on these results in the near future. 

In Section \ref{Results from SWIFT}, we connect the astrophysical environments of GRBs which accounts for phase III of the LCs involving the decay phase after the plateau emission, $\alpha$, and the 3D fundamental plane relation, which considers $T_a$ and $L_a$ at the end of phase II of the LCs.

\section{Fundamental Plane Correlation} 
\label{Results from SWIFT}

Using the classifications from \S \ref{Methodology}, we update the 3D relation
($\log(T_a)$, $\log(L_{\mathrm{peak}})$, $\log(L_a)$), with a plane fitted to the data for all 222 GRBs with redshift. We see the 3D relation still holds with the updated data set. The equation of the plane is given by:
\begin{equation}
\log L_a= C_o + a \log T_a + b \log L_{\mathrm{peak}}\,,
\label{planeequation1}
\end{equation} 
where $C_o = C(\theta, \phi, \sigma_{\mathrm{int}}) + z_0$, represents the normalization of the plane with respect to $\theta$ and $\phi$, as well as $\sigma_{\mathrm{int}}$, the intrinsic scatter of the sample. $z_o$ is a normalization parameter and $C$ is the co-variance function. Furthermore, $a$ and $b$ are both functions of the variables $\theta$ and $\phi$, where $a(\theta,\phi)= -\cos(\phi)\tan({\theta})$ and $b(\theta,\phi) = -\sin(\phi)\tan(\theta)$. Our best-fit plane to the data has $C_o= 8.60 \pm 2.5$, $a = -0.77 \pm 0.06$, and $b = 0.81 \pm 0.05$ with $\sigma=0.52 \pm 0.02$.

\begin{figure}
\centering 
\gridline{\fig{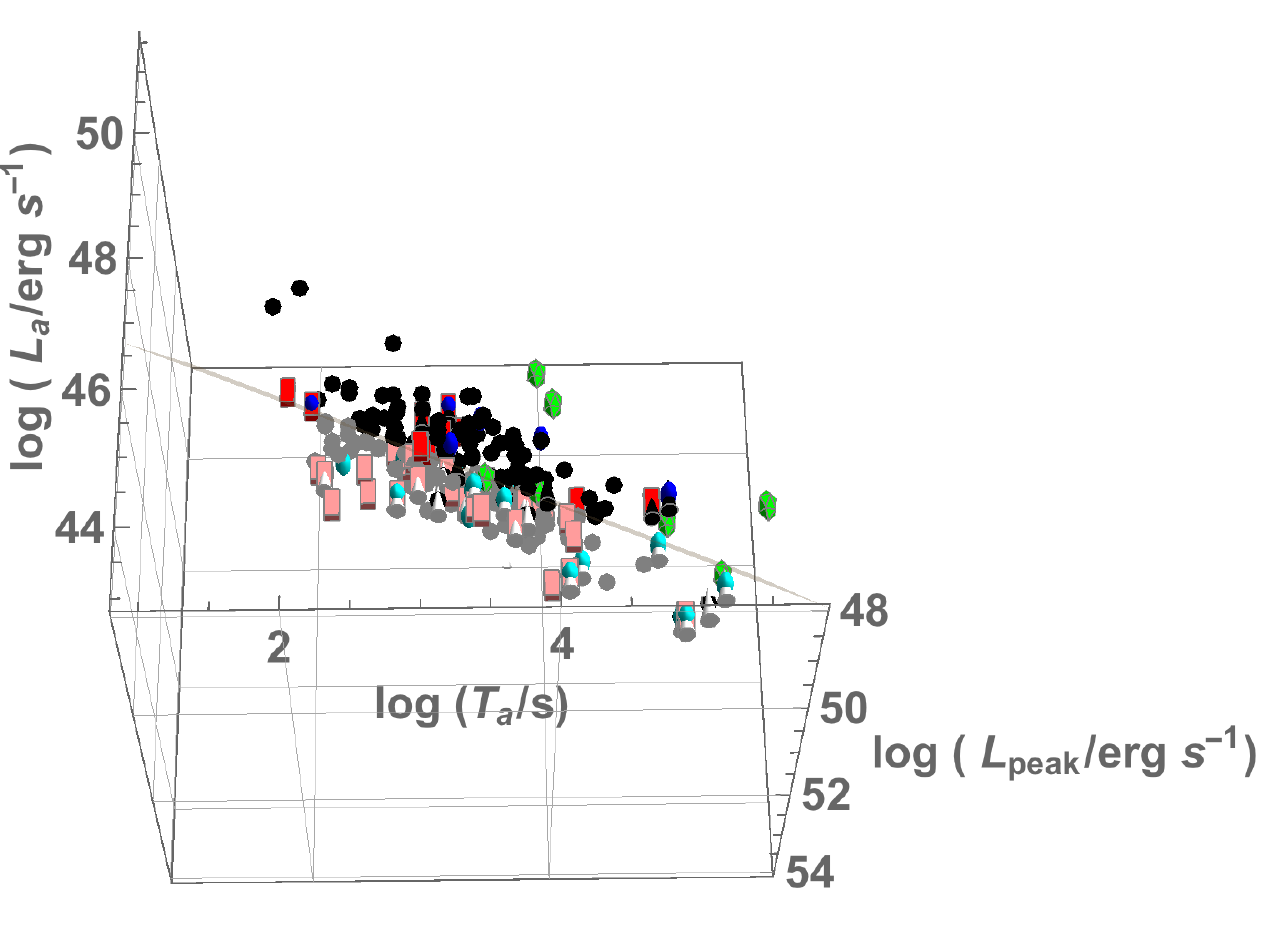}{0.5\textwidth}{(a) Edge-on View}
          \fig{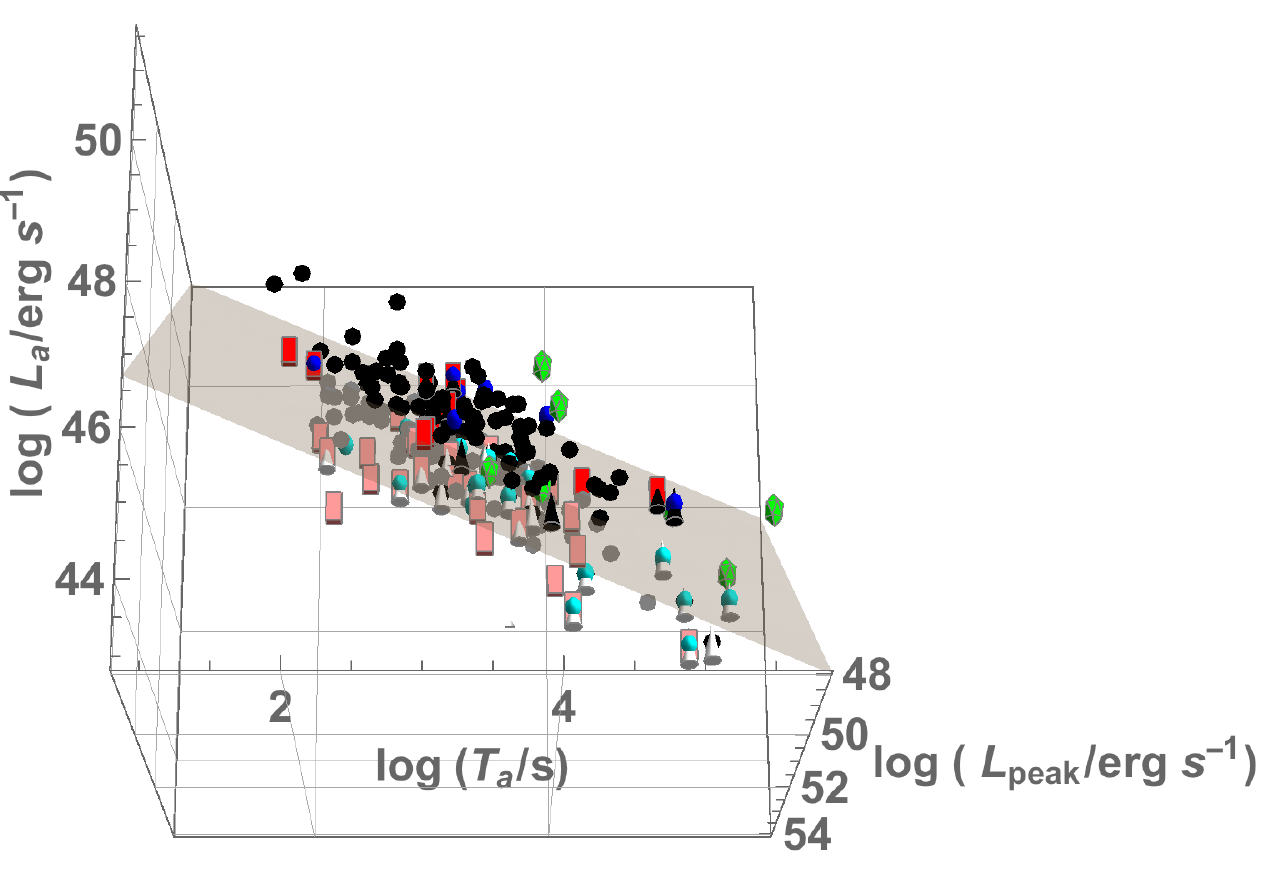}{0.5\textwidth}{(b) Plane View}}
\caption{222 GRBs in the $\log(L_a)-\log(T_a)-\log(L_{\mathrm{peak}})$ space with a plane fitted to the data with the following classifications: GRB-SNe (black cones), XRFs (blue spheres), SEE (cuboids), lGRBs (black circles), UL GRBs (green icosahedrons). The same color coding, but darker colors indicate data points above the plane, while lighter colors indicate GRBs below the plane, except for UL GRBs which are all denoted by bright green truncated icosahedrons.}\label{3-Dplot}
\end{figure}

We show in Figure \ref{3-Dplot} a 3D projection of the relation with the classes of GRBs presented in different shapes and colors: GRB-SNe  (black cones),  XRFs  (blue spheres),  SEE (red cuboids),  lGRBs  (black circles), and UL GRBs  (truncated icosahedron). Darker  colors  indicate  GRBs above the plane, while lighter colors show those below the plane, except for UL GRBs which are all the same shade of green.

\begin{figure}[h!]
\centering
  \includegraphics[width=0.45\columnwidth]{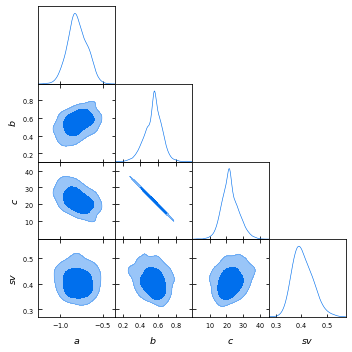}
   \includegraphics[width=0.45\columnwidth]{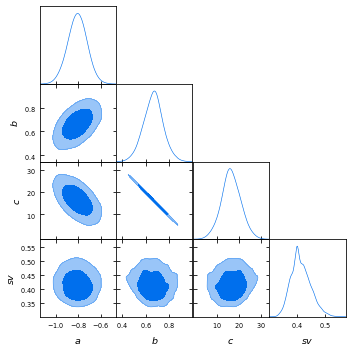}
  \label{gold2}
\caption{Figure showing the contour plots generated by the D'agostini statistical method for the the best-fit parameters and intrinsic scatter for the ``Gold" and ``Gold 2" classes (using the bolometric $K$-correction).}\label{contourgold}
\end{figure}

We test the ``Gold" class using both the old definition from \citet{dainotti16c,Dainotti2017a} along with our new definition for the ``Gold 2" class in order to check if the correlations we derive are more robust than seen in previous literature. There are a total of 69 GRBs in the ``Gold" sample and 100 GRBs in the ``Gold 2" sample, in comparison to 45 GRBs in the ``Gold" set in \citet{Dainotti2017a}.
Using the data points for the ``Gold 2" and the best-fit plane, we are able to derive the $R^2_{adj}$ correlation coefficient for the sample. The ``Gold 2" has a $R^2_{adj}=0.73$, which is 9.9$\%$ lower than that of the ``Gold" in \citet{Dainotti2017a} ($R^2_{adj} = 0.81$). The reasons for this decrease can be due to several factors: (1) the decreased number of data points in the beginning of the plateau emission may give rise to less precise fitting, (2) the different calculation of spectral parameters for the afterglow when compared to previous analysis, (3) the use of a CPL functional form in the prompt emission for GRBs where the necessary parameters are available. The intrinsic scatter of the ``Gold 2" sample ($\sigma = 0.41 \pm 0.03$), calculated through the D'agostini statistical method \citep{DAgostini:95}, is also higher than that of \citet{Dainotti2017a}, where $\sigma=0.33 \pm 0.04$. The current ``Gold" sample in this analysis has $\sigma = 0.40 \pm 0.04$, making it compatible in 1$\sigma$ with ``Gold 2", where the analysis has been done using the $K$-correction computation in the different rest-frame bands of GRBs. It is important to note that the analysis of ``Gold 2" did not increase the $\sigma$, while also increasing the sample size by $45\%$. When we redo the analysis using the bolometric luminosities instead, we obtain  $\sigma=0.40 \pm 0.04$ for the ``Gold" sample and  $\sigma=0.41 \pm 0.04$ for ``Gold 2". This analysis clearly shows that regardless of the different $K$-corrections, the fundamental plane relation results are the same within 1$\sigma$, strengthening our findings toward the use of this relation as a standard candle. The contour plots of the best-fit parameters generated through the \citet{DAgostini:95} method is shown in Figure \ref{gold2}.

\begin{figure}
    \centering
    \includegraphics[width=0.45\columnwidth]{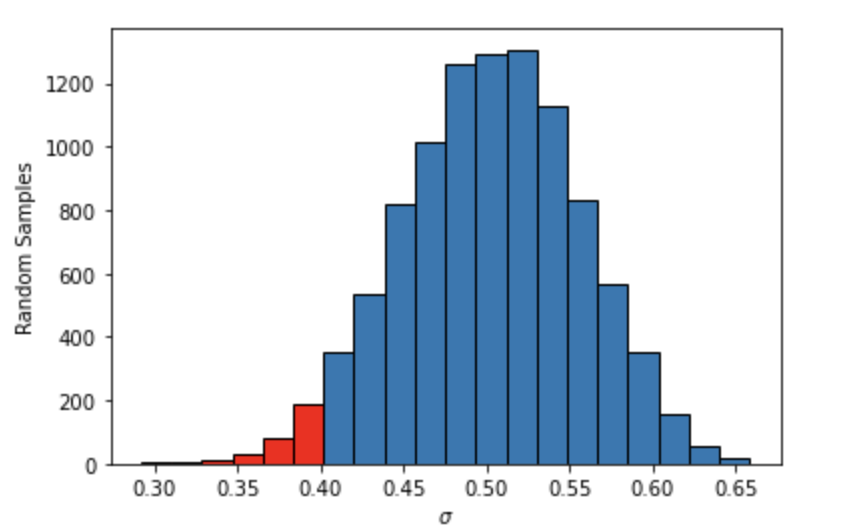}
    \includegraphics[width=0.45\columnwidth]{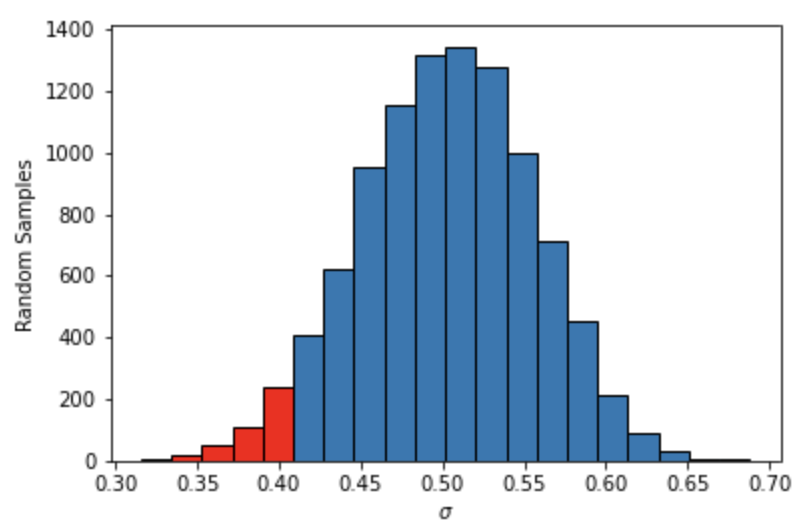}
    \caption{Histograms detailing the random $\sigma$ distributions taken for a set of 69 GRBs (left panel) and 100 GRBs (right panel) 10,000 times. The red bins mark those that are $\leq \sigma$ obtained through analyzing the ``Gold" and ``Gold 2" classes. }
    \label{probhisto}
\end{figure}

Furthermore, in order to determine whether the $\sigma$ we obtain is truly characteristic of the samples rather than due to by chance, we draw a random sample of 69 and 100 GRBs, respectively, corresponding to the ``Gold" and ``Gold 2" classes, out of the total sample of 222 GRBs. We then calculate the best-fit parameters and intrinsic scatter of these samples, and bootstrap the sample 10,000 times, creating histograms representing the distribution of $\sigma$ from the random samples. We see that out of the 10,000 samples, 296 have a $\sigma < 0.40$-- therefore the probability that we randomly obtain a $\sigma < 0.40$ corresponding to the ``Gold" class is 3.0\%. Similarly, out of the 10,000 samples for the ``Gold 2" class, 450 have a $\sigma < 0.41$, corresponding to a probability of 4.9\% of achieving this scatter at random. Because both classes have probabilities of less than 5\% of obtaining their respective intrinsic scatters randomly, it further supports the robustness of our correlations. Histograms of the distributions of the samples are detailed in Figure \ref{probhisto}. 

\FloatBarrier
\subsection{The Fundamental Planes According to the Closure Relations}
\label{Closureplane}
We also group GRBs in terms of their astrophysical environments in relation to Table \ref{CR}, and plot the same 3D relation ($\log(T_a)$, $\log(L_{\mathrm{peak}})$, $\log(L_a)$) to investigate whether grouping GRBs in accordance to their astrophysical environments rather than their classes results in more tighter correlations. We group GRBs with respect to their astrophysical environments taken from Table \ref{CR}, as well as grouping together GRBs that follow a constant-density ISM and wind environment regardless of their cooling regime. The plots of the respective planes are shown in Figure \ref{3-Dplotspecific}.
 The $R^2_{adj}$ correlation coefficients with respect to the planes, the best-fit parameters, and intrinsic scatters are presented in Table \ref{Dagostini}, along with the contour plots in Figure \ref{plotsspecific}.

With the exception of the two fast cooling groups, every other group's $\sigma$ is consistent with the ones of the ``Gold" samples at the 1$\sigma$ level. This is a revealing discovery, since the procedure to create the groups differs enormously. The groups pertaining to the astrophysical environments are derived through checking if they fulfill the theoretical closure relations corresponding to the ES model in Phase III of the LCs, whereas the ``Gold" groups were extracted phenomenologically through fitting the LCs. Furthermore, the GRBs in the ``Gold" groups all display similar intrinsic physical processes, since they are subsamples of lGRBs, whereas GRBs originating from the same environment may not necessarily all have similar intrinsic physical processes. This finding shows that GRBs grouped into their astrophysical environments should continue to be pursued as possible standard candles, as there are intrinsic consistencies in 1$\sigma$ of observable parameters within some of their particular subsamples.
 
\begin{figure}
    \centering
    \includegraphics[width=0.45\columnwidth]{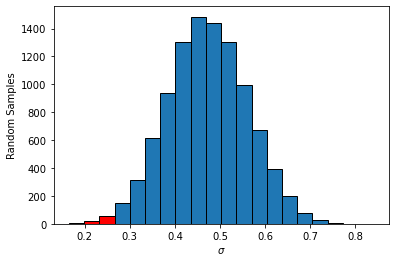}
    \includegraphics[width=0.45\columnwidth]{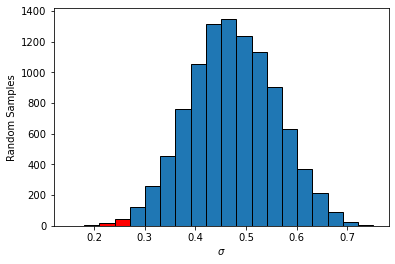}
    \caption{Histograms detailing the random $\sigma$ distributions taken for a set of 31 GRBs (left panel) and 32 GRBs (right panel) 10,000 times. The red bins mark those that are $\leq \sigma$ obtained through analyzing the ISM fast cooling and ISM slow cooling classes.} 
    \label{probhistoCR}
\end{figure}

\begin{figure}[h!]
\centering
\gridline{\fig{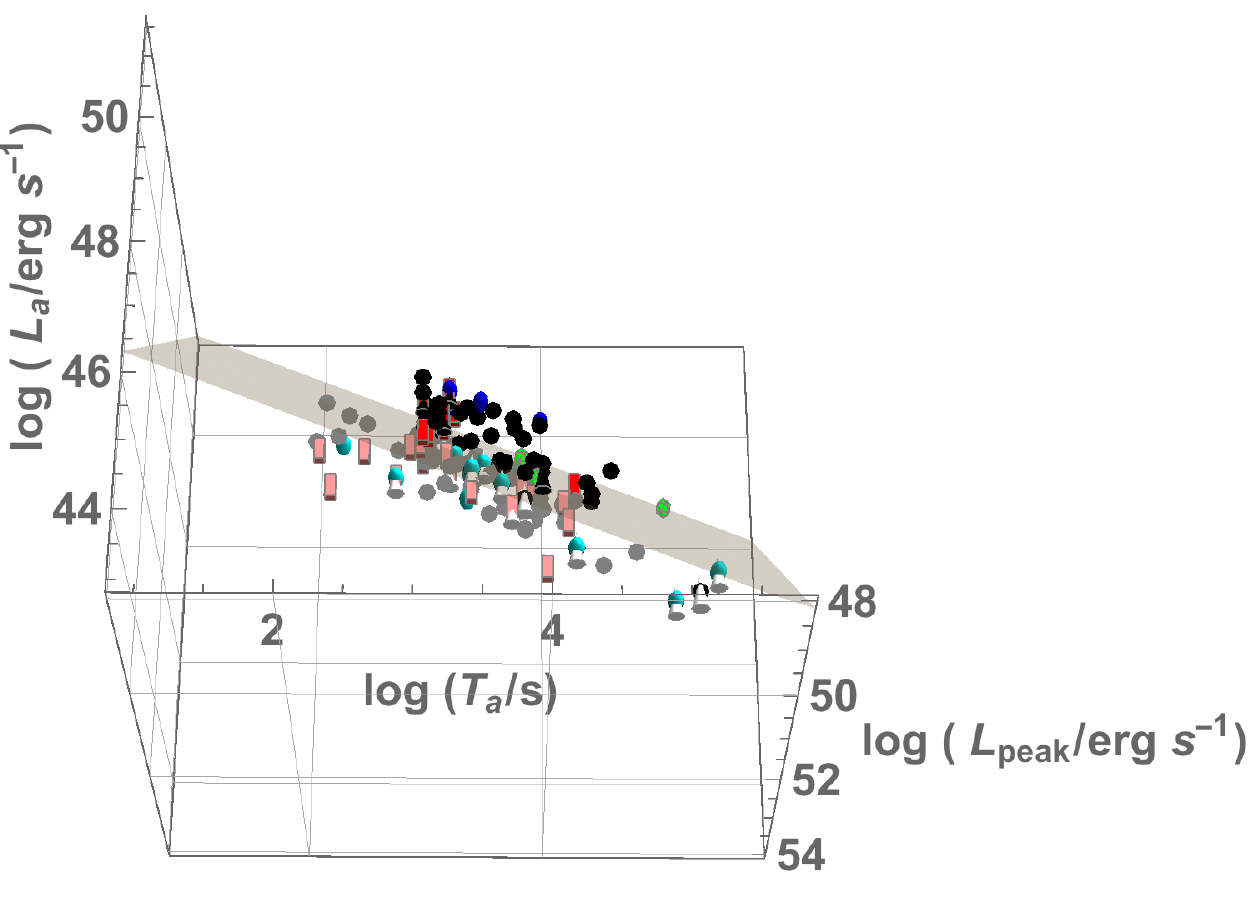}{0.5\textwidth}{(a) 125 GRBs fulfill all ISM environments}
          \fig{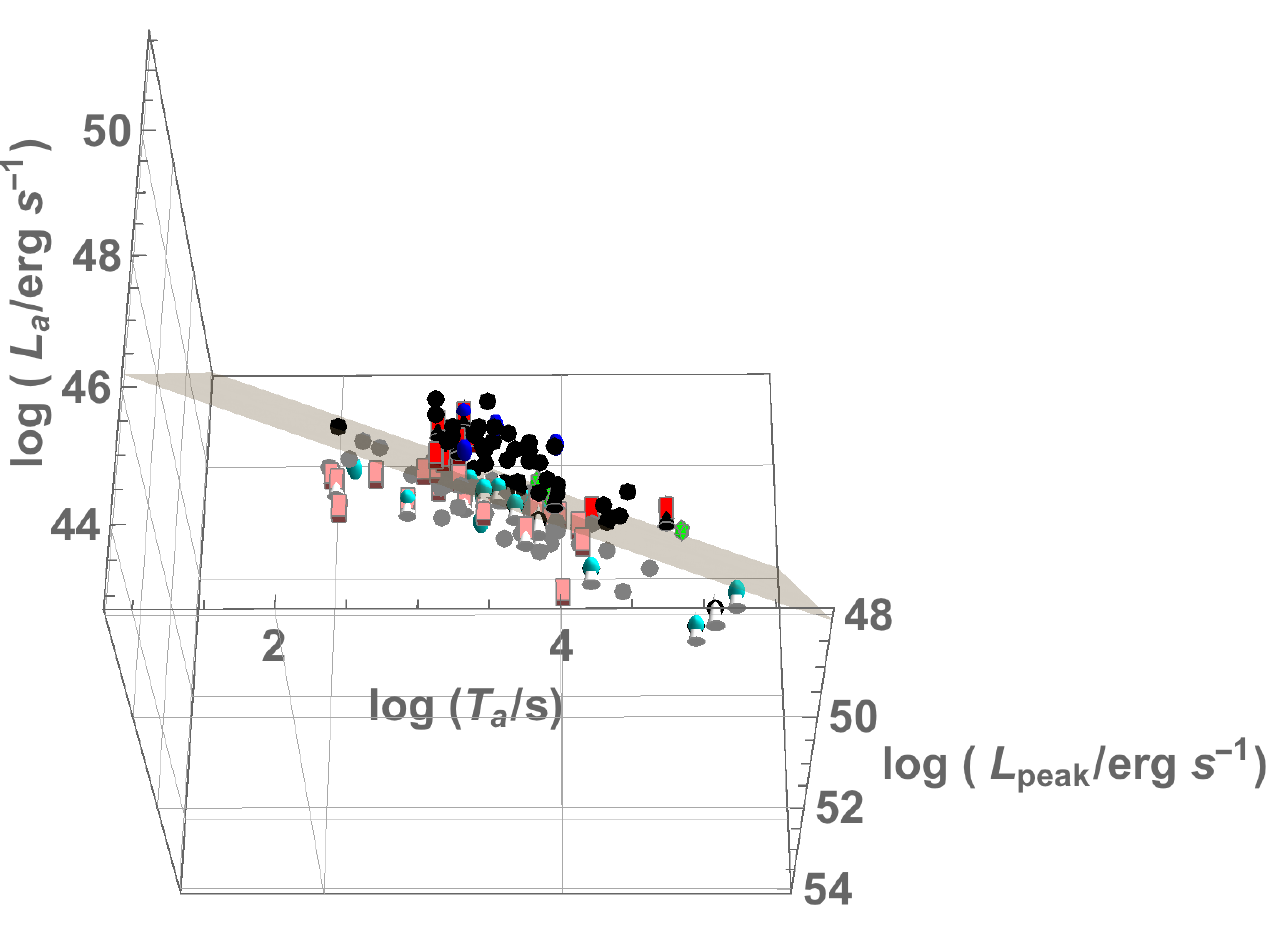}{0.5\textwidth}{(b) 142 GRBs fulfill all wind environments}}
\gridline{\fig{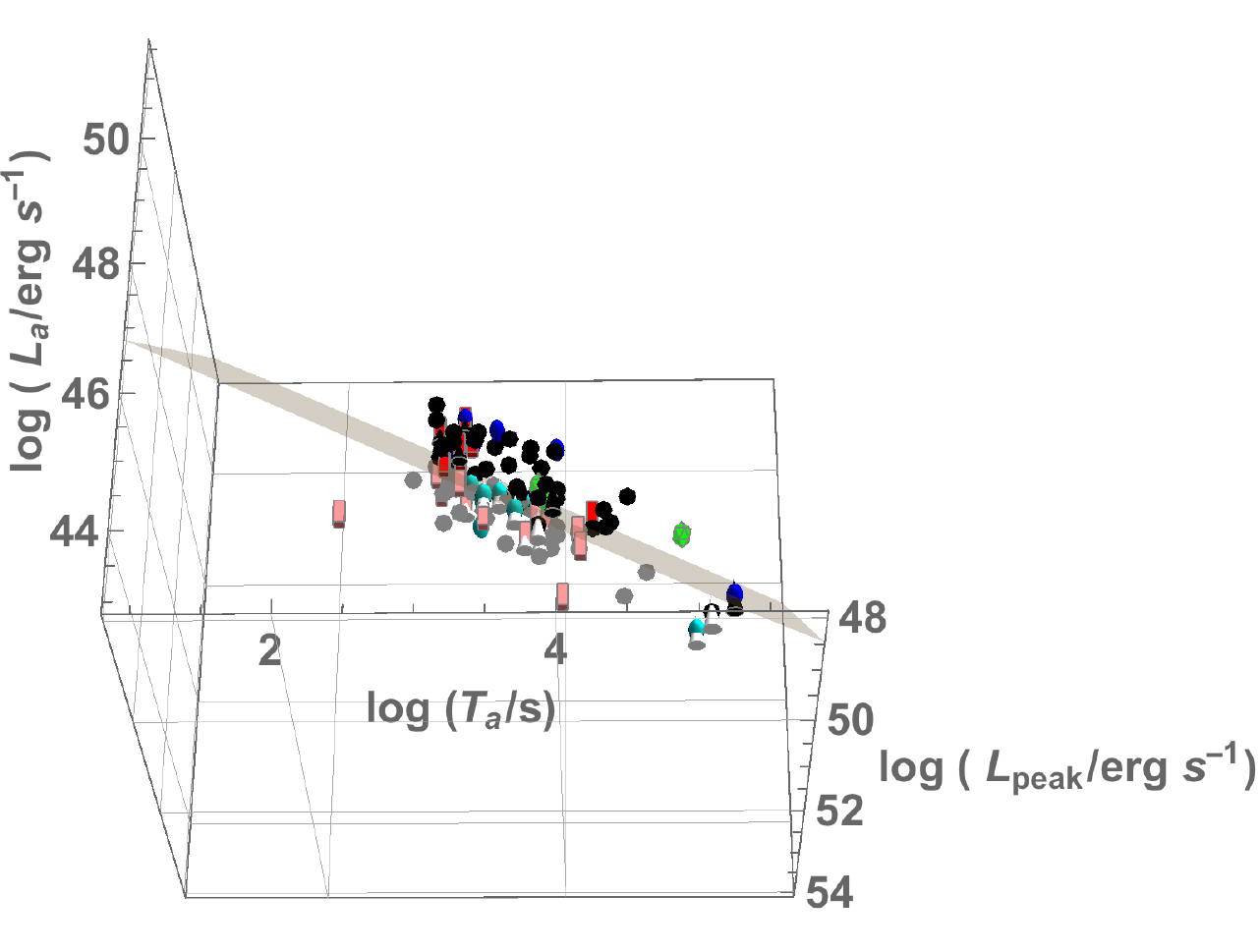}{0.5\textwidth}{(c) 110 GRBs fulfill  ISM slow cooling}
\fig{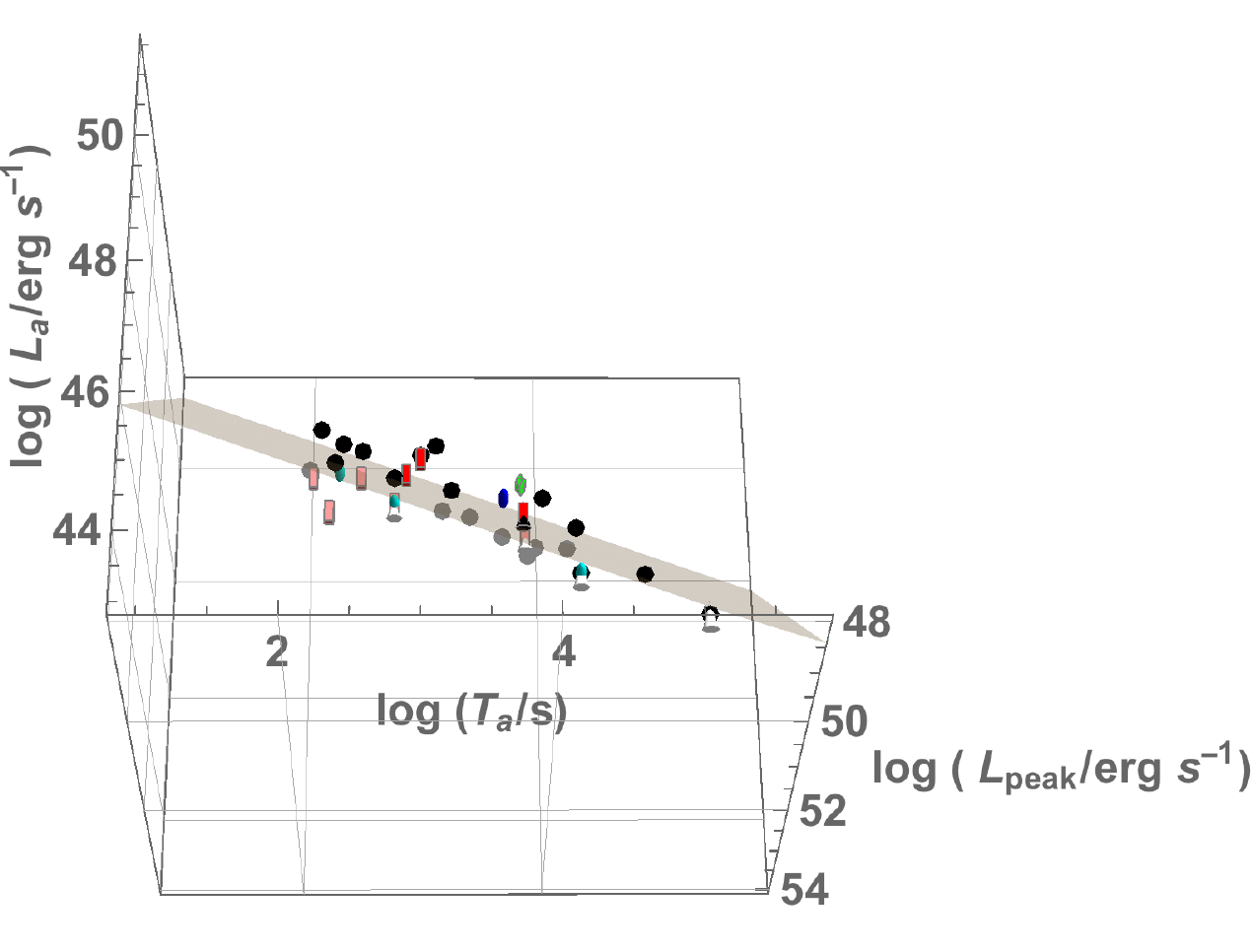}{0.5\textwidth}{(d) 31 GRBs fulfill ISM fast cooling}}
\gridline{\fig{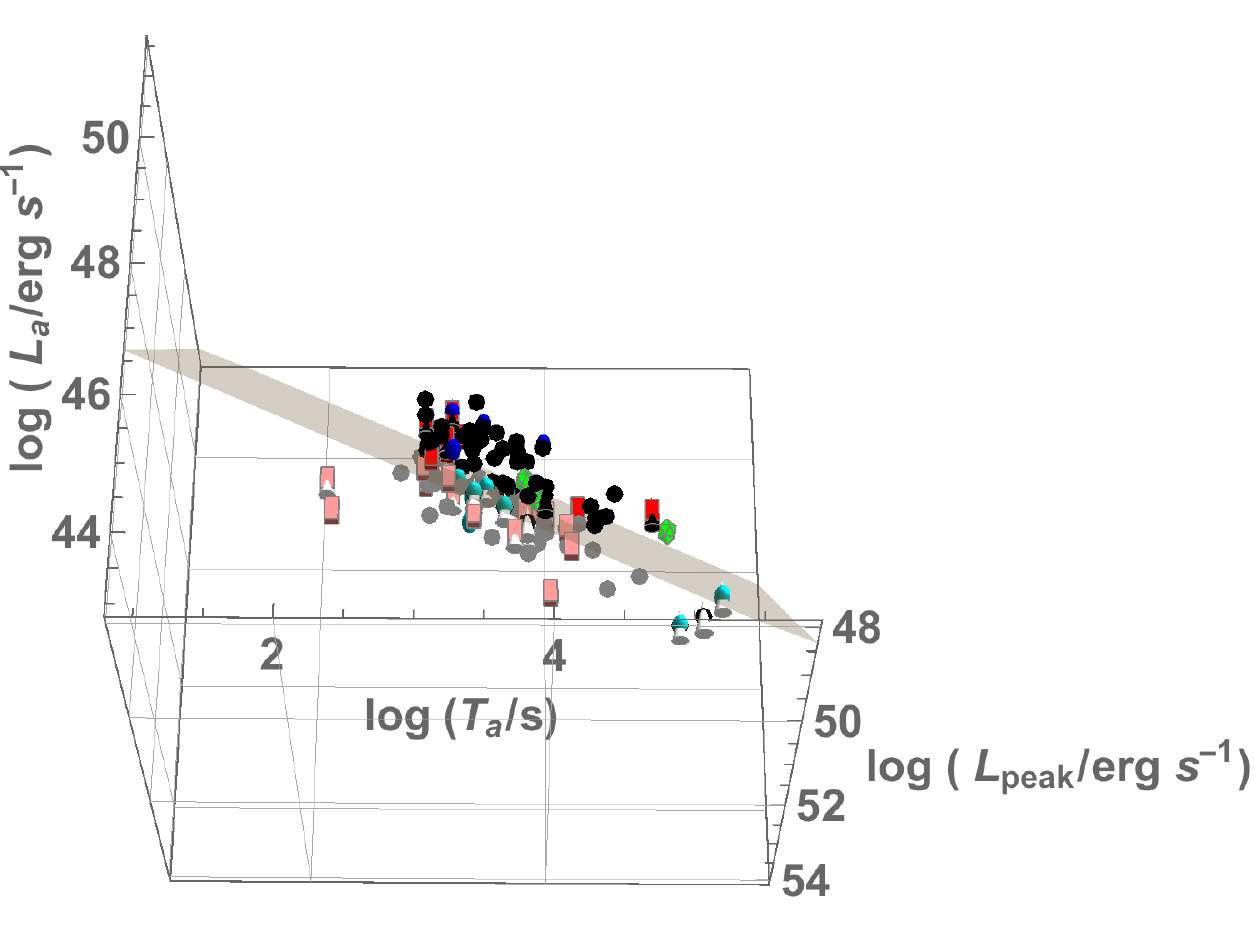}{0.5\textwidth}{(e) 127 GRBs fulfill wind slow cooling}
\fig{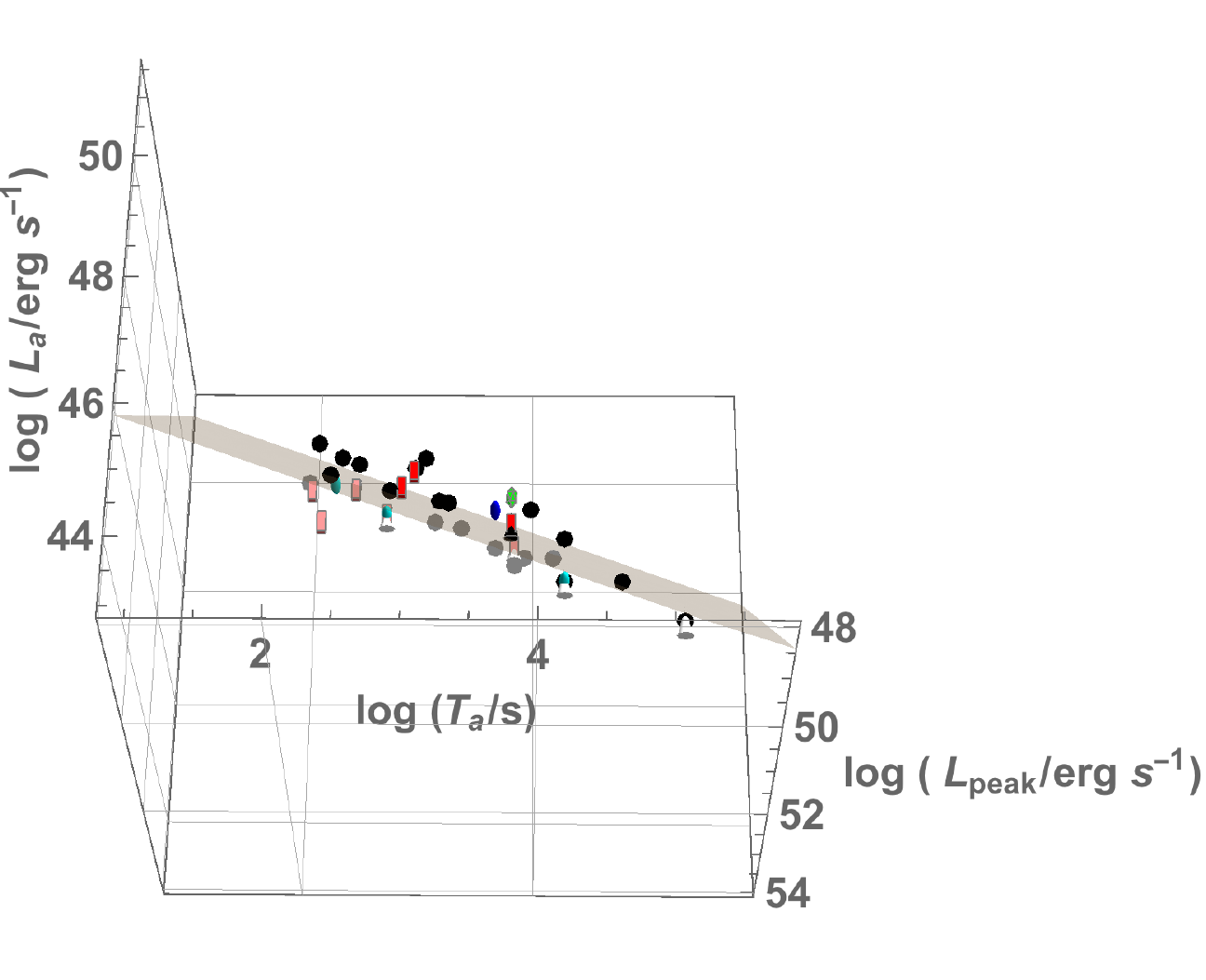}{0.5\textwidth}{(f) 32 GRBs fulfill wind fast cooling}}
\caption{The fundamental planes according to astrophysical environments with same classifications as Figure \ref{3-Dplot}.}\label{3-Dplotspecific}
\end{figure}

\begin{figure}[h!]
\gridline{\fig{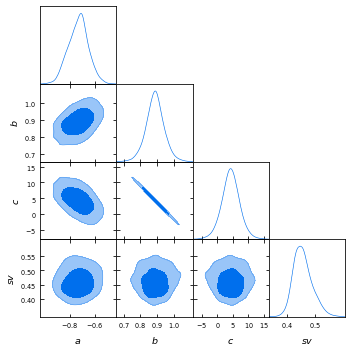}{0.35\textwidth}{(a) All ISM environments}
          \fig{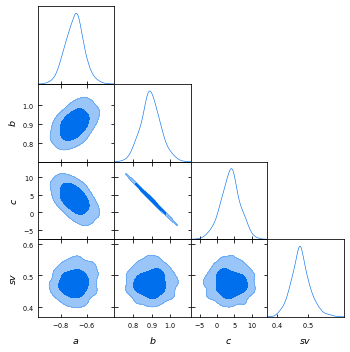}{0.35\textwidth}{(b) All wind environments}}
\gridline{\fig{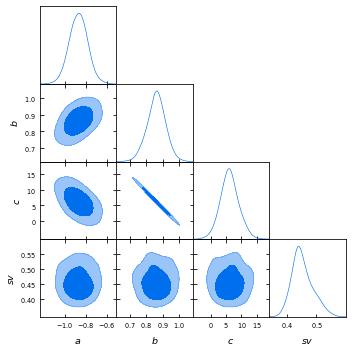}{0.35\textwidth}{(c) ISM slow cooling environments}
          \fig{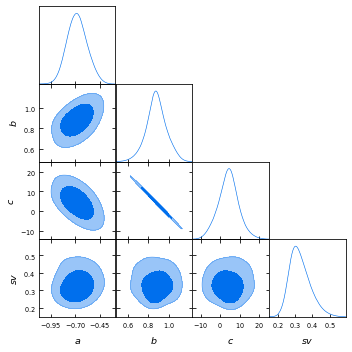}{0.35\textwidth}{(d) ISM fast cooling environments}}
\gridline{\fig{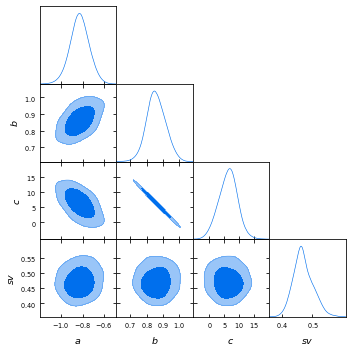}{0.35\textwidth}{(e) Wind slow cooling environments}
          \fig{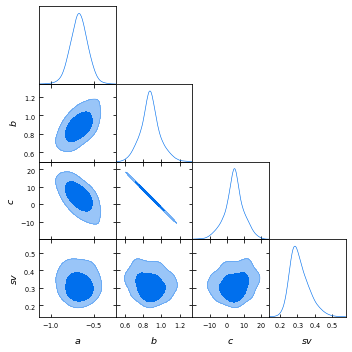}{0.35\textwidth}{(f) Wind fast cooling environments}}
\caption{Contour plots generated by the D'agostini statistical method of the best-fit parameters and $\sigma$ scatter of closure relation groups from Table \ref{CR}.}\label{plotsspecific}
\end{figure}

When looking at the ISM fast cooling and wind fast cooling groups, we obtain a $\sigma$ lower than previously  seen in literature, including the ``Gold" class in \citet{Dainotti2017a}, of $\sigma = 0.29$. In order to verify that these results are not drawn by chance, we again check the probability of obtaining such a $\sigma$ or lower by using the same bootstrapping method that we use previously for the ``Gold" classes. Through this, we determine that the probability of randomly obtaining a $\sigma < 0.29$ is 1.76\% for the ISM fast cooling sample size, and 1.35\% for the wind fast cooling sample. The histograms for the distributions are presented in Figure \ref{probhistoCR}. Thus, the $\sigma$ we obtain is indeed robust, leading us to conclude that the fast cooling groups hold the highest potential to eventually be used as standard candles. 
\begin{table}[!htb]
\begin{center}
\begin{tabular}{|c|c|c|c|c|c|c|}
\hline
Sample & $a$ & $b$ & $c$ & $\sigma$ & $R^{2}_{adj}$ & $N$ \\ \hline
   All ISM   & -0.73 $\pm$ 0.08 &  0.89 $\pm$ 0.05&  4.28 $\pm$ 2.94 &  0.46 $\pm $0.03 & 0.81 & 125  \\
   All Wind   &  -0.70 $\pm$ 0.08 &  0.89 $\pm$ 0.05 &  4.14 $\pm$ 2.91 &  0.47 $\pm$ 0.03 & 0.79 & 142   \\
   ISM Slow Cooling   &   -0.88 $\pm$ 0.09 &  0.85 $\pm$ 0.06 & 6.71 $\pm$  3.05& 0.44 $\pm$ 0.04 & 0.3 & 110   \\
   ISM Fast Cooling   &    -0.68 $\pm$ 0.1&  0.87 $\pm$ 0.10&  4.84 $\pm$ 5.20 & 0.29 $\pm$ 0.06  & 0.88 & 31 \\
   Wind Slow Cooling   &   -0.83 $\pm$ 0.10&  0.86 $\pm$ 0.06 &  6.21 $\pm$ 3.18 &  0.46 $\pm$ 0.03& 0.80 & 127   \\
   Wind Fast Cooling    &  -0.68 $\pm$ 0.1 &  0.87 $\pm$ 0.11 &  4.25 $\pm$ 5.67 &  0.29 $\pm$ 0.06 & 0.88 & 32  \\
   \hline
   ``Gold" (bolometric)     &  -0.80 $\pm$ 0.12 &  0.54 $\pm$ 0.10 &  22.66 $\pm$ 5.25 &  0.40 $\pm$ 0.04 & 0.65 & 69   \\
   ``Gold2" (bolometric)   &  -0.80 $\pm$ 0.09 &  0.68 $\pm$ 0.09 &  15.58 $\pm$ 4.51 &  0.41 $\pm$ 0.04 & 0.73 & 100   \\\hline
\end{tabular}
\end{center}
\caption{Table Indicating the Best-Fit Parameters, Intrinsic Scatter, and Number of GRBs with Redshift Satisfying at Least One Relation from Table \ref{CR} and the ``Gold 2" Class.}
\label{Dagostini}
\end{table}

\FloatBarrier

\FloatBarrier
\section{Summary and Conclusions} \label{conclusion}
In summary, we test whether a set of closure relations corresponding to two distinct astrophysical environments and cooling regimes are fulfilled for 455 Swift X-ray LCs from 2005 January to 2019 August that show a plateau. We employ the closure relations as a quick test on the reliability of the ES emission, and also to infer information about their astrophysical environments. We also confirm the existence of an updated 3D fundamental plane relation between the rest-frame time at the end of the plateau emission, $\log T_a$, the prompt peak luminosity, $\log L_{\mathrm{peak}}$, and luminosity at the end of the plateau emission, $\log L_a$, with two additional years of Swift observations. 
We introduce a new definition of the ``Gold" class, called the ``Gold 2" and compare it to the ``Gold" mentioned in previous papers. Finally, we analyze the 3D fundamental plane for a set of GRBs that fulfill the closure relations and thus reveal peculiar astrophysical environments. 
In conclusion, we have answered the main queries stated in \S \ref{Intro}. 
\begin{enumerate}
\item We find that the ES model works for the majority of Swift GRBs, where 11 out of the 16 closure relation groups we tested have at least 50\% of their GRBs fulfilling the relations. More complex evolution of afterglows, such as considering nonlinear particle acceleration in addition to an energy injection mechanism such as a magnetar or mass accretion onto a black hole, may be able to account for GRBs that do not follow the ES model.

With respect to their astrophysical environments, the majority of GRBs in both the known and unknown redshift data sets fall under an constant-density ISM or wind environment with slow cooling.

\item The 3D fundamental plane relation detailed in \citet{dainotti16c} and \citet{Dainotti2017a} still holds with the addition of two years of {\it Swift} GRBs. We find that the ``Gold 2" class composed of 100 GRBs has a smaller $R_{adj}^2= 0.73$ and a larger intrinsic scatter $\sigma= 0.41$ than the old ``Gold". This is due to the different fitting procedures, calculation of spectral parameters, and the fact that the ``Gold 2" is $2.2$ times larger than the ``Gold" published in \cite{Dainotti2017a}. Furthermore, we redo the analysis of the fundamental plane through using the bolometric luminosities instead, and find that the $\sigma$ for both the ``Gold" and ``Gold 2" classes are equivalent within 1$\sigma$ of \citet{dainotti16c}, \citet{Dainotti2017a}, showing that regardless of the $K$-correction used, the correlations are compatible within 1$\sigma$.

\item We compute the 3D fundamental plane relation with respect to the astrophysical environments and cooling regimes of the GRBs in our data set, and see that the majority of the groups have a consistent intrinsic scatter ($\sigma$) with one another, as well as with the ``Gold" classes. Since the methods we used to determine these groups varies significantly (a theoretical approach for the astrophysical environments and a phenomenological approach for the ``Gold" classes), this also strengthens the argument for pursuing GRBs grouped into their astrophysical environments and cooling regimes as possible standard candles. Furthermore, the two groups of GRBs that satisfy the fast cooling CRs have the lowest $\sigma$ obtained thus far in literature, further strengthening this argument above. Further investigation of these groups will be the topic of further papers, as they have the highest potential to eventually be used as standard candles.
\end{enumerate}

\newpage
\section{Acknowledgements}
G.S. is grateful for the support of the United States Department of Energy in funding the Science Undergraduate Laboratory Internship (SULI) program. M.G. Dainotti is grateful to MINIATURA2, grant No. 2018/02/X/ST9/03673 and the American Astronomical Society Chretienne Fellowship. N.F. acknowledges the support from UNAM-DGAPA-PAPIT through grant IA102019.  X.H. acknowledges support from DGAPAUNAM PAPIIT IN104517 and CONACyT.
S.N. acknowledges the "JSPS Grant-in-Aid for Scientific Research ``KAKENHI" (A) with 
grant No. JP19H00693, the ``Pioneering Program of RIKEN for Evolution of Matter in the universe (r-EMU)", and
``Interdisciplinary Theoretical and Mathematical Sciences Program of RIKEN (iTHEMS)". The authors are grateful for the help of Ray Wynne and Zooey Ngyuen, undergraduate students at the Massachusetts Institute of Technology and University of California Los Angeles, for their help in the fitting of the lightcurves. Authors are particularly grateful to Giuseppe Sarracino and Stefano Savastano for the help in writing the python code for deriving the best-fit parameters of the fundamental plane relation. This work made use of data supplied by the UK Swift Science Data Centre at the University of Leicester.

\bibliography{bib_Swiftpaper.tex}

\begin{thebibliography}{}
\providecommand\natexlab[1]{#1}
\providecommand\JournalTitle[1]{#1}

\bibitem[{{Barthelmy} {et~al.}(2005){Barthelmy}, {Barbier}, {Cummings},
  {Fenimore}, {Gehrels}, {Hullinger}, {Krimm}, {Markwardt}, {Palmer},
  {Parsons}, {Sato}, {Suzuki}, {Takahashi}, {Tashiro}, \&
  {Tueller}}]{Barthelmy:05}
{Barthelmy}, S.~D., {Barbier}, L.~M., {Cummings}, J.~R., {et~al.} 2005,
  \href{http://dx.doi.org/10.1007/s11214-005-5096-3}{\JournalTitle{Space
  Science Reviews}, 120, 143}

\bibitem[{{Beniamini} \& {Giannios}(2017)}]{Beniamini2017}
{Beniamini}, P., \& {Giannios}, D., M. B.~D. 2017,
  \href{http://dx.doi.org/https://ui.adsabs.harvard.edu/abs/2017MNRAS.472.3058B}{\JournalTitle{\mnras},
  472, 3058}

\bibitem[{{Beniamini} \& {Mochkovitch}(2017)}]{BeniaminiandMochkovitch2017}
{Beniamini}, P., \& {Mochkovitch}, R. 2017,
  \href{http://dx.doi.org/10.1051/0004-6361/201730523}{\JournalTitle{\aap},
  605, A60}

\bibitem[{{Bloom} {et~al.}(2001){Bloom}, {Frail}, \& {Sari}}]{Bloom2001}
{Bloom}, J.~S., {Frail}, D.~A., \& {Sari}, R. 2001,
  \href{http://dx.doi.org/10.1086/321093}{\JournalTitle{\aj}, 121, 2879}

\bibitem[{{Burrows} {et~al.}(2005{\natexlab{a}}){Burrows}, {Romano}, {Falcone},
  {Kobayashi}, {Zhang}, {Moretti}, {O'Brien}, {Goad}, {Campana}, {Page},
  {Angelini}, {Barthelmy}, {Beardmore}, {Capalbi}, {Chincarini}, {Cummings},
  {Cusumano}, {Fox}, {Giommi}, {Hill}, {Kennea}, {Krimm}, {Mangano},
  {Marshall}, {M{\'e}sz{\'a}ros}, {Morris}, {Nousek}, {Osborne}, {Pagani},
  {Perri}, {Tagliaferri}, {Wells}, {Woosley}, \& {Gehrels}}]{Burrows05}
{Burrows}, D.~N., {Romano}, P., {Falcone}, A., {et~al.} 2005{\natexlab{a}},
  \href{http://dx.doi.org/10.1126/science.1116168}{\JournalTitle{Science}, 309,
  1833}

\bibitem[{{Burrows} {et~al.}(2005{\natexlab{b}}){Burrows}, {Hill}, {Nousek},
  {Kennea}, {Wells}, {Osborne}, {Abbey}, {Beardmore}, {Mukerjee}, {Short},
  {Chincarini}, {Campana}, {Citterio}, {Moretti}, {Pagani}, {Tagliaferri},
  {Giommi}, {Capalbi}, {Tamburelli}, {Angelini}, {Cusumano}, {Br{\"a}uninger},
  {Burkert}, \& {Hartner}}]{2005SSRv..120..165B}
{Burrows}, D.~N., {Hill}, J.~E., {Nousek}, J.~A., {et~al.} 2005{\natexlab{b}},
  \href{http://dx.doi.org/10.1007/s11214-005-5097-2}{\JournalTitle{Space
  Science Reviews}, 120, 165}

\bibitem[{{Cannizzo} \& {Gehrels}(2009)}]{Cannizzo2009}
{Cannizzo}, J.~K., \& {Gehrels}, N. 2009,
  \href{http://dx.doi.org/10.1088/0004-637X/700/2/1047}{\JournalTitle{\apj},
  700, 1047}

\bibitem[{{Cannizzo} {et~al.}(2011){Cannizzo}, {Troja}, \&
  {Gehrels}}]{cannizzo2011}
{Cannizzo}, J.~K., {Troja}, E., \& {Gehrels}, N. 2011,
  \href{http://dx.doi.org/10.1088/0004-637X/734/1/35}{\JournalTitle{\apj}, 734,
  35}

\bibitem[{{Cardone} {et~al.}(2009){Cardone}, {Capozziello}, \&
  {Dainotti}}]{cardone09}
{Cardone}, V.~F., {Capozziello}, S., \& {Dainotti}, M.~G. 2009,
  \href{http://dx.doi.org/10.1111/j.1365-2966.2009.15456.x}{\JournalTitle{\mnras},
  400, 775}

\bibitem[{{Cardone} {et~al.}(2010){Cardone}, {Dainotti}, {Capozziello}, \&
  {Willingale}}]{cardone10}
{Cardone}, V.~F., {Dainotti}, M.~G., {Capozziello}, S., \& {Willingale}, R.
  2010,
  \href{http://dx.doi.org/10.1111/j.1365-2966.2010.17197.x}{\JournalTitle{\mnras},
  408, 1181}

\bibitem[{{Cavallo} \& {Rees}(1978)}]{Cavallo}
{Cavallo}, G., \& {Rees}, M.~J. 1978,
  \href{http://dx.doi.org/10.1093/mnras/183.3.359}{\JournalTitle{\mnras}, 183,
  359}

\bibitem[{{Chevalier} \& {Li}(2000)}]{Chevalier+00}
{Chevalier}, R.~A., \& {Li}, Z. 2000,
  \href{http://dx.doi.org/10.1086/308914}{\JournalTitle{\apj}, 536, 195}

\bibitem[{{Cucchiara} {et~al.}(2011){Cucchiara}, {Levan}, {Fox}, {Tanvir},
  {Ukwatta}, {Berger}, {Kr{\"u}hler}, {K{\"u}pc{\"u} Yolda{\c s}}, {Wu},
  {Toma}, {Greiner}, {Olivares}, {Rowlinson}, {Amati}, {Sakamoto}, {Roth},
  {Stephens}, {Fritz}, {Fynbo}, {Hjorth}, {Malesani}, {Jakobsson}, {Wiersema},
  {O'Brien}, {Soderberg}, {Foley}, {Fruchter}, {Rhoads}, {Rutledge}, {Schmidt},
  {Dopita}, {Podsiadlowski}, {Willingale}, {Wolf}, {Kulkarni}, \&
  {D'Avanzo}}]{cucchiara11}
{Cucchiara}, A., {Levan}, A.~J., {Fox}, D.~B., {et~al.} 2011,
  \href{http://dx.doi.org/10.1088/0004-637X/736/1/7}{\JournalTitle{\apj}, 736,
  7}

\bibitem[{D'Agostini(1995)}]{DAgostini:95}
D'Agostini, G. 1995,
  \href{http://dx.doi.org/https://doi.org/10.1016/0168-9002(95)00274-X}{\JournalTitle{Nuclear
  Instruments and Methods in Physics Research Section A: Accelerators,
  Spectrometers, Detectors and Associated Equipment}, 362, 487 }

\bibitem[{{Dai} \& {Cheng}(2001)}]{Dai&Cheng2001}
{Dai}, Z.~G., \& {Cheng}, K. 2001,
  \href{http://dx.doi.org/10.1086/308914}{\JournalTitle{\apj}, 558, 109}

\bibitem[{{Dai} \& {Lu}(1998)}]{dai98}
{Dai}, Z.~G., \& {Lu}, T. 1998, \JournalTitle{\aap}, 333, L87

\bibitem[{{Dainotti} {et~al.}(2015{\natexlab{a}}){Dainotti}, {Petrosian},
  {Willingale}, {O'Brien}, {Ostrowski}, \& {Nagataki}}]{Dainotti2015b}
{Dainotti}, M., {Petrosian}, V., {Willingale}, R., {et~al.} 2015{\natexlab{a}},
  \href{http://dx.doi.org/10.1093/mnras/stv1229}{\JournalTitle{\mnras}, 451,
  3898}

\bibitem[{{Dainotti} {et~al.}(2008){Dainotti}, {Cardone}, \&
  {Capozziello}}]{Dainotti2008}
{Dainotti}, M.~G., {Cardone}, V.~F., \& {Capozziello}, S. 2008,
  \href{http://dx.doi.org/10.1111/j.1745-3933.2008.00560.x}{\JournalTitle{\mnras},
  391, L79}

\bibitem[{{Dainotti} {et~al.}(2013){Dainotti}, {Cardone}, {Piedipalumbo}, \&
  {Capozziello}}]{Dainotti2013a}
{Dainotti}, M.~G., {Cardone}, V.~F., {Piedipalumbo}, E., \& {Capozziello}, S.
  2013, \href{http://dx.doi.org/10.1093/mnras/stt1516}{\JournalTitle{\mnras},
  436, 82}

\bibitem[{{Dainotti} \& {Del Vecchio}(2017)}]{Dainotti2017b}
{Dainotti}, M.~G., \& {Del Vecchio}, R. 2017,
  \href{http://dx.doi.org/10.1016/j.newar.2017.04.001}{\JournalTitle{New
  Astronomy Reviews}, 77, 23}

\bibitem[{{Dainotti} {et~al.}(2015{\natexlab{b}}){Dainotti}, {Del Vecchio},
  {Nagataki}, \& {Capozziello}}]{Dainotti15a}
{Dainotti}, M.~G., {Del Vecchio}, R., {Nagataki}, S., \& {Capozziello}, S.
  2015{\natexlab{b}},
  \href{http://dx.doi.org/10.1088/0004-637X/800/1/31}{\JournalTitle{\apj}, 800,
  31}

\bibitem[{{Dainotti} {et~al.}(2011){Dainotti}, {Fabrizio Cardone},
  {Capozziello}, {Ostrowski}, \& {Willingale}}]{dainotti11a}
{Dainotti}, M.~G., {Fabrizio Cardone}, V., {Capozziello}, S., {Ostrowski}, M.,
  \& {Willingale}, R. 2011,
  \href{http://dx.doi.org/10.1088/0004-637X/730/2/135}{\JournalTitle{\apj},
  730, 135}

\bibitem[{{Dainotti} {et~al.}(2017{\natexlab{a}}){Dainotti}, {Hernandez},
  {Postnikov}, {Nagataki}, {O'brien}, {Willingale}, \&
  {Striegel}}]{Dainotti2017c}
{Dainotti}, M.~G., {Hernandez}, X., {Postnikov}, S., {et~al.}
  2017{\natexlab{a}},
  \href{http://dx.doi.org/10.3847/1538-4357/aa8a6b}{\JournalTitle{\apj}, 848,
  88}

\bibitem[{{Dainotti} {et~al.}(2017{\natexlab{b}}){Dainotti}, {Nagataki},
  {Maeda}, {Postnikov}, \& {Pian}}]{Dainotti2017a}
{Dainotti}, M.~G., {Nagataki}, S., {Maeda}, K., {Postnikov}, S., \& {Pian}, E.
  2017{\natexlab{b}},
  \href{http://dx.doi.org/10.1051/0004-6361/201628384}{\JournalTitle{\aap},
  600, A98}

\bibitem[{{Dainotti} {et~al.}(2016){Dainotti}, {Postnikov}, {Hernandez}, \&
  {Ostrowski}}]{dainotti16c}
{Dainotti}, M.~G., {Postnikov}, S., {Hernandez}, X., \& {Ostrowski}, M. 2016,
  \href{http://dx.doi.org/10.3847/2041-8205/825/2/L20}{\JournalTitle{\apjl},
  825, L20}

\bibitem[{{Dainotti} {et~al.}(2010){Dainotti}, {Willingale}, {Capozziello},
  {Fabrizio Cardone}, \& {Ostrowski}}]{dainotti2010}
{Dainotti}, M.~G., {Willingale}, R., {Capozziello}, S., {Fabrizio Cardone}, V.,
  \& {Ostrowski}, M. 2010,
  \href{http://dx.doi.org/10.1088/2041-8205/722/2/L215}{\JournalTitle{\apjl},
  722, L215}

\bibitem[{{Dall'Osso} {et~al.}(2011){Dall'Osso}, {Stratta}, {Guetta}, {Covino},
  {De Cesare}, \& {Stella}}]{dallosso2011}
{Dall'Osso}, S., {Stratta}, G., {Guetta}, D., {et~al.} 2011,
  \href{http://dx.doi.org/10.1051/0004-6361/201014168}{\JournalTitle{\aap},
  526, A121}

\bibitem[{{Del Vecchio} {et~al.}(2016){Del Vecchio}, {Dainotti}, \&
  {Ostrowski}}]{delvecchio16}
{Del Vecchio}, R., {Dainotti}, M.~G., \& {Ostrowski}, M. 2016,
  \href{http://dx.doi.org/10.3847/0004-637X/828/1/36}{\JournalTitle{\apj}, 828,
  36}

\bibitem[{{Evans} {et~al.}(2009){Evans}, {Beardmore}, {Page}, {Osborne},
  {O'Brien}, {Willingale}, {Starling}, {Burrows}, {Godet}, {Vetere}, {Racusin},
  {Goad}, {Wiersema}, {Angelini}, {Capalbi}, {Chincarini}, {Gehrels}, {Kennea},
  {Margutti}, {Morris}, {Mountford}, {Pagani}, {Perri}, {Romano}, \&
  {Tanvir}}]{evans09}
{Evans}, P.~A., {Beardmore}, A.~P., {Page}, K.~L., {et~al.} 2009,
  \href{http://dx.doi.org/10.1111/j.1365-2966.2009.14913.x}{\JournalTitle{\mnras},
  397, 1177}

\bibitem[{{Falcone} {et~al.}(2007){Falcone}, {Morris}, {Racusin}, {Chincarini},
  {Moretti}, {Romano}, {Burrows}, {Pagani}, {Stroh}, {Grupe}, {Campana},
  {Covino}, {Tagliaferri}, {Willingale}, \& {Gehrels}}]{Falcone2007}
{Falcone}, A.~D., {Morris}, D., {Racusin}, J., {et~al.} 2007,
  \href{http://dx.doi.org/10.1086/523296}{\JournalTitle{\apj}, 671, 1921}

\bibitem[{{Fraija} {et~al.}(2020){Fraija}, {Veres}, {Beniamini},
  {Galvan-Gamez}, {Metzger}, {Barniol Duran}, \& {Becerra}}]{Fraija2020}
{Fraija}, N., {Veres}, P., {Beniamini}, P., {et~al.} 2020, \JournalTitle{arXiv
  e-prints}, arXiv:2003.11252

\bibitem[{{Gao} {et~al.}(2013){Gao}, {Lei}, {Zou}, {Wu}, \& {Zhang}}]{gao13}
{Gao}, H., {Lei}, W.-H., {Zou}, Y.-C., {Wu}, X.-F., \& {Zhang}, B. 2013,
  \href{http://dx.doi.org/10.1016/j.newar.2013.10.001}{\JournalTitle{\nar}, 57,
  141}

\bibitem[{{Gehrels} {et~al.}(2004){Gehrels}, {Chincarini}, {Giommi}, {Mason},
  {Nousek}, {Wells}, {White}, {Barthelmy}, {Burrows}, {Cominsky}, {Hurley},
  {Marshall}, {M{\'e}sz{\'a}ros}, {Roming}, {Angelini}, {Barbier}, {Belloni},
  {Campana}, {Caraveo}, {Chester}, {Citterio}, {Cline}, {Cropper}, {Cummings},
  {Dean}, {Feigelson}, {Fenimore}, {Frail}, {Fruchter}, {Garmire}, {Gendreau},
  {Ghisellini}, {Greiner}, {Hill}, {Hunsberger}, {Krimm}, {Kulkarni}, {Kumar},
  {Lebrun}, {Lloyd-Ronning}, {Markwardt}, {Mattson}, {Mushotzky}, {Norris},
  {Osborne}, {Paczynski}, {Palmer}, {Park}, {Parsons}, {Paul}, {Rees},
  {Reynolds}, {Rhoads}, {Sasseen}, {Schaefer}, {Short}, {Smale}, {Smith},
  {Stella}, {Tagliaferri}, {Takahashi}, {Tashiro}, {Townsley}, {Tueller},
  {Turner}, {Vietri}, {Voges}, {Ward}, {Willingale}, {Zerbi}, \&
  {Zhang}}]{Gehrels2004}
{Gehrels}, N., {Chincarini}, G., {Giommi}, P., {et~al.} 2004,
  \href{http://dx.doi.org/10.1086/422091}{\JournalTitle{\apj}, 611, 1005}

\bibitem[{{Goodman}(1986)}]{g86}
{Goodman}, J. 1986,
  \href{http://dx.doi.org/10.1086/184741}{\JournalTitle{\apjl}, 308, L47}

\bibitem[{{Granot} \& {Sari}(2002)}]{Granot+02}
{Granot}, J., \& {Sari}, R. 2002,
  \href{http://dx.doi.org/10.1086/338966}{\JournalTitle{\apj}, 568, 820}

\bibitem[{{Katz} \& {Piran}(1997)}]{Katz+97}
{Katz}, J.~I., \& {Piran}, T. 1997,
  \href{http://dx.doi.org/10.1086/304913}{\JournalTitle{\apj}, 490, 772}

\bibitem[{{Kouveliotou} {et~al.}(1993){Kouveliotou}, {Meegan}, {Fishman},
  {Bhat}, {Briggs}, {Koshut}, {Paciesas}, \& {Pendleton}}]{kouveliotou93}
{Kouveliotou}, C., {Meegan}, C.~A., {Fishman}, G.~J., {et~al.} 1993,
  \href{http://dx.doi.org/10.1086/186969}{\JournalTitle{\apjl}, 413, L101}

\bibitem[{{Kumar} {et~al.}(2008){Kumar}, {Narayan}, \& {Johnson}}]{Kumar2008}
{Kumar}, P., {Narayan}, R., \& {Johnson}, J.~L. 2008,
  \href{http://dx.doi.org/10.1126/science.1159003}{\JournalTitle{Science}, 321,
  376}

\bibitem[{{Kumar} \& {Panaitescu}(2000)}]{Kumar+00}
{Kumar}, P., \& {Panaitescu}, A. 2000,
  \href{http://dx.doi.org/10.1086/312905}{\JournalTitle{\apjl}, 541, L51}

\bibitem[{{Levan} {et~al.}(2007){Levan}, {Jakobsson}, {Hurkett}, {Tanvir},
  {Gorosabel}, {Vreeswijk}, {Rol}, {Chapman}, {Gehrels}, {O'Brien}, {Osborne},
  {Priddey}, {Kouveliotou}, {Starling}, {vanden Berk}, \&
  {Wiersema}}]{Levan2007}
{Levan}, A.~J., {Jakobsson}, P., {Hurkett}, C., {et~al.} 2007,
  \href{http://dx.doi.org/10.1111/j.1365-2966.2007.11879.x}{\JournalTitle{\mnras},
  378, 1439}

\bibitem[{{Levan} {et~al.}(2014){Levan}, {Tanvir}, {Starling}, {Wiersema},
  {Page}, {Perley}, {Schulze}, {Wynn}, {Chornock}, {Hjorth}, {Cenko},
  {Fruchter}, {O'Brien}, {Brown}, {Tunnicliffe}, {Malesani}, {Jakobsson},
  {Watson}, {Berger}, {Bersier}, {Cobb}, {Covino}, {Cucchiara}, {de Ugarte
  Postigo}, {Fox}, {Gal-Yam}, {Goldoni}, {Gorosabel}, {Kaper}, {Kr{\"u}hler},
  {Karjalainen}, {Osborne}, {Pian}, {S{\'a}nchez-Ram{\'{\i}}rez}, {Schmidt},
  {Skillen}, {Tagliaferri}, {Th{\"o}ne}, {Vaduvescu}, {Wijers}, \&
  {Zauderer}}]{levan14}
{Levan}, A.~J., {Tanvir}, N.~R., {Starling}, R.~L.~C., {et~al.} 2014,
  \href{http://dx.doi.org/10.1088/0004-637X/781/1/13}{\JournalTitle{\apj}, 781,
  13}

\bibitem[{{Liang} \& {Zhang}(2006)}]{liang06}
{Liang}, E., \& {Zhang}, B. 2006,
  \href{http://dx.doi.org/10.1111/j.1745-3933.2006.00169.x}{\JournalTitle{\mnras},
  369, L37}

\bibitem[{{Liang} {et~al.}(2007){Liang}, {Zhang}, \& {Zhang}}]{liang2007}
{Liang}, E.-W., {Zhang}, B.-B., \& {Zhang}, B. 2007,
  \href{http://dx.doi.org/10.1086/521870}{\JournalTitle{\apj}, 670, 565}

\bibitem[{{L{\"u}} \& {Zhang}(2014)}]{lu2014}
{L{\"u}}, H.-J., \& {Zhang}, B. 2014,
  \href{http://dx.doi.org/10.1088/0004-637X/785/1/74}{\JournalTitle{\apj}, 785,
  74}

\bibitem[{{L{\"u}} {et~al.}(2015){L{\"u}}, {Zhang}, {Wei-Hua}, {Li}, \&
  {Lasky}}]{lu2015}
{L{\"u}}, H.-J., {Zhang}, B., {Wei-Hua}, L., {Li}, Y., \& {Lasky}, P.-D. 2015,
  \href{http://dx.doi.org/10.1088/0004-637X/805/2/89}{\JournalTitle{\apj}, 805,
  89}

\bibitem[{{MacFadyen}(2001)}]{macfadyen01}
{MacFadyen}, A.~I. 2001, \href{http://dx.doi.org/10.1063/1.1368288}{in
  {American Institute of Physics Conference Series}, Vol. 556, {Explosive
  Phenomena in Astrophysical Compact Objects}, ed. H.-Y. {Chang}, C.-H. {Lee},
  M.~{Rho}, \& I.~{Yi}}, 313

\bibitem[{{M\'esz\'aros}(1998)}]{Meszaros1998}
{M\'esz\'aros}, P.~{Rees}, J. 1998,
  \href{http://dx.doi.org/10.1046/j.1365-8711.1998.01992.x}{\JournalTitle{MNRAS},
  299, L10}

\bibitem[{{M{\'e}sz{\'a}ros}(2002)}]{m02}
{M{\'e}sz{\'a}ros}, P. 2002,
  \href{http://dx.doi.org/10.1146/annurev.astro.40.060401.093821}{\JournalTitle{\araa},
  40, 137}

\bibitem[{Meszaros \& Rees(1997)}]{meszaros97}
Meszaros, P., \& Rees, M.~J. 1997,
  \href{http://dx.doi.org/10.1086/303625}{\JournalTitle{The Astrophysical
  Journal}, 476, 232}

\bibitem[{{Metzger} {et~al.}(2018){Metzger}, {Beniamini}, \&
  {Giannios}}]{Metzger2018}
{Metzger}, B.~D., {Beniamini}, P., \& {Giannios}, D. 2018,
  \href{http://dx.doi.org/10.3847/1538-4357/aab70c}{\JournalTitle{\apj}, 857,
  95}

\bibitem[{{Nakauchi} {et~al.}(2013){Nakauchi}, {Kashiyama}, {Suwa}, \&
  {Nakamura}}]{Nakauchi2013}
{Nakauchi}, D., {Kashiyama}, K., {Suwa}, Y., \& {Nakamura}, T. 2013,
  \href{http://dx.doi.org/10.1088/0004-637X/778/1/67}{\JournalTitle{\apj}, 778,
  67}

\bibitem[{{Norris} \& {Bonnell}(2006)}]{Norris:06}
{Norris}, J.~P., \& {Bonnell}, J.~T. 2006,
  \href{http://dx.doi.org/10.1086/502796}{\JournalTitle{\apj}, 643, 266}

\bibitem[{{Norris} {et~al.}(2010){Norris}, {Gehrels}, \&
  {Scargle}}]{Norris2010}
{Norris}, J.~P., {Gehrels}, N., \& {Scargle}, J.~D. 2010,
  \href{http://dx.doi.org/10.1088/0004-637X/717/1/411}{\JournalTitle{\apj},
  717, 411}

\bibitem[{{Nousek} {et~al.}(2006){Nousek}, {Kouveliotou}, {Grupe}, {Page},
  {Granot}, {Ramirez-Ruiz}, {Patel}, {Burrows}, {Mangano}, {Barthelmy},
  {Beardmore}, {Campana}, {Capalbi}, {Chincarini}, {Cusumano}, {Falcone},
  {Gehrels}, {Giommi}, {Goad}, {Godet}, {Hurkett}, {Kennea}, {Moretti},
  {O'Brien}, {Osborne}, {Romano}, {Tagliaferri}, \& {Wells}}]{Nousek2006}
{Nousek}, J.~A., {Kouveliotou}, C., {Grupe}, D., {et~al.} 2006,
  \href{http://dx.doi.org/10.1086/500724}{\JournalTitle{\apj}, 642, 389}

\bibitem[{{O'Brien} {et~al.}(2006){O'Brien}, {Willingale}, {Osborne}, {Goad},
  {Page}, {Vaughan}, {Rol}, {Beardmore}, {Godet}, {Hurkett}, {Wells}, {Zhang},
  {Kobayashi}, {Burrows}, {Nousek}, {Kennea}, {Falcone}, {Grupe}, {Gehrels},
  {Barthelmy}, {Cannizzo}, {Cummings}, {Hill}, {Krimm}, {Chincarini},
  {Tagliaferri}, {Campana}, {Moretti}, {Giommi}, {Perri}, {Mangano}, \&
  {LaParola}}]{OBrien2006}
{O'Brien}, P.~T., {Willingale}, R., {Osborne}, J., {et~al.} 2006,
  \href{http://dx.doi.org/10.1086/505457}{\JournalTitle{\apj}, 647, 1213}

\bibitem[{{Paczynski}(1986)}]{paczynski86}
{Paczynski}, B. 1986,
  \href{http://dx.doi.org/10.1086/184740}{\JournalTitle{\apjl}, 308, L43}

\bibitem[{{Paczynski} \& {Rhoads}(1993)}]{Paczynski1993}
{Paczynski}, B., \& {Rhoads}, J.~E. 1993,
  \href{http://dx.doi.org/10.1086/187102}{\JournalTitle{\apjl}, 418, L5}

\bibitem[{{Piran}(2004)}]{piran04}
{Piran}, T. 2004,
  \href{http://dx.doi.org/10.1103/RevModPhys.76.1143}{\JournalTitle{Reviews of
  Modern Physics}, 76, 1143}

\bibitem[{{Qin} {et~al.}(2004){Qin}, {Zhang}, {Zhang}, \& {Cui}}]{Qin2004}
{Qin}, Y.-P., {Zhang}, Z.-B., {Zhang}, F.-W., \& {Cui}, X.-H. 2004,
  \href{http://dx.doi.org/10.1086/425335}{\JournalTitle{\apj}, 617, 439}

\bibitem[{{Racusin} {et~al.}(2009){Racusin}, {Liang}, {Burrows}, {Falcone},
  {Sakamoto}, {Zhang}, {Zhang}, {Evans}, \& {Osborne}}]{Racusin+09}
{Racusin}, J.~L., {Liang}, E.~W., {Burrows}, D.~N., {et~al.} 2009,
  \href{http://dx.doi.org/10.1088/0004-637X/698/1/43}{\JournalTitle{\apj}, 698,
  43}

\bibitem[{{Rea} {et~al.}(2015){Rea}, {Gull{\'o}n}, {Pons}, {Perna}, {Dainotti},
  {Miralles}, \& {Torres}}]{rea15}
{Rea}, N., {Gull{\'o}n}, M., {Pons}, J.~A., {et~al.} 2015,
  \href{http://dx.doi.org/10.1088/0004-637X/813/2/92}{\JournalTitle{\apj}, 813,
  92}

\bibitem[{{Rees} \& {M{\'e}sz{\'a}ros}(1998)}]{rees98}
{Rees}, M.~J., \& {M{\'e}sz{\'a}ros}, P. 1998,
  \href{http://dx.doi.org/10.1086/311244}{\JournalTitle{\apjl}, 496, L1}

\bibitem[{{Rhoads}(1999)}]{Rhoads99}
{Rhoads}, J.~E. 1999,
  \href{http://dx.doi.org/10.1086/307907}{\JournalTitle{\apj}, 525, 737}

\bibitem[{{Rodney} {et~al.}(2015){Rodney}, {Riess}, {Scolnic}, {Jones},
  {Hemmati}, {Molino}, {McCully}, {Mobasher}, {Strolger}, {Graur}, {Hayden}, \&
  {Casertano}}]{Rodney2015}
{Rodney}, S.~A., {Riess}, A.~G., {Scolnic}, D.~M., {et~al.} 2015,
  \href{http://dx.doi.org/10.1088/0004-6256/150/5/156}{\JournalTitle{\aj}, 150,
  156}

\bibitem[{{Roming} {et~al.}(2005){Roming}, {Kennedy}, {Mason}, {Nousek}, {Ahr},
  {Bingham}, {Broos}, {Carter}, {Hancock}, {Huckle}, {Hunsberger}, {Kawakami},
  {Killough}, {Koch}, {McLelland}, {Smith}, {Smith}, {Soto}, {Boyd},
  {Breeveld}, {Holland}, {Ivanushkina}, {Pryzby}, {Still}, \&
  {Stock}}]{2005SSRv..120...95R}
{Roming}, P.~W.~A., {Kennedy}, T.~E., {Mason}, K.~O., {et~al.} 2005,
  \href{http://dx.doi.org/10.1007/s11214-005-5095-4}{\JournalTitle{Space
  Science Reviews}, 120, 95}

\bibitem[{{Rowlinson} {et~al.}(2014){Rowlinson}, {Gompertz}, {Dainotti},
  {O'Brien}, {Wijers}, \& {van der Horst}}]{rowlinson14}
{Rowlinson}, A., {Gompertz}, B.~P., {Dainotti}, M., {et~al.} 2014,
  \href{http://dx.doi.org/10.1093/mnras/stu1277}{\JournalTitle{\mnras}, 443,
  1779}

\bibitem[{{Rowlinson} {et~al.}(2013){Rowlinson}, {O'Brien}, {Metzger},
  {Tanvir}, \& {Levan}}]{rowlinson2013}
{Rowlinson}, A., {O'Brien}, P.~T., {Metzger}, B.~D., {Tanvir}, N.~R., \&
  {Levan}, A.~J. 2013,
  \href{http://dx.doi.org/10.1093/mnras/sts683}{\JournalTitle{\mnras}, 430,
  1061}

\bibitem[{{Sakamoto} {et~al.}(2007){Sakamoto}, {Hill}, {Yamazaki}, {Angelini},
  {Krimm}, {Sato}, {Swindell}, {Takami}, \& {Osborne}}]{sakamoto07}
{Sakamoto}, T., {Hill}, J.~E., {Yamazaki}, R., {et~al.} 2007,
  \href{http://dx.doi.org/10.1086/521640}{\JournalTitle{\apj}, 669, 1115}

\bibitem[{{Sakamoto} {et~al.}(2011){Sakamoto}, {Barthelmy}, {Baumgartner},
  {Cummings}, {Fenimore}, {Gehrels}, {Krimm}, {Markwardt}, {Palmer}, {Parsons},
  {Sato}, {Stamatikos}, {Tueller}, {Ukwatta}, \& {Zhang}}]{Sakamoto+11}
{Sakamoto}, T., {Barthelmy}, S.~D., {Baumgartner}, W.~H., {et~al.} 2011,
  \href{http://dx.doi.org/10.1088/0067-0049/195/1/2}{\JournalTitle{\apjs}, 195,
  2}

\bibitem[{{Sari} \& {M{\'e}sz{\'a}ros}(2000)}]{sari2000}
{Sari}, R., \& {M{\'e}sz{\'a}ros}, P. 2000,
  \href{http://dx.doi.org/10.1086/312689}{\JournalTitle{\apjl}, 535, L33}

\bibitem[{{Sari} \& {Piran}(1999)}]{sari99}
{Sari}, R., \& {Piran}, T. 1999,
  \href{http://dx.doi.org/10.1086/312039}{\JournalTitle{\apjl}, 517, L109}

\bibitem[{{Sari} {et~al.}(1998){Sari}, {Piran}, \& {Narayan}}]{Sari+98}
{Sari}, R., {Piran}, T., \& {Narayan}, R. 1998,
  \href{http://dx.doi.org/10.1086/311269}{\JournalTitle{\apjl}, 497, L17}

\bibitem[{{Schaefer}(2007)}]{schaefer2007}
{Schaefer}, B.~E. 2007,
  \href{http://dx.doi.org/10.1086/511742}{\JournalTitle{\apj}, 660, 16}

\bibitem[{{Stratta} {et~al.}(2018){Stratta}, {Dainotti}, {Dall'Osso},
  {Hernandez}, \& {De Cesare, G.}}]{Stratta2018}
{Stratta}, G., {Dainotti}, M., {Dall'Osso}, S., {Hernandez}, X., \& {De Cesare,
  G.} 2018,
  \href{http://dx.doi.org/10.3847/1538-4357/aadd8f}{\JournalTitle{\apj}, 859,
  155}

\bibitem[{{Stratta} {et~al.}(2013){Stratta}, B., {Atteia}, {Boer}, {Coward},
  {De Pasquale}, {Howell}, {Klotz}, {Oates}, \& L.}]{Stratta2013}
{Stratta}, G., B., G., {Atteia}, J.~L., {et~al.} 2013,
  \href{http://dx.doi.org/10.1088/0004-637X/779/1/66}{\JournalTitle{\apj}, 779,
  66}

\bibitem[{{Tagliaferri} {et~al.}(2005){Tagliaferri}, {Goad}, {Chincarini},
  {Moretti}, {Campana}, {Burrows}, {Perri}, {Barthelmy}, {Gehrels}, {Krimm},
  {Sakamoto}, {Kumar}, {M{\'e}sz{\'a}ros}, {Kobayashi}, {Zhang}, {Angelini},
  {Banat}, {Beardmore}, {Capalbi}, {Covino}, {Cusumano}, {Giommi}, {Godet},
  {Hill}, {Kennea}, {Mangano}, {Morris}, {Nousek}, {O'Brien}, {Osborne},
  {Pagani}, {Page}, {Romano}, {Stella}, \& {Wells}}]{Tagliaferri2005}
{Tagliaferri}, G., {Goad}, M., {Chincarini}, G., {et~al.} 2005,
  \href{http://dx.doi.org/10.1038/nature03934}{\JournalTitle{\nat}, 436, 985}

\bibitem[{{Toma} {et~al.}(2007){Toma}, {Ioka}, {Sakamoto}, \&
  {Nakamura}}]{Toma2007}
{Toma}, K., {Ioka}, K., {Sakamoto}, T., \& {Nakamura}, T. 2007,
  \href{http://dx.doi.org/10.1086/512481}{\JournalTitle{\apj}, 659, 1420}

\bibitem[{{Troja} {et~al.}(2007){Troja}, {Cusumano}, {O'Brien}, {Zhang},
  {Sbarufatti}, {Mangano}, {Willingale}, {Chincarini}, {Osborne}, {Marshall},
  {Burrows}, {Campana}, {Gehrels}, {Guidorzi}, {Krimm}, {La Parola}, {Liang},
  {Mineo}, {Moretti}, {Page}, {Romano}, {Tagliaferri}, {Zhang}, {Page}, \&
  {Schady}}]{troja07}
{Troja}, E., {Cusumano}, G., {O'Brien}, P.~T., {et~al.} 2007,
  \href{http://dx.doi.org/10.1086/519450}{\JournalTitle{\apj}, 665, 599}

\bibitem[{{Uhm} \& {Zhang}(2014)}]{Uhm13}
{Uhm}, Z.~L., \& {Zhang}, B. 2014,
  \href{http://dx.doi.org/10.1088/0004-637X/780/1/82}{\JournalTitle{\apj}, 780,
  82}

\bibitem[{{Wang} {et~al.}(2015){Wang}, {Zhang}, {Liang}, {Gao}, {Li}, {Deng},
  {Qin}, {Tang}, {Kann}, {Ryde}, \& {Kumar}}]{wang15}
{Wang}, X.~G., {Zhang}, B., {Liang}, E.~W., {et~al.} 2015,
  \href{http://dx.doi.org/10.1088/0067-0049/219/1/9}{\JournalTitle{\apjs}, 219,
  9}

\bibitem[{{Warren} {et~al.}(2017){Warren}, {Ellison}, {Barkov}, \&
  {Nagataki}}]{2017ApJ...835..248W}
{Warren}, D.~C., {Ellison}, D.~C., {Barkov}, M.~V., \& {Nagataki}, S. 2017,
  \href{http://dx.doi.org/10.3847/1538-4357/aa56c3}{\JournalTitle{\apj}, 835,
  248}

\bibitem[{{Willingale} {et~al.}(2007){Willingale}, {O'Brien}, {Osborne},
  {Godet}, {Page}, {Goad}, {Burrows}, {Zhang}, {Rol}, {Gehrels}, \&
  {Chincarini}}]{Willingale2007}
{Willingale}, R., {O'Brien}, P.~T., {Osborne}, J.~P., {et~al.} 2007,
  \href{http://dx.doi.org/10.1086/517989}{\JournalTitle{\apj}, 662, 1093}

\bibitem[{{Xiao} \& {Schaefer}(2009)}]{xiao09}
{Xiao}, L., \& {Schaefer}, B.~E. 2009,
  \href{http://dx.doi.org/10.1088/0004-637X/707/1/387}{\JournalTitle{\apj},
  707, 387}

\bibitem[{{Zhang} {et~al.}(2006{\natexlab{a}}){Zhang}, {Fan}, {Dyks},
  {Kobayashi}, {M{\'e}sz{\'a}ros}, {Burrows}, {Nousek}, \&
  {Gehrels}}]{Zhang2006}
{Zhang}, B., {Fan}, Y.~Z., {Dyks}, J., {et~al.} 2006{\natexlab{a}},
  \href{http://dx.doi.org/10.1086/500723}{\JournalTitle{\apj}, 642, 354}

\bibitem[{{Zhang} \& {M{\'e}sz{\'a}ros}(2001)}]{zhang2001}
{Zhang}, B., \& {M{\'e}sz{\'a}ros}, P. 2001,
  \href{http://dx.doi.org/10.1086/320255}{\JournalTitle{\apjl}, 552, L35}

\bibitem[{{Zhang} \& {M{\'e}sz{\'a}ros}(2004)}]{Zhang2004}
---. 2004,
  \href{http://dx.doi.org/10.1142/S0217751X0401746X}{\JournalTitle{International
  Journal of Modern Physics A}, 19, 2385}

\bibitem[{{Zhang} {et~al.}(2007){Zhang}, {Liang}, {Page}, {Grupe}, {Zhang},
  {Barthelmy}, {Burrows}, {Campana}, {Chincarini}, {Gehrels}, {Kobayashi},
  {M{\'e}sz{\'a}ros}, {Moretti}, {Nousek}, {O'Brien}, {Osborne}, {Roming},
  {Sakamoto}, {Schady}, \& {Willingale}}]{zhang07b}
{Zhang}, B., {Liang}, E., {Page}, K.~L., {et~al.} 2007,
  \href{http://dx.doi.org/10.1086/510110}{\JournalTitle{\apj}, 655, 989}

\bibitem[{{Zhang} {et~al.}(2006{\natexlab{b}}){Zhang}, {Xie}, {Deng}, \&
  {Jin}}]{zhang06}
{Zhang}, Z., {Xie}, G.~Z., {Deng}, J.~G., \& {Jin}, W. 2006{\natexlab{b}},
  \href{http://dx.doi.org/10.1111/j.1365-2966.2006.11058.x}{\JournalTitle{\mnras},
  373, 729}

\bibitem[{Zhao {et~al.}(2019)Zhao, Zhang, Gao, Lan, L{\"u}, \&
  Zhang}]{Zhang2019}
Zhao, L., Zhang, B., Gao, H., {et~al.} 2019,
  \href{http://dx.doi.org/10.3847/1538-4357/ab38c4}{\JournalTitle{The
  Astrophysical Journal}, 883, 97}

\end{thebibliography}
\bibliographystyle{yahapj}

\typeout{get arXiv to do 4 passes: Label(s) may have changed. Rerun}
\end{document}